\documentclass[english,notitlepage,nofootinbib]{revtex4-1}
\usepackage[T1]{fontenc}
\usepackage[latin9]{inputenc}
\usepackage{babel}
\usepackage{amsmath}
\usepackage{amssymb}
\usepackage{graphicx}
\PassOptionsToPackage{normalem}{ulem}
\usepackage{ulem}
\usepackage[unicode=true,
 bookmarks=false,
 breaklinks=false,pdfborder={0 0 1},backref=false,colorlinks=false]
 {hyperref}

\makeatletter

\usepackage{babel}

\usepackage[]{ulem}

\usepackage[]{ulem}

\makeatother

\begin{document}
\title{Exclusive photoproduction of $D$-meson pairs with large invariant
mass}
\author{Marat Siddikov, Iván Schmidt}
\affiliation{Departamento de Física, Universidad Técnica Federico Santa María,~~~~~~~~\\
 y Centro Científico - Tecnológico de Valparaíso, Casilla 110-V, Valparaíso,
Chile}
\begin{abstract}
In this paper we analyze the exclusive photoproduction of the $D$-meson
pairs with large invariant mass. We perform evaluations in the collinear
factorization framework and in the leading order of the strong coupling
$\alpha_{s},$ expressing the cross-section in terms of generalized
parton distributions (GPDs) of different parton flavors in the proton.
We focus on the photoproduction of the pseudoscalar-vector pairs,
like \emph{e.g}. $D^{\pm}D^{*\mp}$, $D^{0}\overline{D}^{*0}$, $D_{s}^{+}D_{s}^{*-}$,
which gets the dominant contribution from the chiral even GPDs of
the target, and estimate the cross-section in the kinematics of the
future Electron Ion Collider (EIC). In all channels the amplitude
of the process obtains comparable contributions from gluons and only
one of the light quark flavors. This finding signals that the process
potentially could be used to single out the contributions of the individual
chiral even GPDs of light flavors. We found that the process is mostly
sensitive to the behavior of GPDs in the so-called Efremov-Radyushkin-Brodsky-Lepage
(ERBL) region. Numerically, the cross-section of the process is sufficiently
large for experimental studies and thus can be used as a complementary
probe for studies of the GPDs.
\end{abstract}
\maketitle

\section{Introduction}

In the last three decades the Generalized Parton Distributions (GPDs)
of the nucleon turned into a standardized tool to encode information
about the nonperturbative interactions of individual partons in the
hadronic target~\cite{Diehl:2000xz,Goeke:2001tz,Diehl:2003ny,Guidal:2013rya,Boer:2011fh,Burkert:2022hjz},
and for this reason have been in the center of theoretical and experimental
studies. The GPDs allow to understand the contributions of different
parton flavors to various observables which characterize the hadronic
target. At present it is not possible to evaluate the GPDs directly
from first principles, and for this reason studies of these objects
rely on phenomenological extractions from experimental data or results
of lattice simulations~\cite{Egerer:2021ymv,Karpie:2021pap,Bhattacharya:2022aob,Bhattacharya:2023ays,Bhattacharya:2023nmv}.
However, the existing lattice studies, due to technical challenges,
at present mostly focus on the special zero-skewedness ($\xi=0$)
limit and studies of some moments of GPDs, whereas phenomenological
extractions suffer from various uncertainties, even for the cleanest
and best understood channels~\cite{Kumericki:2016ehc}. This motivates
the search for new processes which could be used for extractions of
the GPDs~\cite{Pire:2015iza,Pire:2017lfj,Pire:2017yge,Pire:2021dad}.

The expected high-luminosity experiments at the future Electron Ion
Collider stimulated interest in various channels, which were previously
disregarded due to the smallness of the cross-section. For studies
of GPDs, a special interest is present in exclusive $2\to3$ processes,
which have already been analyzed in the literature~\cite{GPD2x3:9,GPD2x3:8,GPD2x3:7,GPD2x3:6,GPD2x3:5,GPD2x3:4,GPD2x3:3,GPD2x3:2,GPD2x3:1,Duplancic:2022wqn,ElBeiyad:2010pji,Boussarie:2016qop,Pedrak:2017cpp,Pedrak:2020mfm}.
Due to their different kinematic structure, these new probes could
complement existing studies and provide new independent constraints
on existing phenomenological models of GPDs. For amplitudes of such
processes, the factorization theorem has been proven in the kinematics
case when all the produced hadrons are well-separated kinematically,
i.e. the pairwise invariant masses ($\approx$relative velocities
of the produced hadrons) are sufficiently large to avoid soft final-state
interactions~\cite{GPD2x3:10,GPD2x3:11}. Most of these studies focused
on the production of pairs of light mesons and photons, and it has
been discussed in detail how these novel channels could help to access
new information about the GPDs~\cite{LehmannDronke:1999vvq,LehmannDronke:2000hlo,ElBeiyad:2010pji,Boussarie:2016qop,Clerbaux:2000hb}.

Potentially the production of the heavier meson pairs, like $D$-mesons
and quarkonia, might be also used for the same purpose: in the kinematics
where factorization theorems are applicable, the overall cross-section
suppression by the heavy quark mass is comparable to the suppression
by a large invariant mass. However, for these mesons the theoretical
treatment should be adjusted, since the heavy quark masses cannot
be disregarded, and should be considered as one of the heavy scales
in the problem. This breaks the conventional twist suppression used
for light quarks, and leads to new probes of the GPDs (compared to
light meson production channels with the same quantum numbers). Furthermore,
due to lack of substantial contributions from intrinsic heavy flavors
in the proton, the heavy meson production channels might be used to
disentangle the flavor structure of the GPDs, thus avoiding the usual
superposition of GPDs of all light flavors. For heavy quarkonia pair
production, this allows to single out the contribution of the gluon
GPDs~\cite{Goncalves:2015sfy,Goncalves:2019txs,Goncalves:2006hu,Baranov:2012vu,Yang:2020xkl,Goncalves:2016ybl,Andrade:2022rbn,Siddikov:2022bku}.
For the $D$-mesons, the cross-sections get an additional contribution
from one of the light flavors~\cite{Pire:2015iza,Pire:2017lfj,Pire:2017yge,Pire:2021dad},
which potentially allows to test individually the GPDs of light flavors.
In this paper we will focus on the production of scalar-vector mesons
pairs, like $D^{+}D^{*-}$,~$D^{0}\overline{D}^{*0}$ and $D_{s}^{+}D_{s}^{*-}$
which has not been discussed so far in the literature and might be
used as a complementary probe of the target GPDs. This choice of quantum
numbers also allows to avoid the contribution of the photon-photon
fusion mechanism, which has been discussed previously in~\cite{Luszczak:2011js},
and the contribution of the poorly known chiral odd transversity GPDs
of the target, discussed in~\cite{ElBeiyad:2010pji,Boussarie:2016qop}.
We will analyze this process in the conventional collinear factorization
approach, although maintaining the heavy quark mass as a hard scale,
and will provide numerical predictions in the kinematics of low- and
middle-energy electron-proton collisions at the forthcoming Electron
Ion Collider (EIC)~\cite{Accardi:2012qut,DOEPR,BNLPR,AbdulKhalek:2021gbh}.

The paper is structured as follows. In the next Section~\ref{sec:Formalism}
we introduce the framework and provide analytical expressions for
the amplitude and cross-section of the process. In Section~\ref{sec:Numer}
we estimate numerically the cross-sections, using publicly available
parametrizations of the proton GPDs and $D$-meson distribution amplitudes.
Finally, in Section~\ref{sec:Conclusions} we draw conclusions.

\section{Exclusive photoproduction of meson pairs}

\label{sec:Formalism}

Below, in Section~\ref{subsec:Kinematics}, we define the kinematic
variables, discuss their typical ranges in EIC kinematics and introduce
the light-cone decomposition for momenta of all particles, which will
be used later. In the next Section~\ref{subsec:Amplitudes}, we evaluate
analytically the amplitude of the process in the collinear factorization
framework.

\subsection{Kinematics of the process}

\label{subsec:Kinematics} For our evaluations we will use the photon-proton
collision frame, in which the photon and proton move along the axis
$z$, so the light-cone decomposition of their momenta are given by

\begin{align}
q & =\,\left(-\frac{Q^{2}}{2q^{-}},\,q^{-},\,\,\boldsymbol{0}_{\perp}\right),\quad q^{-}=E_{\gamma}+\sqrt{E_{\gamma}^{2}+Q^{2}}\label{eq:qPhoton-2}\\
P & =\left(P^{+},\,\frac{m_{N}^{2}}{2P^{+}},\,\,\boldsymbol{0}_{\perp}\right),\quad P^{+}=E_{p}+\sqrt{E_{p}^{2}-m_{N}^{2}}\label{eq:PMomentum}
\end{align}
where the shorthand notation $q$ stands for the momentum of the photon,
$Q^{2}=-q^{2}$ is its virtuality, and $P$, $P'$, are the proton
momenta before and after the collision. For the production of light
mesons and meson pairs the analysis is frequently done in the so-called
Bjorken kinematics~\cite{LehmannDronke:1999vvq,LehmannDronke:2000hlo,Clerbaux:2000hb,Diehl:1999cg,ZEUS:1998xpo},
when the hard scale is set by the photon virtuality $Q^{2}$, exceeding
significantly the nucleon mass $m_{N}^{2}$, as well as all light
quark masses. For processes involving heavy mesons this regime is
hardly achievable experimentally, due to the rapid decrease of the
flux of equivalent photons as a function of $Q$. Furthermore, due
to the expected smallness of the cross-sections, most of the detected
events will proceed via quasi-real photons with $Q\approx0$. For
the sake of generality, till the end of this section we will keep
$Q\not=0$, assuming that the virtuality $Q$ is bound by $\Lambda_{{\rm QCD}}^{2}\ll Q^{2}\lesssim m_{Q}^{2}$,
although eventually we'll take $Q=0$ in the numerical estimates.
Since the spectrum of equivalent quasi-real photons emitted from the
electron falls off rapidly as a function of the transverse photon
momentum, at high energies the photon-proton frame should be close
to the laboratory frame, in which case the electron-proton collision
axis points in the direction of the axis $\hat{z}$. In the limit
$Q\to0$, this frame, up to a trivial longitudinal boost~\footnote{In the $Q=0$ limit the correspondence with previous papers~\cite{GPD2x3:9,GPD2x3:8,GPD2x3:7,GPD2x3:6,GPD2x3:5,GPD2x3:4,GPD2x3:3,GPD2x3:2,GPD2x3:1,Duplancic:2022wqn}
might be achieved making longitudinal boost of all vectors $k^{\mu}$
as $k^{+}\to k^{+}\Lambda$, $k^{-}\to k^{-}/\Lambda,$ where $\Lambda=q^{-}\sqrt{2/s}$,
and substituting $P^{+}\to s(1+\xi)/\left(2q^{-}\right)$, where $\xi$
is the skewedness variable and $s=\left(W^{2}-m_{N}^{2}\right)/(1+\xi)$.
We prefer to maintain our notations, in order to have a better understanding
how the different observables behave as functions of the lab-frame
rapidities $y_{1},y_{2}$, defined in~(\ref{eq:MesonLC-2}).}, coincides with the frame used in earlier studies of exclusive photoproduction
$\gamma p\to\gamma Mp$~\cite{GPD2x3:9,GPD2x3:8,GPD2x3:7,GPD2x3:6,GPD2x3:5,GPD2x3:4,GPD2x3:3,GPD2x3:2,GPD2x3:1,Duplancic:2022wqn}.
In this frame, the polarization vectors of the virtual photons may
be chosen as 
\begin{equation}
\varepsilon_{L}=\left(\frac{Q}{q^{-}},\,0,\,\boldsymbol{0}_{\perp}\right),\quad\varepsilon_{T}^{(\pm)}=\left(0,\,0,\frac{1}{\sqrt{2}},\pm\frac{i}{\sqrt{2}}\right).\label{eq:PolVector}
\end{equation}
for the longitudinal and transverse polarizations respectively. The
4-momenta $p_{1},\,p_{2}$ of the produced heavy $D$-mesons can be
parametrized in terms of the rapidities $y_{a}$ and transverse momenta
$\boldsymbol{p}_{a}^{\perp}$ of these heavy mesons as 
\begin{align}
p_{a} & =\left(\frac{M_{a}^{\perp}}{2}\,e^{-y_{a}}\,,\,M_{a}^{\perp}e^{y_{a}},\,\,\boldsymbol{p}_{a}^{\perp}\right),\quad M_{a}^{\perp}\equiv\sqrt{M_{a}^{2}+\left(\boldsymbol{p}_{a}^{\perp}\right)^{2}},\quad a=1,2,\label{eq:MesonLC-2}
\end{align}
where the positive rapidity is chosen in the photon direction. As
we will see below, the cross-section falls rapidly as a function of
transverse momenta; for this reason the dominant contribution to the
cross-section in EIC kinematics comes from the region of relatively
small momenta $p_{a}^{\perp}$. In these notations, the 4-vector $\Delta$
of momentum transfer to the target is given by 
\begin{align}
\Delta & =P'-P=q-p_{1}-p_{2}=\left(\Delta^{+},\,\Delta^{-},\,\,\boldsymbol{\Delta}^{\perp}\right),
\end{align}
\begin{align}
 & \Delta^{+}=-\frac{Q^{2}}{2q^{-}}-\frac{M_{1}^{\perp}e^{-y_{1}}}{2}-\frac{M_{2}^{\perp}e^{-y_{2}}}{2},\quad\Delta^{-}=q^{-}-M_{1}^{\perp}\,e^{y_{1}}-M_{2}^{\perp}\,e^{y_{2}},\quad\boldsymbol{\Delta}_{\perp}=-\boldsymbol{p}_{1}^{\perp}-\boldsymbol{p}_{2}^{\perp}
\end{align}
and the Mandelstam invariant $t\equiv\Delta^{2}$ is parametrized
as 
\begin{align}
t & =\Delta^{2}=-\left(q^{-}-M_{1}^{\perp}\,e^{y_{1}}-M_{2}^{\perp}\,e^{y_{2}}\right)\left(\frac{Q^{2}}{q^{-}}+M_{1}^{\perp}e^{-y_{1}}+M_{2}^{\perp}e^{-y_{2}}\right)-\left(\boldsymbol{p}_{1}^{\perp}+\boldsymbol{p}_{2}^{\perp}\right)^{2}\label{eq:tDef}\\
 & =-Q^{2}+M_{1}^{2}+M_{2}^{2}-q^{-}\left(M_{1}^{\perp}e^{-y_{1}}+M_{2}^{\perp}e^{-y_{2}}\right)+\frac{Q^{2}}{q^{-}}\left(M_{1}^{\perp}\,e^{y_{1}}+M_{2}^{\perp}\,e^{y_{2}}\right)+\nonumber \\
 & +2\left(M_{1}^{\perp}M_{2}^{\perp}\cosh\Delta y-\boldsymbol{p}_{1}^{\perp}\cdot\boldsymbol{p}_{2}^{\perp}\right).\nonumber 
\end{align}
The 4-momentum of the recoil proton 
\begin{equation}
P'=P+\Delta=\left(q^{-}+\frac{m_{N}^{2}}{2P^{+}}-M_{1}^{\perp}\,e^{y_{1}}-M_{2}^{\perp}\,e^{y_{2}}\,,\,P^{+}-\frac{Q^{2}}{2q^{-}}-\frac{M_{1}^{\perp}e^{-y_{1}}+M_{2}^{\perp}e^{-y_{2}}}{2},-\boldsymbol{p}_{1}^{\perp}-\boldsymbol{p}_{2}^{\perp}\right),
\end{equation}
should satisfy the onshellness condition $\left(P+\Delta\right)^{2}=m_{N}^{2}$,
which provides an additional constraint 
\begin{align}
q^{-}P^{+} & =P^{+}\left(M_{1}^{\perp}\,e^{y_{1}}+M_{2}^{\perp}\,e^{y_{2}}\right)-\frac{m_{N}^{2}+t}{2}+\frac{m_{N}^{2}}{4P^{+}}\left(M_{1}^{\perp}e^{-y_{1}}+M_{2}^{\perp}e^{-y_{2}}+\frac{Q^{2}}{q^{+}}\right).\label{qPlus}
\end{align}
The Equation~(\ref{qPlus}) may be solved with respect to $q^{-}$,
yielding 
\begin{align}
q^{-} & =\frac{M_{1}^{\perp}\,e^{y_{1}}+M_{2}^{\perp}\,e^{y_{2}}-\frac{m_{N}^{2}+t}{2P^{+}}+\frac{m_{N}^{2}}{4\left(P^{+}\right)^{2}}\left(M_{1}^{\perp}e^{-y_{1}}+M_{2}^{\perp}e^{-y_{2}}\right)}{2}\pm\label{qPlus-1}\\
 & +\frac{1}{2}\,\sqrt{\left(M_{1}^{\perp}\,e^{y_{1}}+M_{2}^{\perp}\,e^{y_{2}}-\frac{m_{N}^{2}+t}{2P^{+}}+\frac{m_{N}^{2}}{4\left(P^{+}\right)^{2}}\left(M_{1}^{\perp}e^{-y_{1}}+M_{2}^{\perp}e^{-y_{2}}\right)\right)^{2}+\frac{Q^{2}m_{N}^{2}}{\left(P^{+}\right)^{2}}},\nonumber 
\end{align}
which, together with~(\ref{eq:qPhoton-2}), allows to find the energy
of the photon $E_{\gamma}\approx q^{-}/2$ in terms of the kinematic
variables $\left(y_{a},\,\boldsymbol{p}_{a}^{\perp}\right)$ of the
produced $D$-mesons. In terms of these variables, the invariant energy
$W$ of the $\gamma p$ collision and the invariant mass ${\mathcal{M}}_{12}$
of the produced heavy quarkonia pair can be rewritten as 
\begin{equation}
W^{2}\equiv s_{\gamma p}=\left(q+P\right)^{2}=-Q^{2}+m_{N}^{2}+2q\cdot P,\label{eq:W2}
\end{equation}
and 
\begin{align}
{\mathcal{M}}_{12}^{2} & =\left(p_{1}+p_{2}\right)^{2}=M_{1}^{2}+M_{2}^{2}+2\left(M_{1}^{\perp}M_{2}^{\perp}\cosh\Delta y-\boldsymbol{p}_{1}^{\perp}\cdot\boldsymbol{p}_{2}^{\perp}\right)\label{eq:M12}
\end{align}
respectively. In the high-energy limit $q^{-},P^{+}\gg Q,\,M_{a}\gg\{m_{N},\,\sqrt{|t|}\}$,
the results found earlier in this section may be simplified to 
\begin{align}
 & q^{-}\approx M_{1}^{\perp}\,e^{y_{1}}+M_{2}^{\perp}\,e^{y_{2}},\quad t\approx-\left(\boldsymbol{p}_{1}^{\perp}+\boldsymbol{p}_{2}^{\perp}\right)^{2},\quad W^{2}\approx2q^{-}p^{+}.\label{eq:qPlus}
\end{align}
In this kinematics the rapidities $y_{1},\,y_{2}$ and their difference
can be rewritten in terms of the invariant Mandelstam variable(s)
$u_{1},u_{2}$, defined as 
\begin{align}
u_{a} & \equiv\left(p_{a}-P\right)^{2}=m_{N}^{2}+M_{a}^{2}-M_{a}^{\perp}\left(2P^{+}e^{y_{a}}-e^{-y_{a}}\frac{m_{N}^{2}}{2P^{+}}\right)\approx m_{N}^{2}+M_{a}^{2}-2P^{+}M_{a}e^{y_{a}},\quad a=1,2,\\
y_{a} & \approx\ln\left(\frac{m_{N}^{2}+M_{a}^{2}-u_{a}}{2P^{+}M_{a}}\right),\qquad\Delta y=y_{1}-y_{2}\approx\ln\left(\frac{m_{N}^{2}+M_{1}^{2}-u_{1}}{m_{N}^{2}+M_{2}^{2}-u_{2}}\right).
\end{align}

\begin{figure}
\includegraphics[width=8cm]{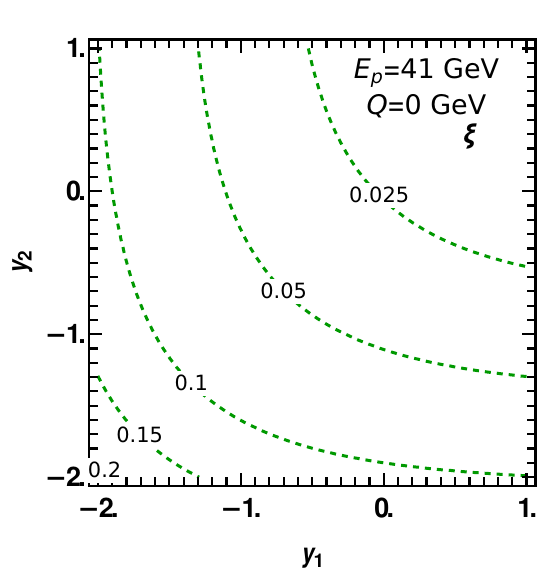}\includegraphics[width=8cm]{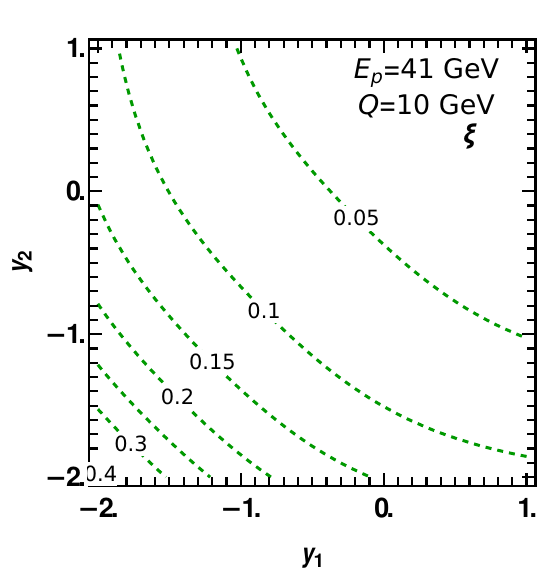}

\includegraphics[width=8cm]{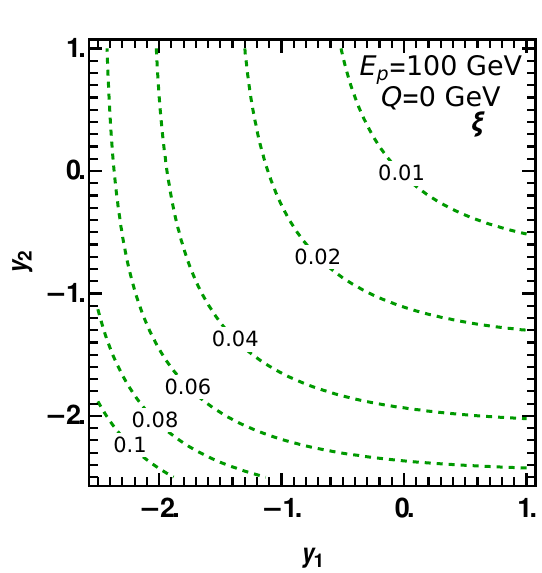}\includegraphics[width=8cm]{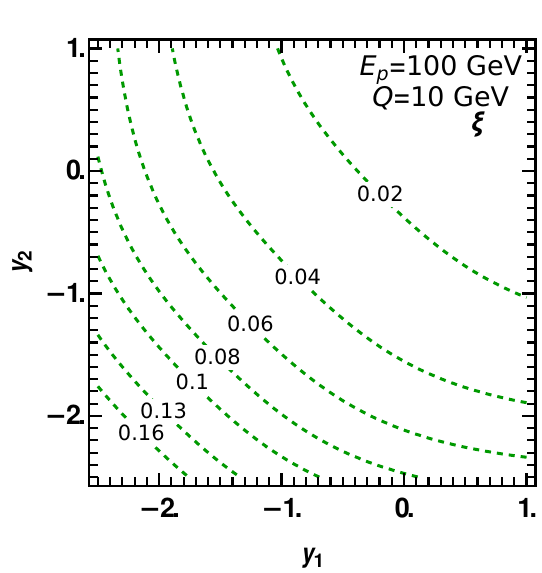}

\caption{\label{fig:xiRapidity}The contour plots illustrate the dependence
of the skewedness variable $\xi=-\left(P_{f}^{+}-P_{i}^{+}\right)/\left(P_{f}^{+}+P_{i}^{+}\right)$
on the rest-frame rapidities $y_{1},y_{2}$, of the $D$-mesons, for
different fixed photon virtualities $Q$ and proton energies $E_{p}$.
Each dashed line corresponds to a line $\xi={\rm const}$ in the $y_{1},y_{2}$
plane, with value of constant $\xi$ shown as a label on the contour
line. For simplicity, we disregard the transverse momenta of the produced
$D$-mesons. The upper and lower rows differ by the choice of the
proton energy $E_{p}$ (41 and 100 GeV respectively).}
\end{figure}

In these notations, the conventional Bjorken variable $x_{B}$ can
be represented as 
\begin{align}
x_{B} & =\frac{Q^{2}+\mathcal{M}_{{\rm 12}}^{2}}{Q^{2}+W_{\gamma p}^{2}-m_{N}^{2}}\approx\frac{Q^{2}}{2q^{-}P^{+}}+\frac{M_{1\perp}}{P^{+}}e^{-y_{1}}+\frac{M_{2\perp}}{P^{+}}e^{-y_{2}}\label{eq:xB-1}
\end{align}
In the literature sometimes the variable $x_{B}$ is replaced by the
so-called skewedness variable $\xi$, which is defined as~\cite{Diehl:2003ny}
\begin{equation}
\xi=\frac{x_{B}}{2-x_{B}},\quad x_{B}=\frac{2\xi}{1+\xi}.\label{eq:XiDef}
\end{equation}
This variable is directly related to the longitudinal light-cone momentum
transfer $\xi=-\Delta^{+}/2\bar{P}^{+}=-\Delta^{+}/\left(2P^{+}+\Delta^{+}\right)$.
Using Eq.~(\ref{eq:xB-1}), it is possible to express $\xi$ in terms
of the rapidities $y_{1},\,y_{2}$. As we can see from Figure~\ref{fig:xiRapidity},
the typical values of $\xi,x_{B}$ are relatively small even for the
lowest energy electron-proton beam at EIC; for this reason, we cannot
disregard the gluon GPDs contributions. However, the kinematics of
interest is still far from the saturation regime $x_{B}\lll1$, and
for this reason we may disregard saturation effects in our analysis.

The meson pair production in $ep$ collisions is dominated by single
photon exchange between the leptonic and hadronic parts; for this
reason the cross-section of the process can be expressed as 
\begin{equation}
\frac{d\sigma_{ep\to eM_{1}M_{2}p}}{d\ln x_{B}dQ^{2}\,d\Omega_{h}}=\frac{\alpha_{{\rm em}}}{\pi\,Q^{2}}\,\left[\left(1-y\right)\frac{d\sigma_{\gamma p\to M_{1}M_{2}p}^{(L)}}{d\Omega_{h}}+\left(1-y+\frac{y^{2}}{2}\right)\frac{d\sigma_{\gamma p\to M_{1}M_{2}p}^{(T)}}{d\Omega_{h}}\right],\label{eq:LTSep}
\end{equation}
where $d\sigma^{(T)},\,d\sigma^{(L)}$ are the photoproduction cross-sections
of the transversely and longitudinally polarized virtual photons,
$y$ is the inelasticity (ratio of the energies of the virtual photon
and electron), and $d\Omega_{h}$ the phase volume of the produced
heavy meson pair, which will be specified below.

\begin{figure}
\includegraphics[height=8.5cm]{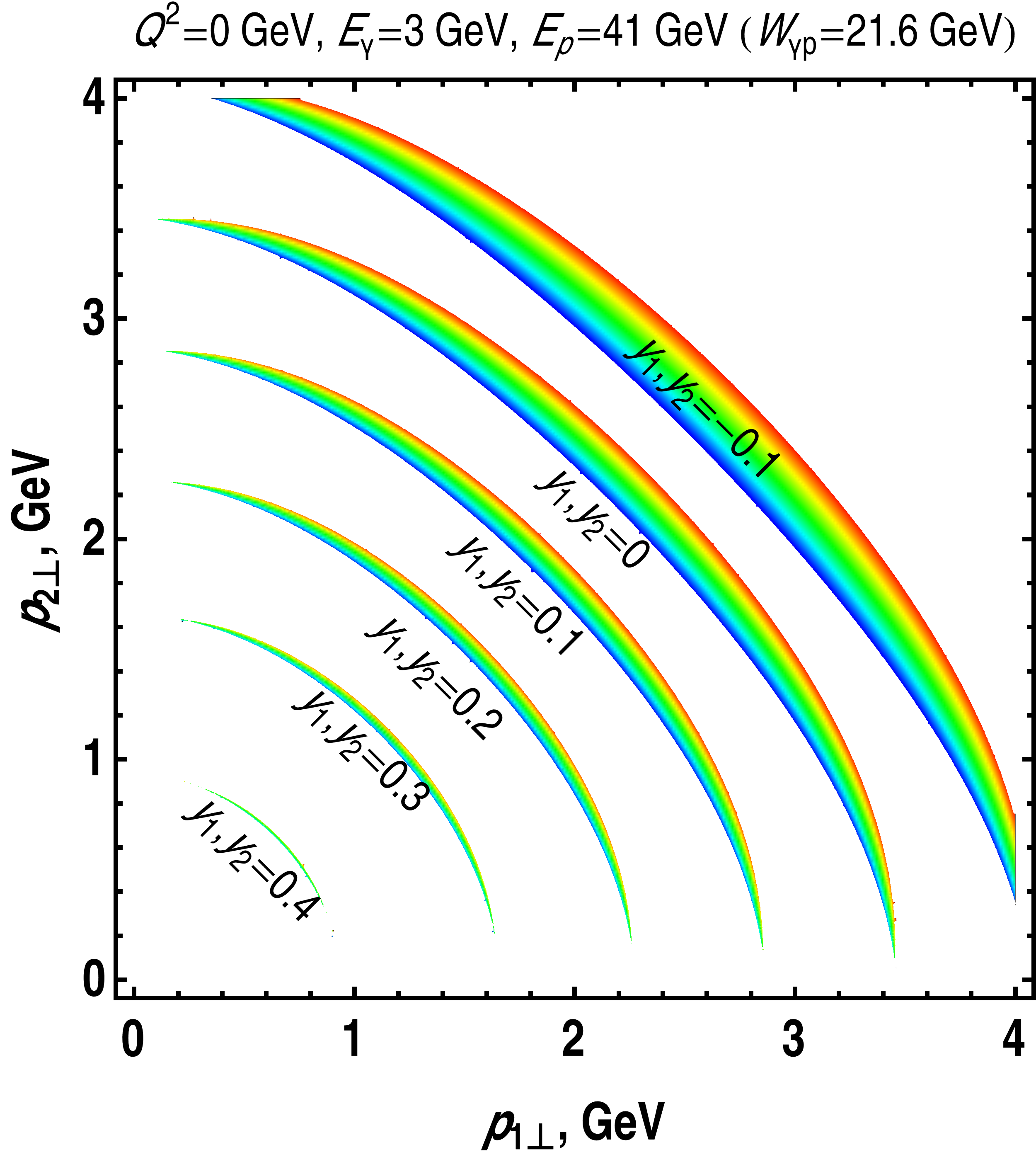}\includegraphics[height=8.5cm]{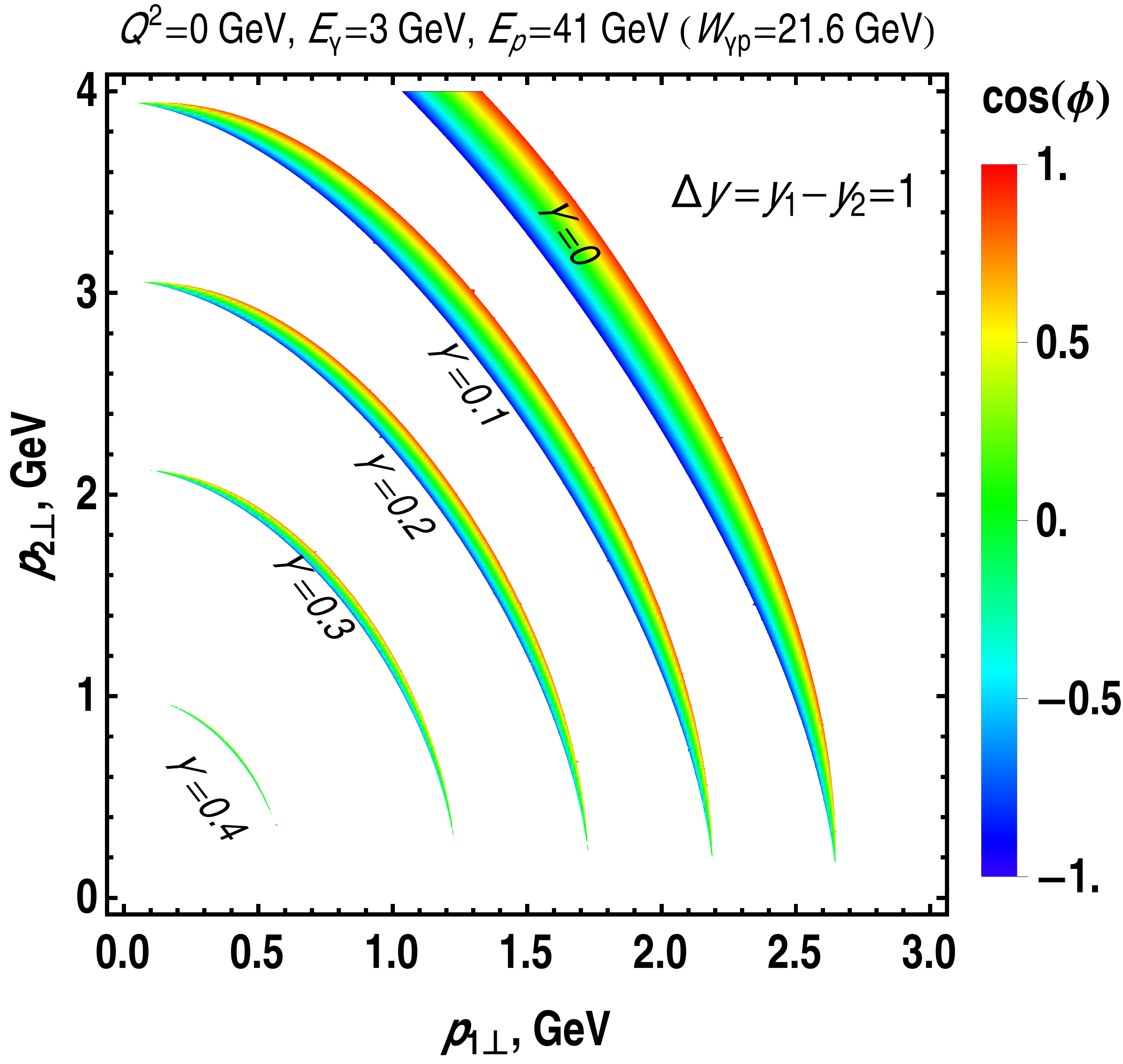}

\includegraphics[height=8.5cm]{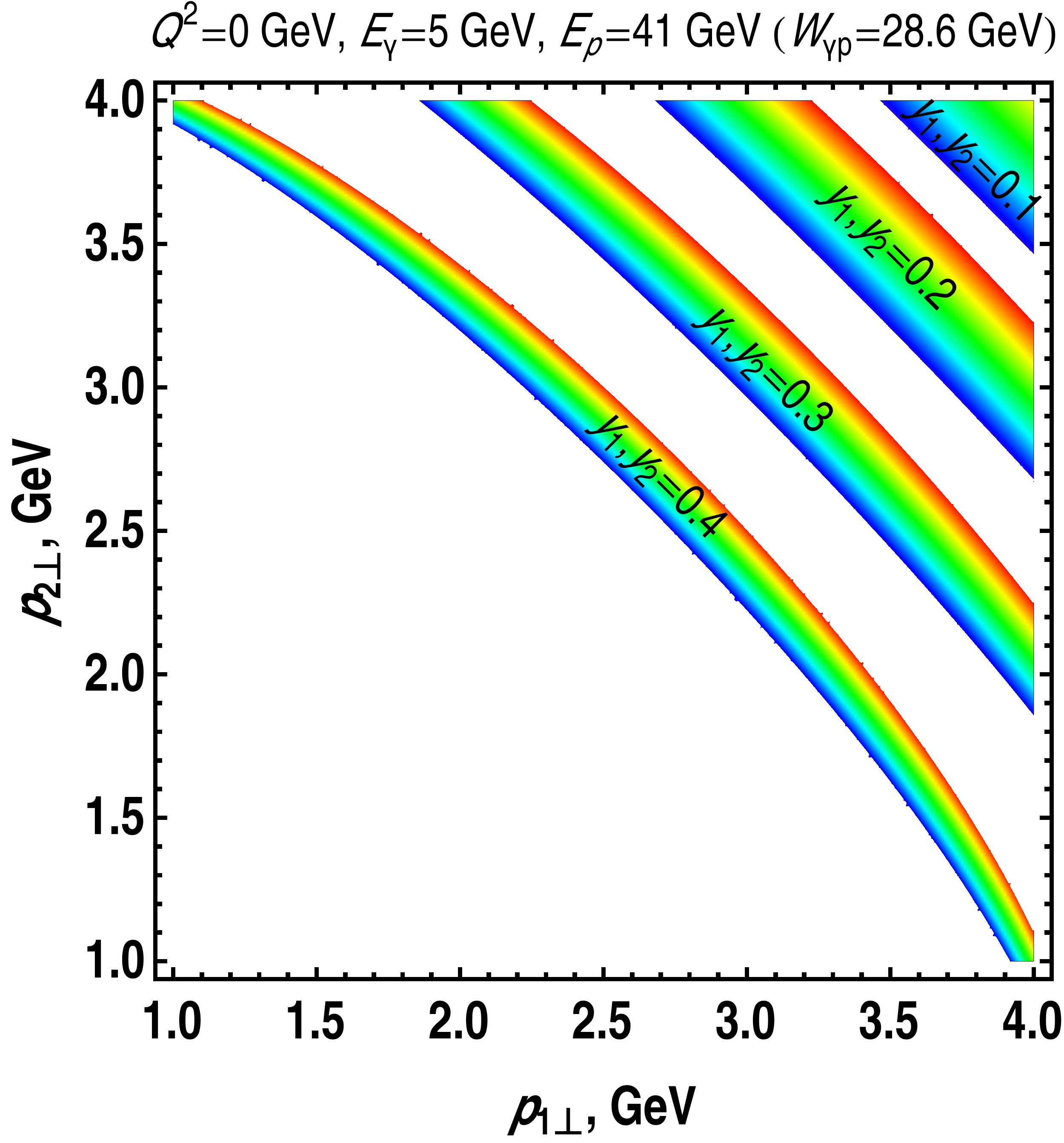}\includegraphics[height=8.5cm]{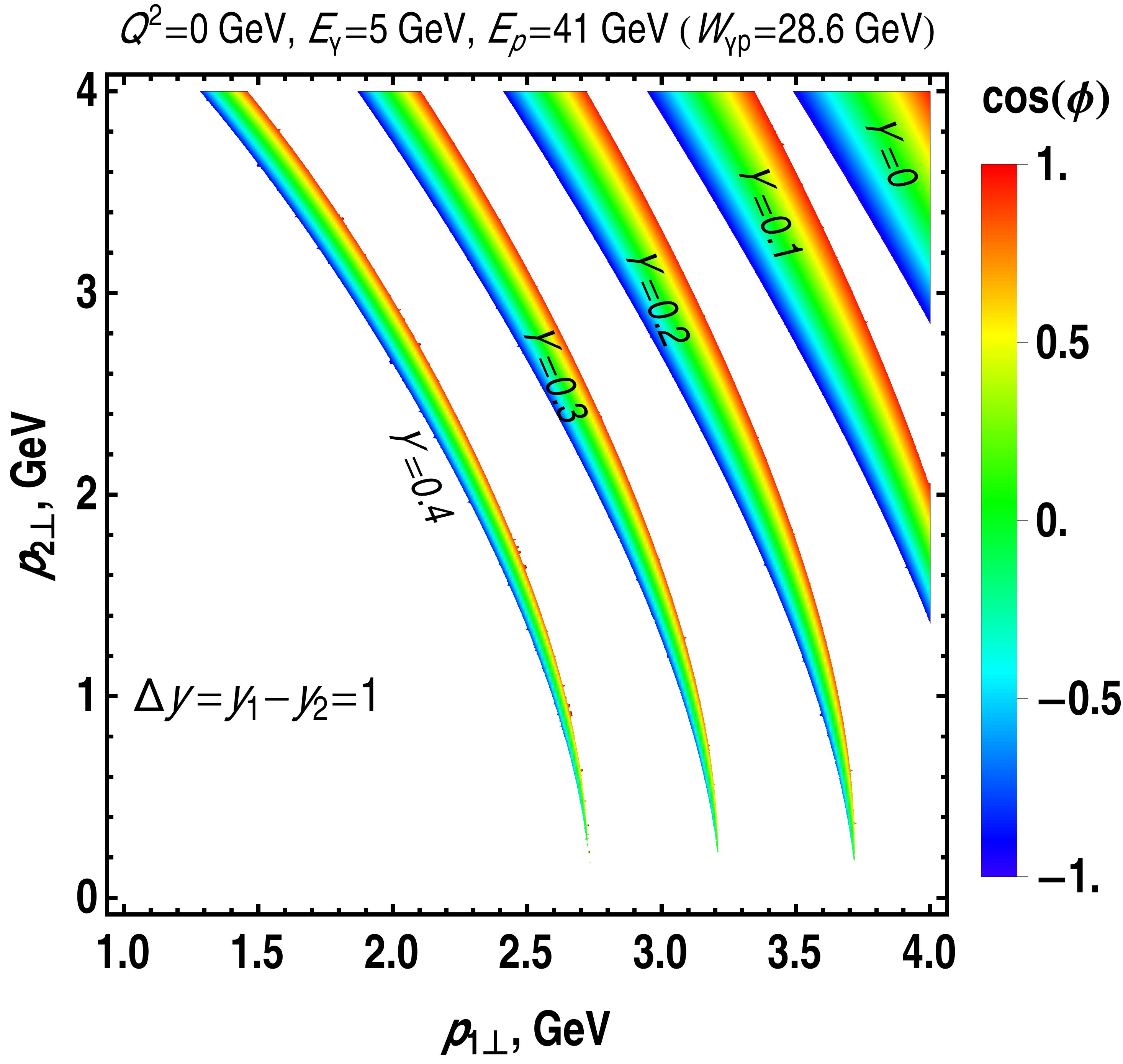}

\caption{\label{fig:Domain}(Color online) The colored bands show the kinematically
permitted regions for $D$-meson pair production, for fixed rapidities
$y_{1},y_{2}$, \uline{at fixed photon energy} $E_{\gamma}$, virtuality
$Q$ and proton energy $E_{p}$. The color of each pixel reflects
the cosine of the angle $\phi$ (azimuthal angle between the transverse
momenta of the $D$-mesons), fixed from~(\ref{eq:cos}). The variable
$Y=(y_{1}+y_{2})/2$ is the average rapidity of the $D$-mesons. The
left and right columns differ by the choice of the rapidity difference
($\Delta y=0$ and $\Delta y=1$ respectively). The upper and lower
rows differ by the choice of the photon energy $E_{\gamma}$ in the
lab frame (3 and 5 GeV respectively). The borders of the kinematic
domains have very mild dependence on $Q^{2}$ (up to 2-3 units of
$M_{D}^{2}$) and mild sensitivity to the proton energy $E_{p}$.}
\end{figure}

\begin{figure}
\includegraphics[height=8.5cm]{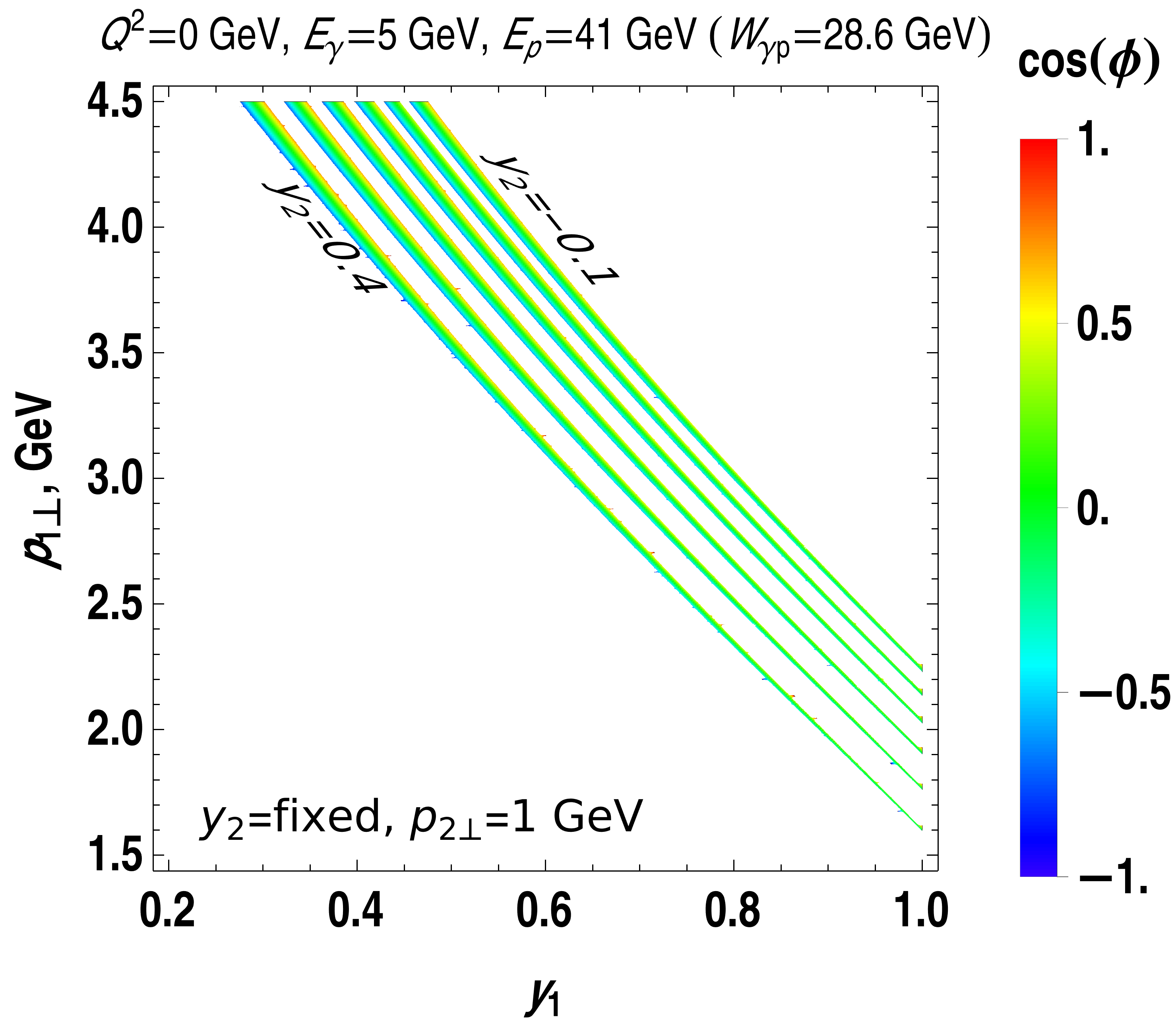}\includegraphics[height=8.5cm]{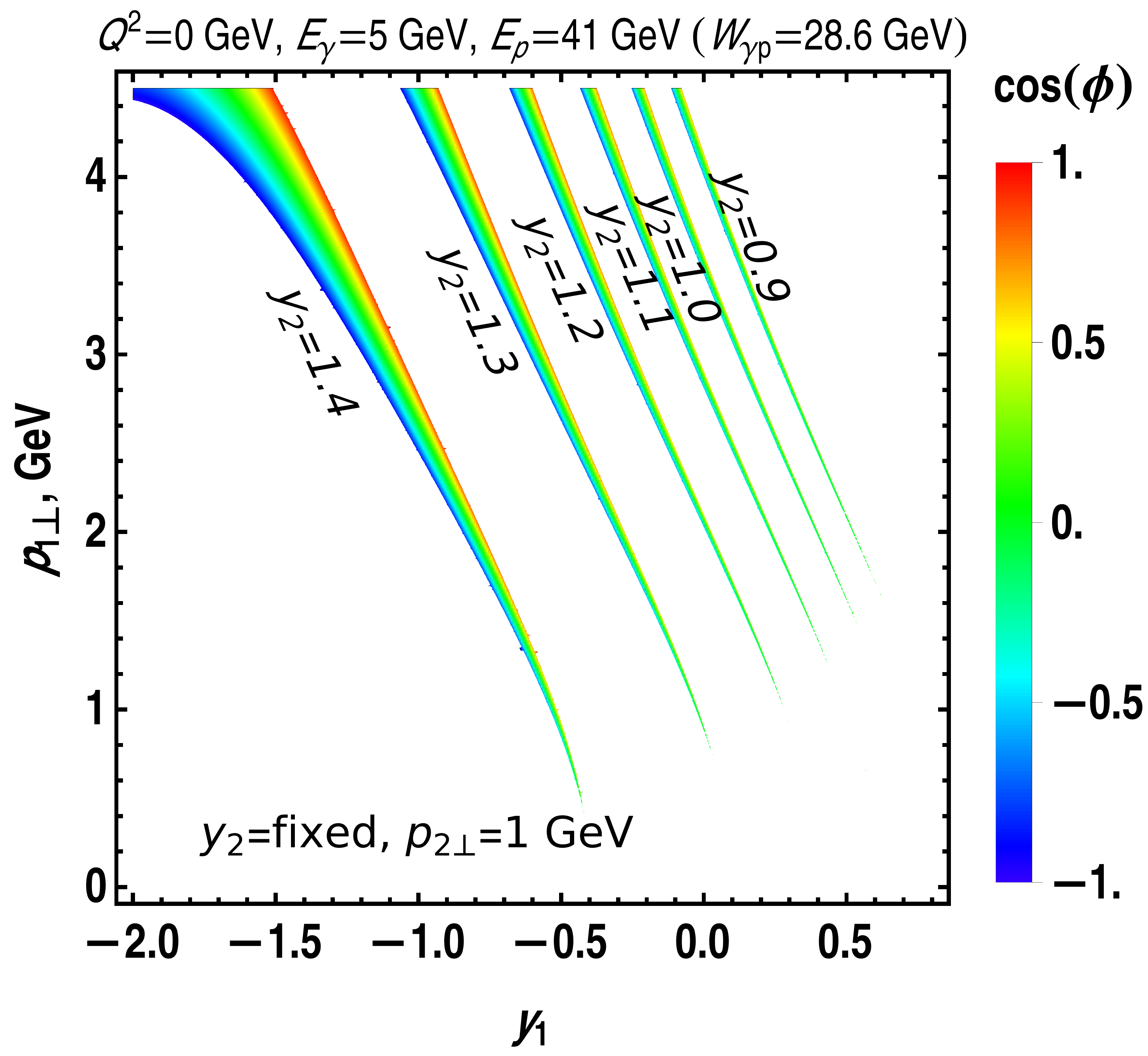}

\caption{\label{fig:Domain-1}(Color online) Kinematic constraints on the rapidity
and transverse momenta of one of the $D$-mesons, when the momentum
of the other (its rapidity $y_{2}$, transverse momentum $p_{2\perp}$)
is fixed. The photon virtuality $Q$ and \uline{energy }%
\mbox{%
$E_{\gamma}$%
}\uline{, as well as proton energy} $E_{p}$ are fixed to values
shown in the upper part of the Figure. The color of each pixel reflects
the value of the angle $\phi$ (azimuthal angle between the transverse
momenta of the $D$-mesons), fixed from~(\ref{eq:cos}). The left
and right columns differ by the choice of the rapidity interval for
$y_{2}$.}
\end{figure}

The cross-section of the photoproduction process is related to the
corresponding amplitude by 
\begin{equation}
d\sigma_{\gamma p\to M_{1}M_{2}p}^{(L,T)}=\frac{dy_{1}dp_{1\perp}^{2}dy_{2}dp_{2\perp}^{2}d\phi\left|\mathcal{A}_{\gamma p\to M_{1}M_{2}p}^{(L,T)}\right|^{2}}{4\left(2\pi\right)^{4}\sqrt{\left(W^{2}+Q^{2}-m_{N}^{2}\right)^{2}+4Q^{2}m_{N}^{2}}}\delta\left(\left(q+P_{1}-p_{1}-p_{2}\right)^{2}-m_{N}^{2}\right)\label{eq:Photo}
\end{equation}
where the $\delta$-function in the right-hand side of~(\ref{eq:Photo})
reflects the onshellness of the recoil proton. Using a light-cone
decomposition~(\ref{eq:qPhoton-2}-\ref{eq:MesonLC-2}), the argument
of the $\delta$-function can be rewritten as 
\begin{align}
 & \left(q+P_{1}-p_{1}-p_{2}\right)^{2}-m_{N}^{2}=\left(q+P_{1}-p_{1}^{||}-p_{2}^{||}\right)^{2}-\left(\boldsymbol{p}_{1}^{\perp}+\boldsymbol{p}_{2}^{\perp}\right)^{2}-m_{N}^{2}\label{eq:delta}\\
 & =\left(q+P_{1}-p_{1}^{||}-p_{2}^{||}\right)^{2}-\left(\left(p_{1}^{\perp}\right)^{2}+\left(p_{2}^{\perp}\right)^{2}+2p_{1}^{\perp}p_{2}^{\perp}\cos\phi\right)-m_{N}^{2}\nonumber 
\end{align}
where $\phi$ is the azimuthal angle between the transverse momenta
$\boldsymbol{p}_{1}^{\perp},\boldsymbol{p}_{2}^{\perp}$, of the produced
$D$-mesons. This allows to rewrite the $\delta$-function in~(\ref{eq:delta})
as 
\begin{align}
 & \delta\left(\left(q+P_{1}-p_{1}-p_{2}\right)^{2}-m_{N}^{2}\right)=\frac{\delta\left(\phi-\phi_{0}\right)}{2p_{1\perp}p_{2\perp}\left|\sin\phi_{0}\right|},\\
 & \phi_{0}=\arccos\left[\frac{\left(q+P_{1}-p_{1}^{||}-p_{2}^{||}\right)^{2}-\left(\left(p_{1}^{\perp}\right)^{2}+\left(p_{2}^{\perp}\right)^{2}+m_{N}^{2}\right)}{2p_{1}^{\perp}p_{2}^{\perp}}\right],\label{eq:cos}
\end{align}
which permits to integrate out the dependence on $\phi$. The condition
$\left|\cos\phi_{0}\right|\le1$ at fixed invariant energies $W$
of the photon-proton collision leads to a nontrivial constraint on
the possible rapidities and transverse momenta of the produced $D$-mesons.
In Figures~(\ref{fig:Domain},\ref{fig:Domain-1}) we illustrate
the kinematically allowed domains for transverse momenta of the $D$-mesons,
at several fixed rapidities of the $D$-mesons and photon-proton energies.
From Figure~(\ref{fig:Domain}) we may see that an increase of $D$-meson
rapidities $y_{1},y_{2}$ leads to an increase of their longitudinal
momenta, and due to energy conservation, this decreases the allowed
transverse momenta of the produced mesons. These qualitative explanation
also allows to understand Figure~(\ref{fig:Domain-1}): increasing
the rapidity or the transverse momentum of one of the $D$-mesons
at fixed total invariant energy $W$ inevitably decreases the allowed
rapidities or transverse momenta of the other meson. The color coding
in all plots reflects the value of $\cos\phi_{0}$, fixed from~(\ref{eq:cos}).
In the heavy quark mass limit, the difference of masses of the various
$D$-mesons is suppressed as $M_{1}-M_{2}\sim\mathcal{O}\left(\Lambda^{2}/m_{Q}\right)$,
where $\Lambda$ is some soft scale, and $m_{Q}$ is the mass of the
heavy quark~\cite{Neubert:1993mb}. For simplicity we disregard this
difference altogether, assuming $M_{1}=M_{2}=\left(M_{D^{+}}+M_{D^{*+}}\right)/2\approx1.9\,{\rm GeV}$;
for this reason the plots in the left column of Figure~(\ref{fig:Domain})
are symmetric w.r.t. permutation of the transverse momenta $p_{1\perp},\,p_{2\perp}$.
In the experiment the invariant energy $W$ and the kinematics of
the $D$-mesons are measured with finite precision; for this reason
the narrow domains shown in Figures~(\ref{fig:Domain},~\ref{fig:Domain-1})
will be smeared, with the values of $\cos\phi$ distributed over some
interval, which depends on the width of the bins in rapidity $(\Delta y)$
and transverse momenta $(\Delta p_{\perp})$.

The complexity of the above-mentioned kinematic restrictions is a
consequence of fixing the invariant energy $W$. In view of the symmetry
of the final-state $D$-meson pair w.r.t. permutations of both mesons,
in electroproduction experiments it might be easier not to impose
conventional constraints on $W$ and work with $D$-meson momenta
as independent unconstrained variables. The energy $W$ in each event
can be rewritten in terms of these variables. The $\delta$-function
in the right-hand side of~(\ref{eq:Photo}) may be represented as
\begin{align}
 & \delta\left(\left(q+P_{1}-p_{1}-p_{2}\right)^{2}-m_{N}^{2}\right)=\delta\left(W^{2}+{\mathcal{M}}_{12}^{2}-2\left(q+P_{1}\right)\cdot\left(p_{1}+p_{2}\right)-m_{N}^{2}\right)=\frac{\delta\left(W-W_{0}\right)+\delta\left(W+W_{0}\right)}{2W_{0}},\\
 & W_{0}^{2}=2\left(q+P_{1}\right)\cdot\left(p_{1}+p_{2}\right)+m_{N}^{2}-{\mathcal{M}}_{12}^{2}=\\
 & =\left(q^{-}+\frac{m_{N}^{2}}{2P^{+}}\right)\cdot\left(M_{1}^{\perp}e^{-y_{1}}+M_{2}^{\perp}e^{-y_{2}}\right)+2\left(P^{+}-\frac{Q^{2}}{2q^{-}}\right)\cdot\left(M_{1}^{\perp}e^{y_{1}}+M_{2}^{\perp}e^{y_{2}}\right)+m_{N}^{2}-{\mathcal{M}}_{12}^{2}.\nonumber 
\end{align}
The integration of~(\ref{eq:LTSep}), over all $x_{B}\sim1/\left(W^{2}+Q^{2}-m_{N}^{2}\right)$,
allows to rewrite the electroproduction cross-section as

\begin{equation}
\frac{d\sigma_{ep\to eM_{1}M_{2}p}}{dQ^{2}\,d\Omega_{h}}=\frac{\alpha_{{\rm em}}}{4\pi\,Q^{2}}\,\left[\left(1-y\right)\frac{d\bar{\sigma}_{\gamma p\to M_{1}M_{2}p}^{(L)}}{d\Omega_{h}}+\left(1-y+\frac{y^{2}}{2}\right)\frac{d\bar{\sigma}_{\gamma p\to M_{1}M_{2}p}^{(T)}}{d\Omega_{h}}\right],\label{eq:LTSep-2}
\end{equation}
\begin{equation}
d\bar{\sigma}_{\gamma p\to M_{1}M_{2}p}^{(L,T)}=\frac{dy_{1}dp_{1\perp}^{2}dy_{2}dp_{2\perp}^{2}d\phi\left|\mathcal{A}_{\gamma p\to M_{1}M_{2}p}^{(L,T)}\right|^{2}}{4\left(2\pi\right)^{4}W_{0}^{2}\sqrt{\left(W_{0}^{2}+Q^{2}-m_{N}^{2}\right)^{2}+4Q^{2}m_{N}^{2}}}\label{eq:Photo-1}
\end{equation}
where the variables ($y_{1},p_{1\perp},y_{2},p_{2\perp},\phi$) fully
characterize the kinematics of the process, and $d\bar{\sigma}_{\gamma p\to M_{1}M_{2}p}^{(L,T)}$
are the photoproduction cross-sections of longitudinal and transverse
photons, for photons energy given by~(\ref{qPlus-1}).

\subsection{Amplitudes of the meson pair production process}

\label{subsec:Amplitudes}As we have seen in the previous section,
in the kinematics that we consider here, the typical values of $x_{B},\xi$
are small, although still far from the saturation regime, and the
dominant contribution to the cross-section comes from the region of
small transverse momenta of $D$-mesons, $p_{\perp}\lesssim M_{D}$.
In this kinematic regime, it is convenient to use a collinear factorization
framework for the evaluation of the amplitudes $\mathcal{A}_{\gamma p\to M_{1}M_{2}p}^{(\mathfrak{a})}$,
and express the latter as a convolution of perturbative coefficient
functions with distribution amplitudes of produced $D$-mesons and
the GPDs of the target~\cite{Diehl:2000xz,Goeke:2001tz,Diehl:2003ny,Guidal:2013rya,Boer:2011fh,Burkert:2022hjz}.
The natural hard scales in this approach are the heavy quark mass
$m_{Q}$ and the invariant mass $\mathcal{M}_{12}\sim m_{Q}$ of the
produced $D$-meson pair. In order to avoid nonperturbative final
state interactions, we'll assume additionally that all hadrons in
tthe final state are kinematically well-separated from each other,
having sufficiently large (frame-invariant) relative velocities 
\begin{align}
 & v_{{\rm rel}}=\sqrt{1-\frac{p_{1}^{2}p_{2}^{2}}{\left(p_{1}\cdot p_{2}\right)^{2}}}=\sqrt{1-\frac{4M_{1}^{2}M_{2}^{2}}{\left({\mathcal{M}}_{12}^{2}-M_{1}^{2}-M_{2}^{2}\right)^{2}}}\gtrsim2\alpha_{s}\left(m_{c}\right)\approx0.7\label{eq:constraint}
\end{align}
for each pair of final-state hadrons. For production of meson pairs
at central rapidities, this constraint is satisfied almost everywhere,
except in the small near-threshold region ${\mathcal{M}}_{12}\sim M_{1}+M_{2}$.
Since in the evaluation of the coefficient functions the transverse
momenta of partons are disregarded, in order to guarantee the validity
of~(\ref{eq:constraint}), we will only consider the kinematics when
the mesons are separated from each other at least by one unit of rapidity,
$\Delta y\gtrsim1$.

\begin{figure}
\includegraphics[scale=0.4]{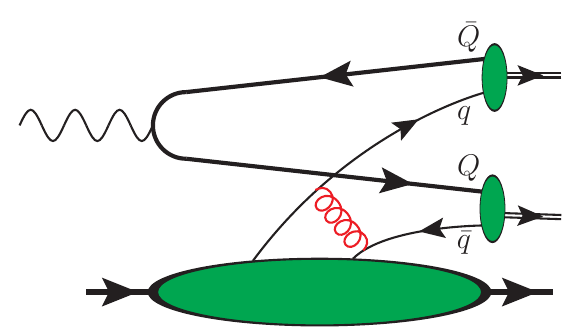}\includegraphics[scale=0.4]{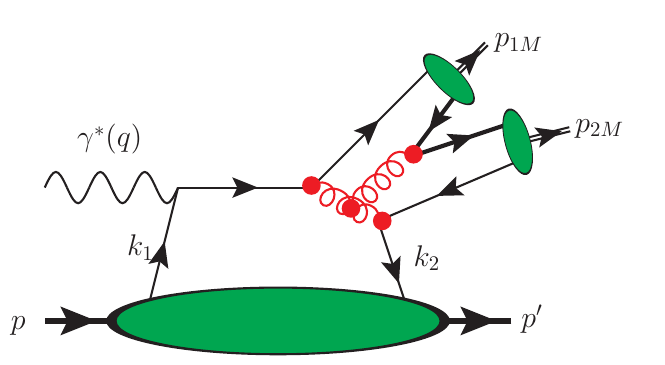}\includegraphics[scale=0.4]{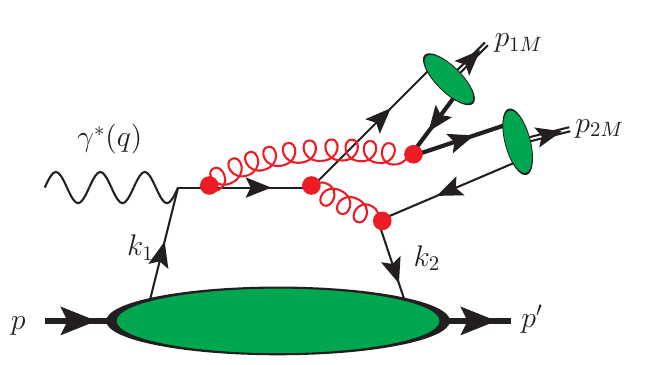}\includegraphics[scale=0.4]{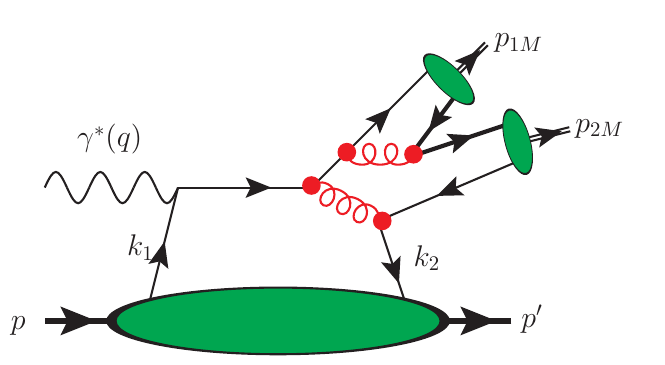}

\includegraphics[scale=0.4]{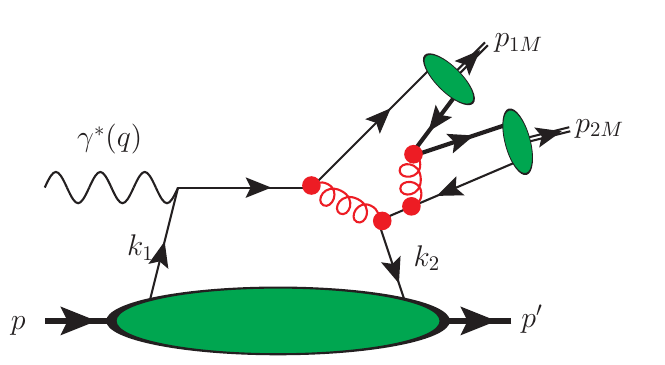}\includegraphics[scale=0.4]{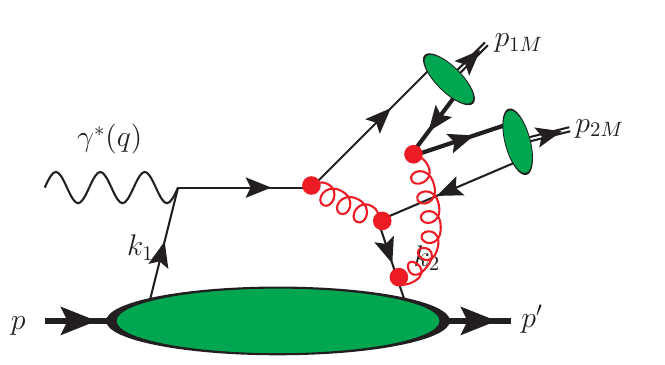}\includegraphics[scale=0.4]{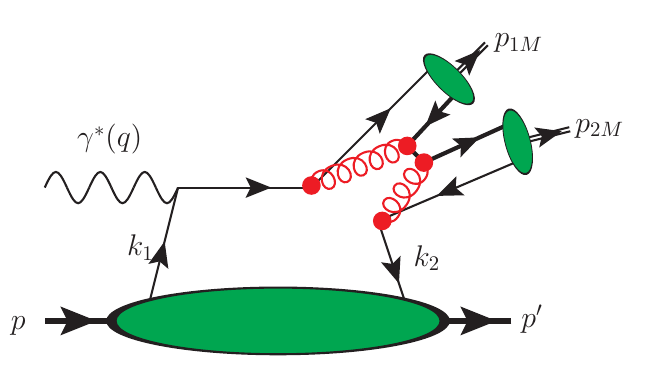}\includegraphics[scale=0.4]{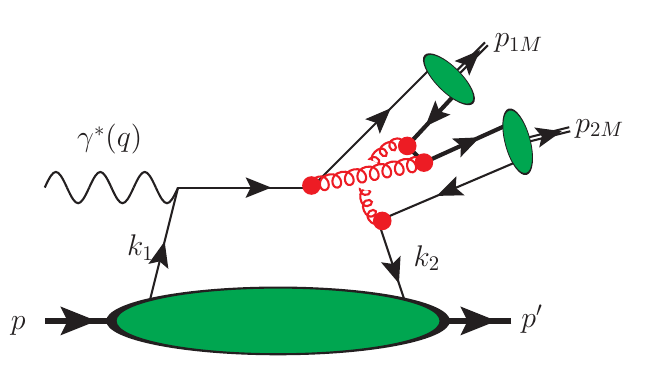}

\caption{\label{fig:Photoproduction-B}First diagram in the upper row: the
leading-order $\sim\mathcal{O}\left(\alpha_{s}\right)$-diagram, which
describes the (light) quark contributions to the $D$-meson pair production.
As explained in the text, a diagram of this type does not contribute
in the collinear approximation. Other diagrams: representative (sub)leading
$\sim\mathcal{O}\left(\alpha_{s}^{2}\right)$ diagrams, which describe
the (light) quark contributions to the $D$-meson pair production.
The thin and thick lines correspond to light and heavy quarks respectively.
Each diagram should be understood as a sum of diagrams with all possible
permutations of the photon coupling to quark lines at fixed gluon
vertices (so the second diagram in the upper row corresponds to 6
different diagrams, whereas each of the other diagrams should be understood
as a sum of 7 different diagrams). We disregard the diagrams with
heavy quark lines attached to the proton, since the intrinsic heavy
flavors in the proton are negligibly small.}
\end{figure}

\begin{figure}
\includegraphics[scale=0.4]{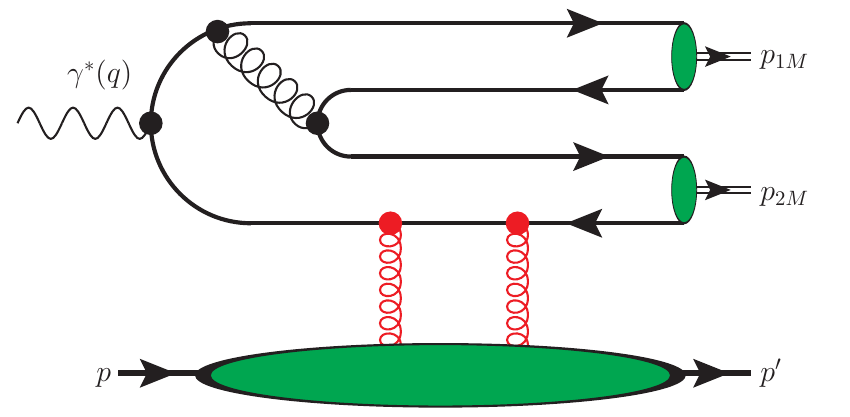}\includegraphics[scale=0.4]{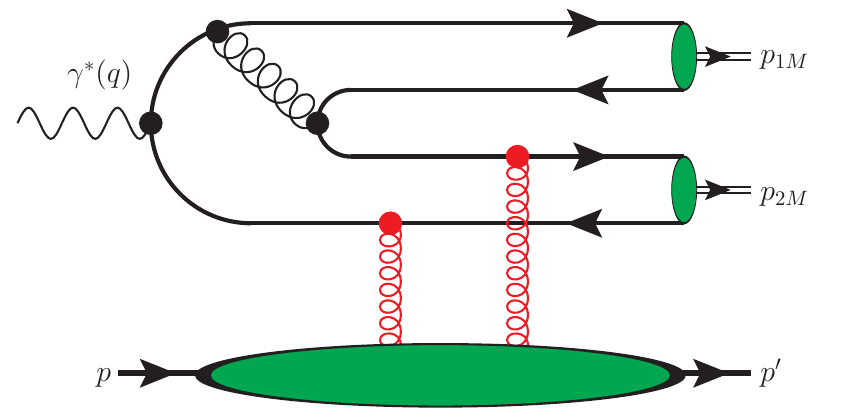}\includegraphics[scale=0.4]{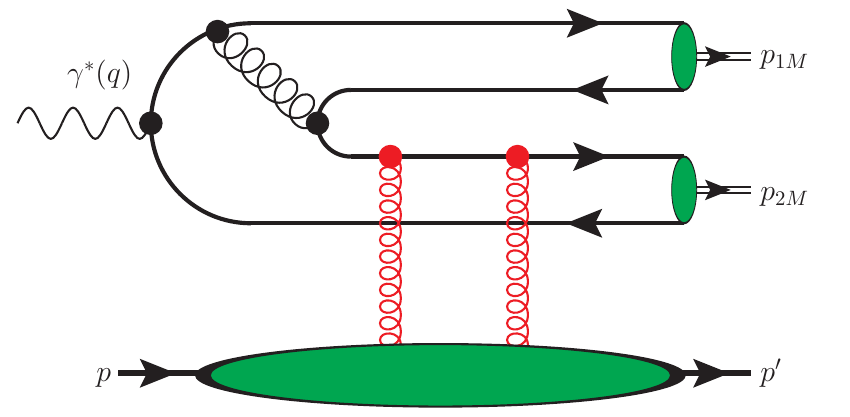}

\includegraphics[scale=0.4]{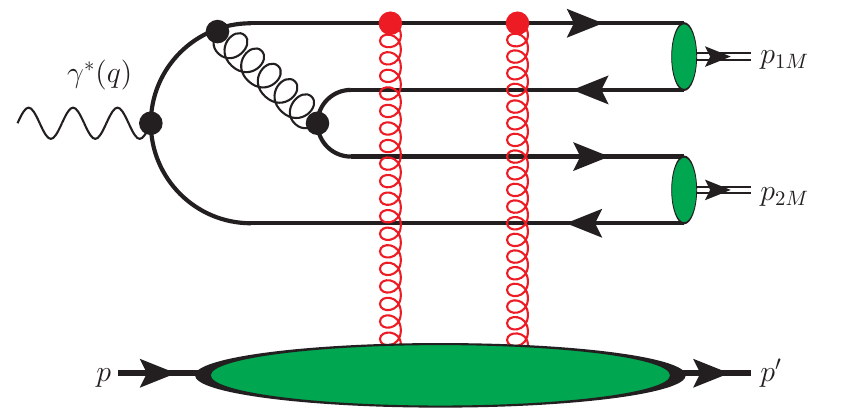}\includegraphics[scale=0.4]{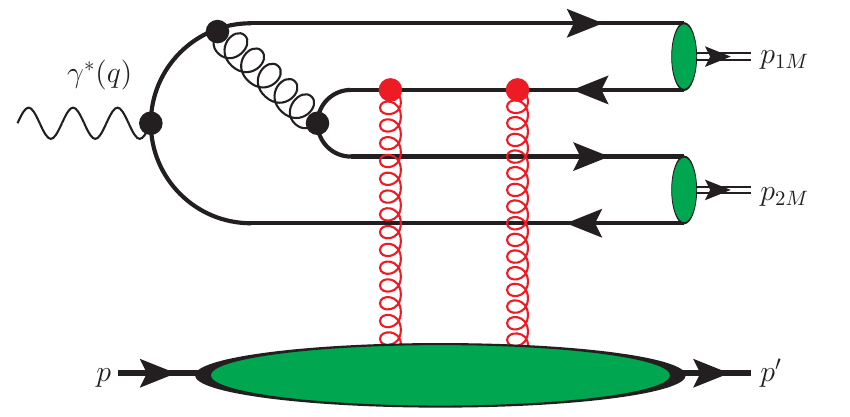}\includegraphics[scale=0.4]{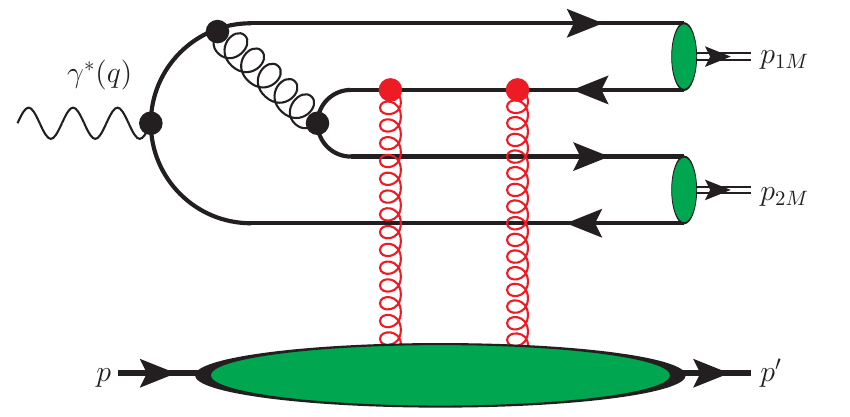}

\includegraphics[scale=0.4]{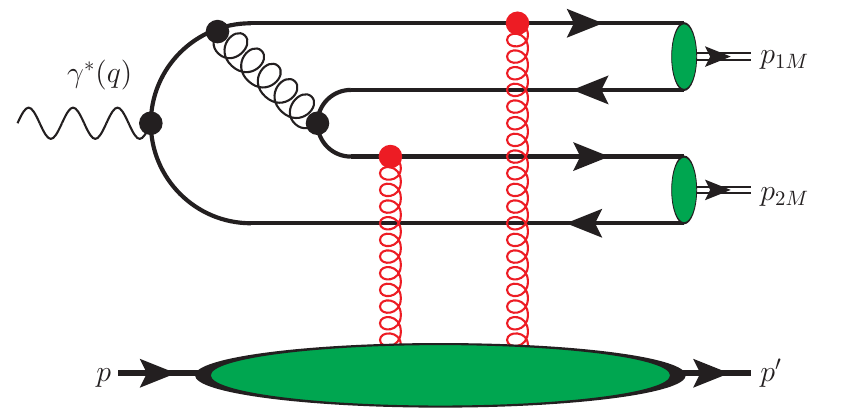}\includegraphics[scale=0.4]{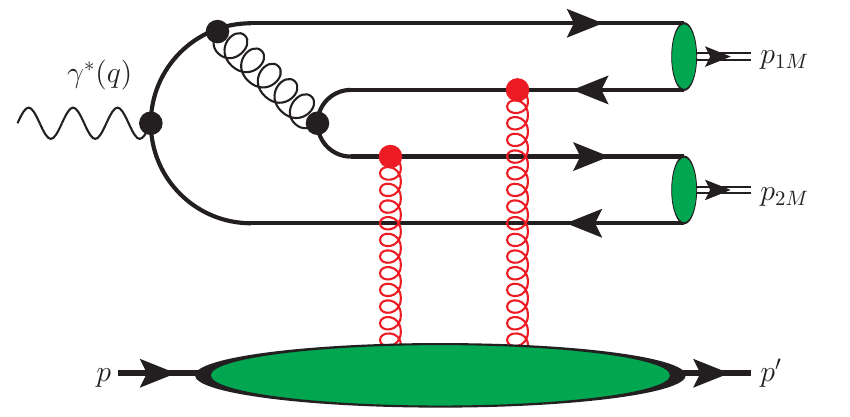}\includegraphics[scale=0.4]{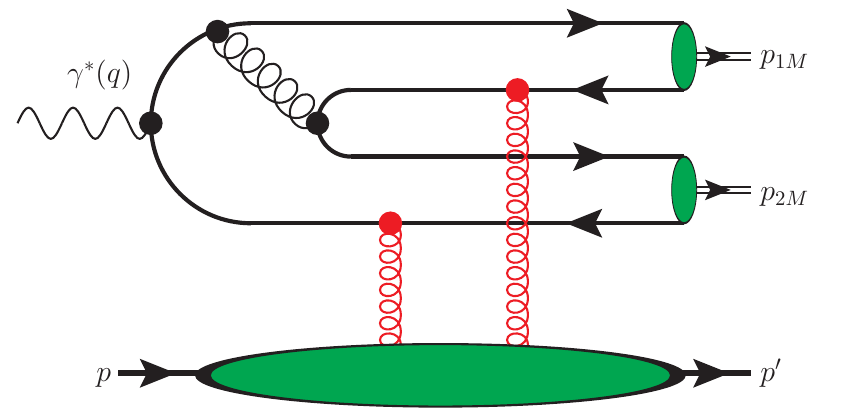}

\includegraphics[scale=0.4]{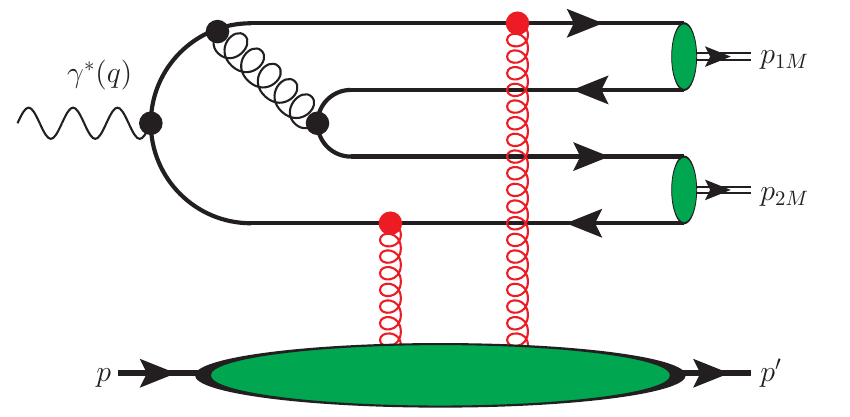}\includegraphics[scale=0.4]{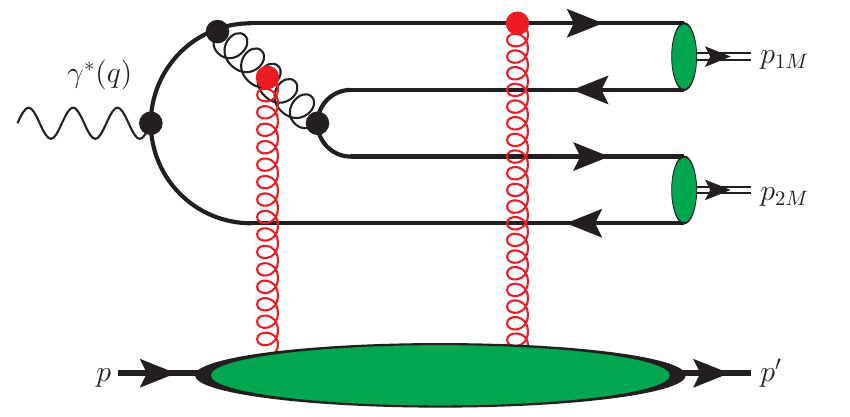}\includegraphics[scale=0.4]{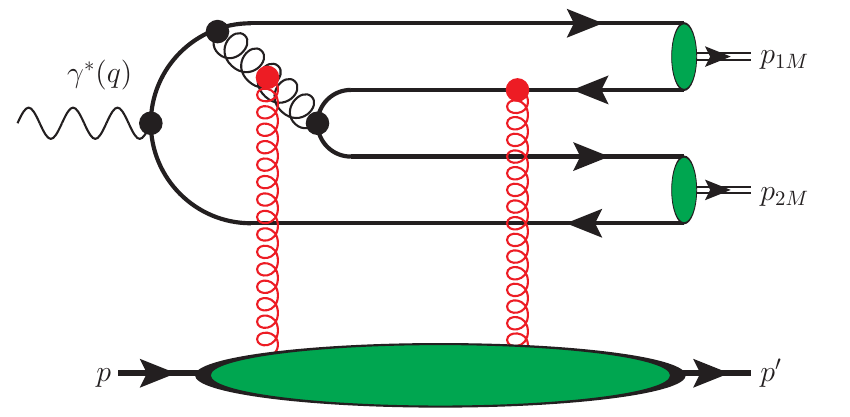}

\includegraphics[scale=0.4]{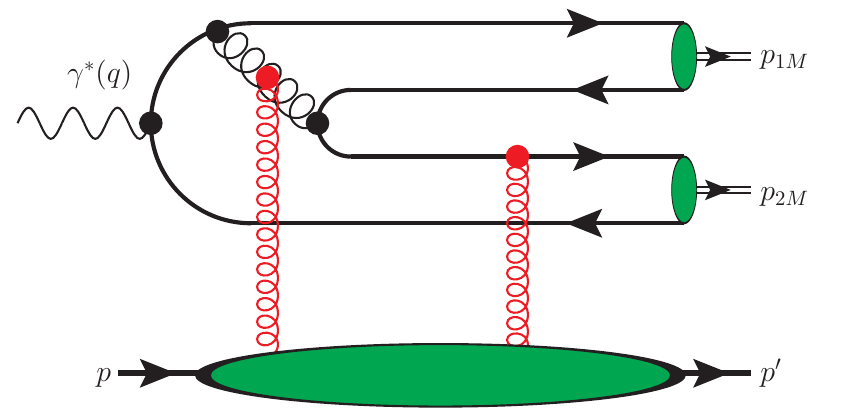}\includegraphics[scale=0.4]{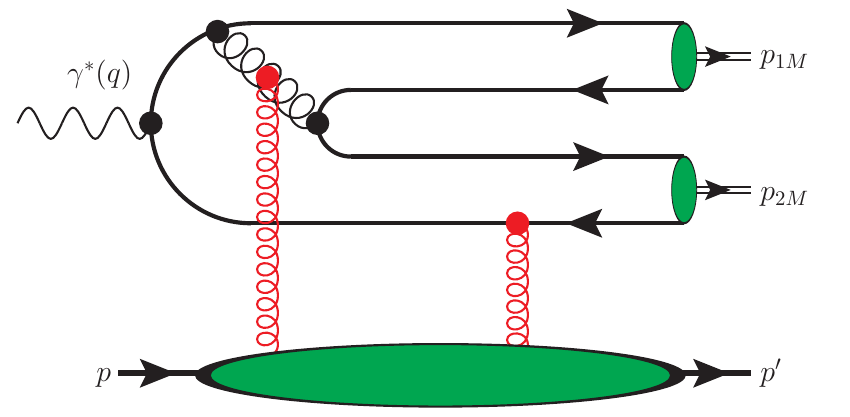}\includegraphics[scale=0.4]{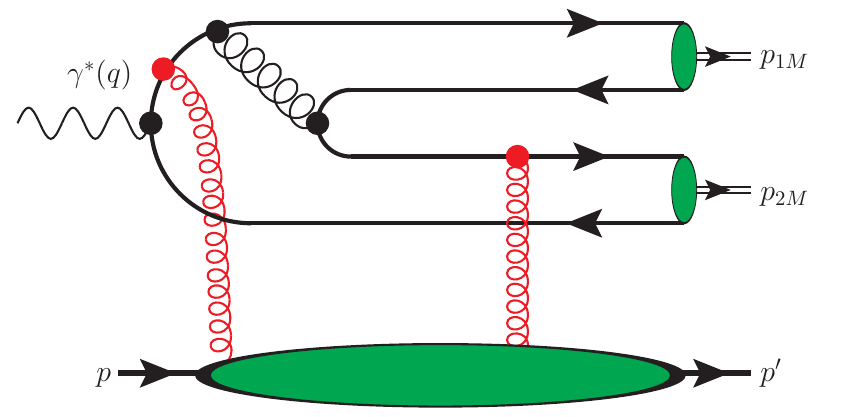}

\includegraphics[scale=0.4]{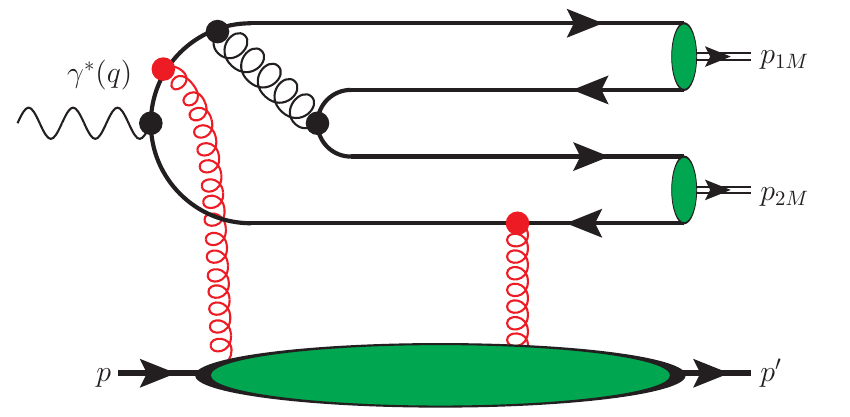}\includegraphics[scale=0.4]{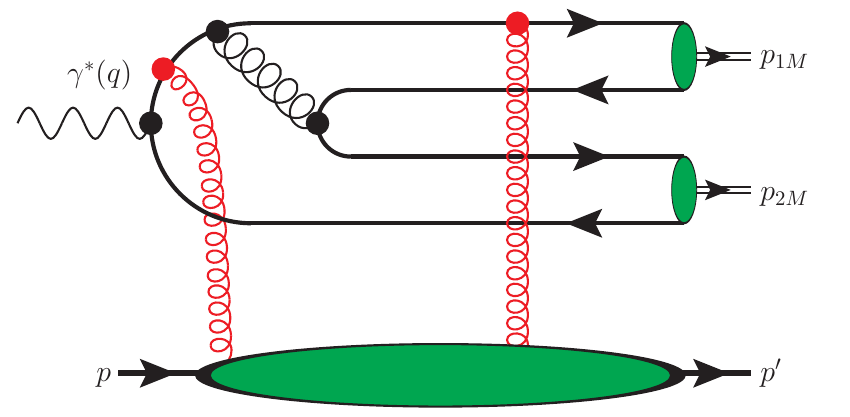}\includegraphics[scale=0.4]{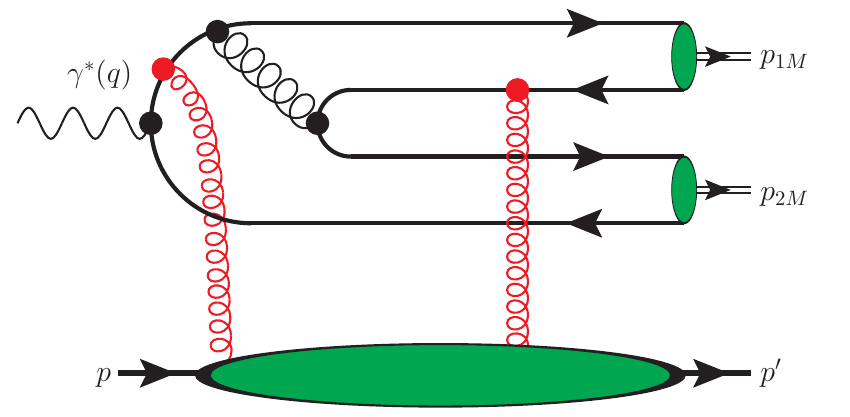}

\includegraphics[scale=0.4]{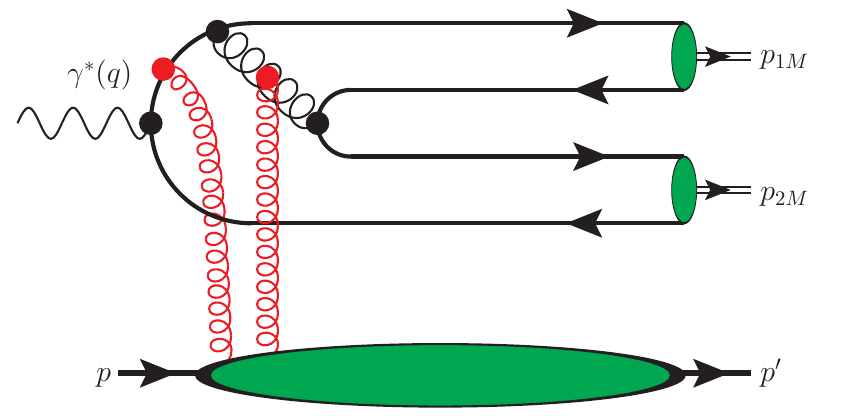}\includegraphics[scale=0.4]{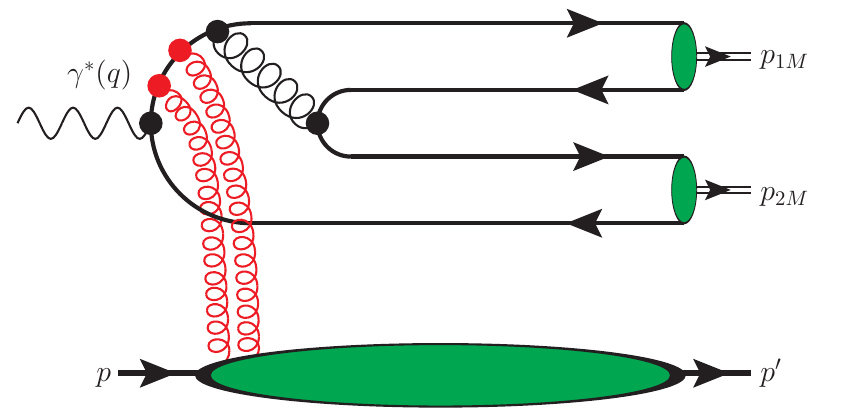}\includegraphics[scale=0.4]{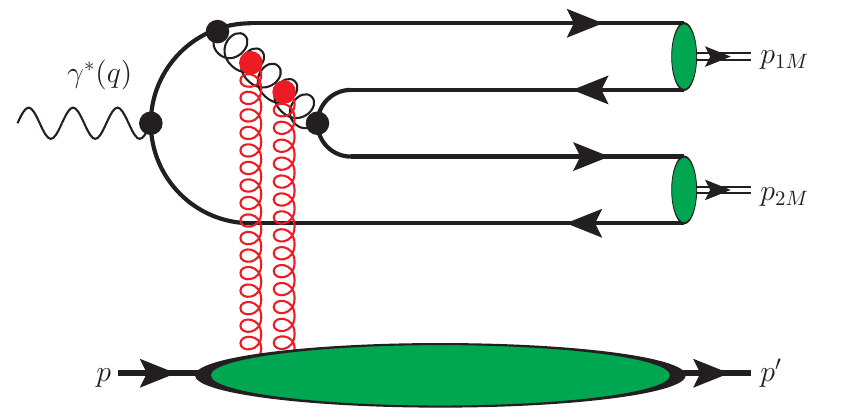}

\caption{\label{fig:Photoproduction-A}The leading order diagrams, which describe
the contribution of the gluons to $D$-meson pair production. Each
diagram should be understood as a sum of two diagrams with complementary
assignment of quark flavors (heavy-light and light-heavy) .}
\end{figure}

In the collinear factorization picture all hadrons can be replaced
with collinear partons, convoluted with nonperturbative distribution
amplitudes which describe the momentum sharing between these hadrons,
and contracted with appropriate spin projectors. For the proton target,
in leading twist the Fock state is dominated by the two-partonic component,
which is described by the quark and gluon GPDs. In what follows we
disregard the contribution of the intrinsic heavy flavors in the proton,
since they should be very small according to phenomenological estimates.
Furthermore, we also do not take into account the contributions of
the poorly known transversity GPDs $H_{T},\,E_{T},\,\tilde{H}_{T},\,\tilde{E}_{T}$,
since their contributions are accompanied by momentum transfer to
the target $\Delta^{\mu}$ and thus should be small for unpolarized
observables in the small-$\Delta$ (small-$t$) kinematics which we
study here (see ~\cite{Pire:2017yge,Goloskokov:2013mba} for more
details). In the chiral even sector we should take into account the
contributions of both quarks and gluons. The other (chiral-even) GPDs
contribute to the amplitude of unpolarized process in the combination

\begin{align}
\sum_{{\rm spins}}\left|\mathcal{A}_{\gamma p\to M_{1}M_{2}p}^{(\mathfrak{a})}\right|^{2} & =\frac{1}{\left(2-x_{B}\right)^{2}}\left[4\left(1-x_{B}\right)\left(\mathcal{H}_{\mathfrak{a}}\mathcal{H}_{\mathfrak{a}}^{*}+\tilde{\mathcal{H}}_{\mathfrak{a}}\tilde{\mathcal{H}}_{\mathfrak{a}}^{*}\right)-x_{B}^{2}\left(\mathcal{H}_{\mathfrak{a}}\mathcal{E}_{\mathfrak{a}}^{*}+\mathcal{E}_{\mathfrak{a}}\mathcal{H}_{\mathfrak{a}}^{*}+\tilde{\mathcal{H}}_{\mathfrak{a}}\tilde{\mathcal{E}}_{\mathfrak{a}}^{*}+\tilde{\mathcal{E}}_{\mathfrak{a}}\tilde{\mathcal{H}}_{\mathfrak{a}}^{*}\right)\right.\label{eq:AmpSq}\\
 & \qquad\left.-\left(x_{B}^{2}+\left(2-x_{B}\right)^{2}\frac{t}{4m_{N}^{2}}\right)\mathcal{E}_{\mathfrak{a}}\mathcal{E}_{\mathfrak{a}}^{*}-x_{B}^{2}\frac{t}{4m_{N}^{2}}\tilde{\mathcal{E}}_{\mathfrak{a}}\tilde{\mathcal{E}}_{\mathfrak{a}}^{*}\right],\qquad\mathfrak{a}=L,T\nonumber 
\end{align}
where the index $\mathfrak{a}$ distinguishes the transverse and longitudinal
polarizations of the photons, and, inspired by the previous studies
of DVCS and DVMP~\cite{Belitsky:2001ns,Belitsky:2005qn}, we introduced
the double meson form factors

\begin{align}
 & \left.\begin{array}{c}
\mathcal{H}_{\mathfrak{a}}\left(\xi,\,\Delta y,t\right)\\
\mathcal{E}_{\mathfrak{a}}\left(\xi,\,\Delta y,t\right)
\end{array}\right\} =\sum_{\kappa=q,g}\int_{-1}^{1}dx\int_{0}^{1}dz_{1}\int_{0}^{1}dz_{2}\,\varphi_{D_{1}}\left(z_{1}\right)\varphi_{D_{2}}\left(z_{2}\right)C_{\mathfrak{a}}^{(\kappa)}\left(x,\,\xi,\,\Delta y,z_{1},\,z_{2}\right)\times\left\{ \begin{array}{c}
H_{\kappa}\left(x,\xi,t\right)\\
E_{\kappa}\left(x,\xi,t\right)
\end{array}\right.,\label{eq:Ha}\\
 & \left.\begin{array}{c}
\tilde{\mathcal{H}}_{\mathfrak{a}}\left(\xi,\,\Delta y,t\right)\\
\tilde{\mathcal{E}}_{\mathfrak{a}}\left(\xi,\,\Delta y,t\right)
\end{array}\right\} =\sum_{\kappa=q,g}\int_{-1}^{1}dx\int_{0}^{1}dz_{1}\int_{0}^{1}dz_{2}\varphi_{D_{1}}\left(z_{1}\right)\varphi_{D_{2}}\left(z_{2}\right)\,\tilde{C}_{\mathfrak{a}}^{(\kappa)}\left(x,\,\xi,\,\Delta y,t\right)\times\left\{ \begin{array}{c}
\tilde{H}_{\kappa}\left(x,\xi,t\right)\\
\tilde{E}_{\kappa}\left(x,\xi,t\right)
\end{array}\right..\label{eq:ETildeA}
\end{align}
The variables $z_{1},\,z_{2}$, are the light-cone fractions of the
total momentum carried by the quarks in the $D$-mesons. In the evaluation
of~(\ref{eq:Ha},~\ref{eq:ETildeA}) we took into account that the
final-state $D$-mesons are kinematically separated from each other;
for this reason the Fock state of the final system is a direct product
of Fock states of individual $D$-mesons, which in the heavy quark
mass limit might be described by $D$-meson distribution amplitudes~$\varphi_{D}\left(z\right)$.
The detailed definitions of these distributions and discussion of
their parametrizations may be found in Appendix~\ref{sec:DMesonDAs}.
The remaining partonic amplitudes $C_{\mathfrak{a}}^{(\kappa)},\,\tilde{C}_{\mathfrak{a}}^{(\kappa)}$
can be evaluated perturbatively, taking into account the diagrams
shown in Figures~\ref{fig:Photoproduction-B},~\ref{fig:Photoproduction-A}.
The applicability of the perturbation theory is justified for this
purpose in the heavy quark mass limit and large invariant mass $\mathcal{M}_{12}$
of the produced $D$-mesons. The first diagram in the upper row of
Figure~\ref{fig:Photoproduction-B} presents the dominant $\mathcal{O}\left(\alpha_{s}\right)$
contribution; however, it is forbidden kinematically: the heavy quarks
in the produced $D$-mesons carry significant fractions of the momentum,
and thus are expected to have a large positive invariant mass $\left(p_{c}+p_{\bar{c}}\right)^{2}\gtrsim4m_{c}^{2}$,
whereas the photon in the electroproduction process has $q^{2}=-Q^{2}\lesssim0$.
For the same reason, in the next-to-leading order there is no contributions
which merely renormalize the propagators and vertices in this forbidden
``leading order'' diagram. Since we are mostly interested in the
photoproduction regime, in what follows we will focus on the contribution
of the transversely polarized photons; the contribution of the longitudinal
photons is suppressed as $\sim Q/m_{Q}$ and thus may be disregarded.
The complete expressions for $C_{\mathfrak{a}}^{(\kappa)},\,\tilde{C}_{\mathfrak{a}}^{(\kappa)}$
and some technical details of their evaluation may be found in Appendix~\ref{sec:CoefFunction}.

\begin{figure}
\includegraphics[width=8.5cm]{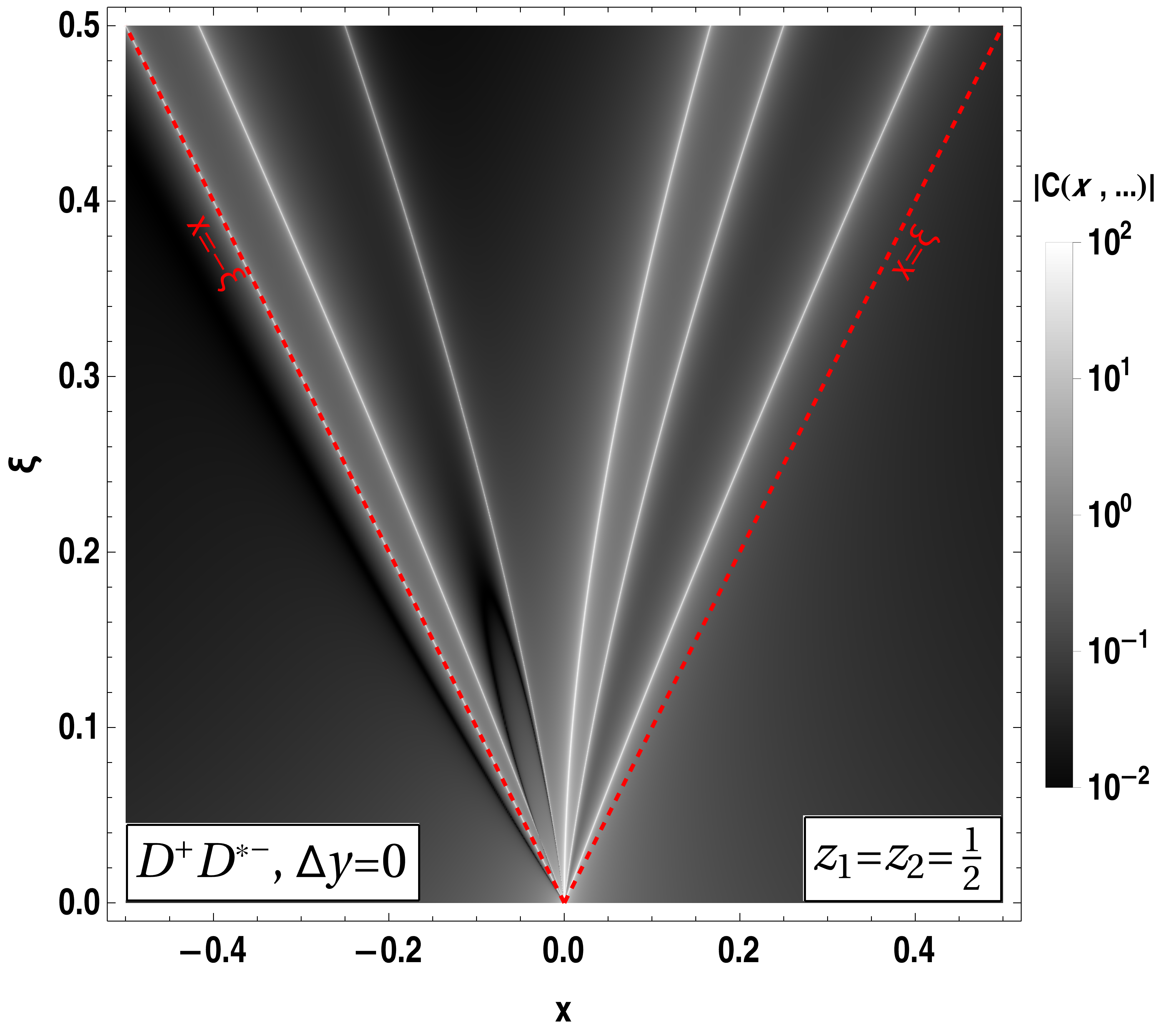}\includegraphics[width=8.5cm]{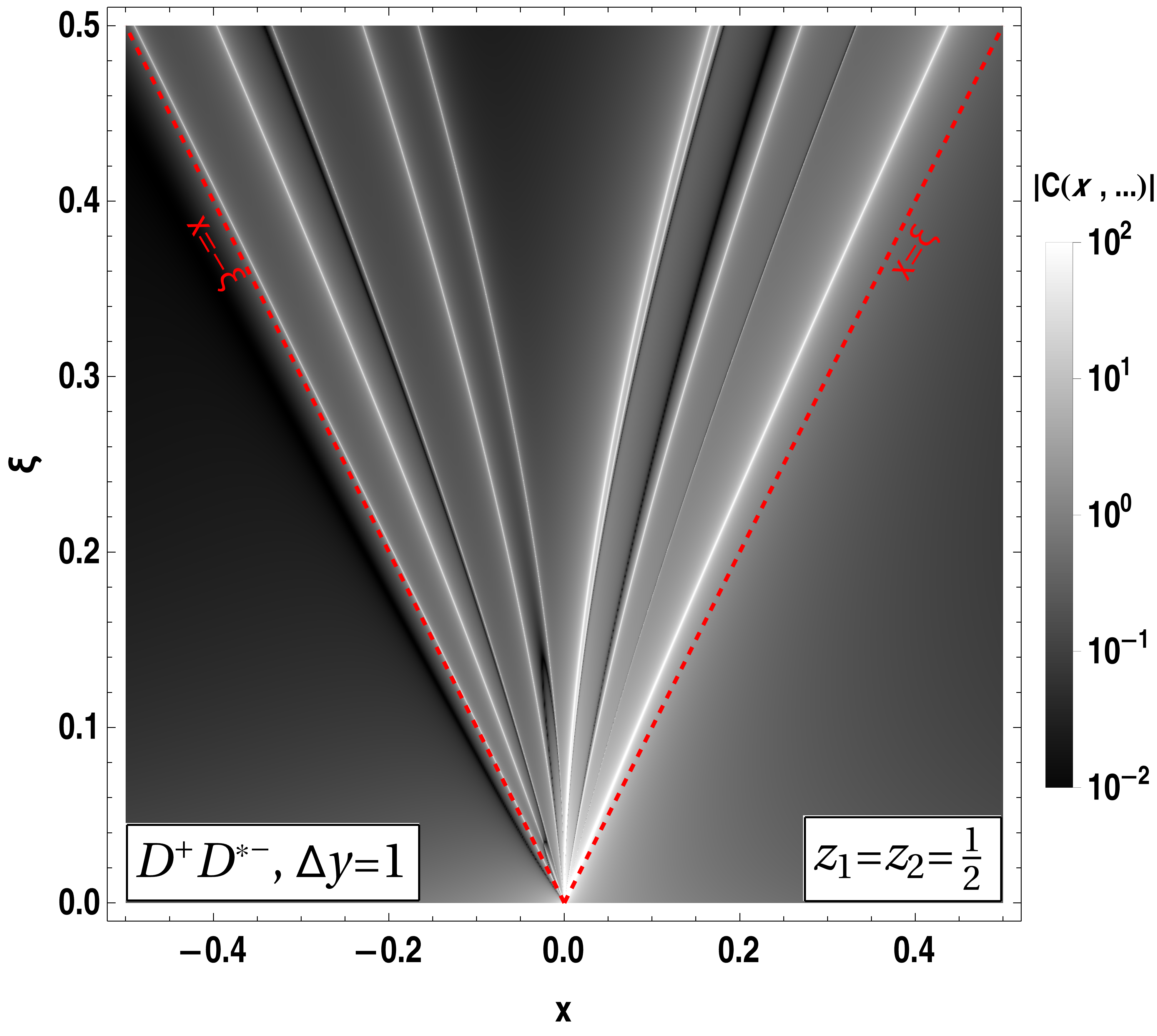}

\includegraphics[width=8.5cm]{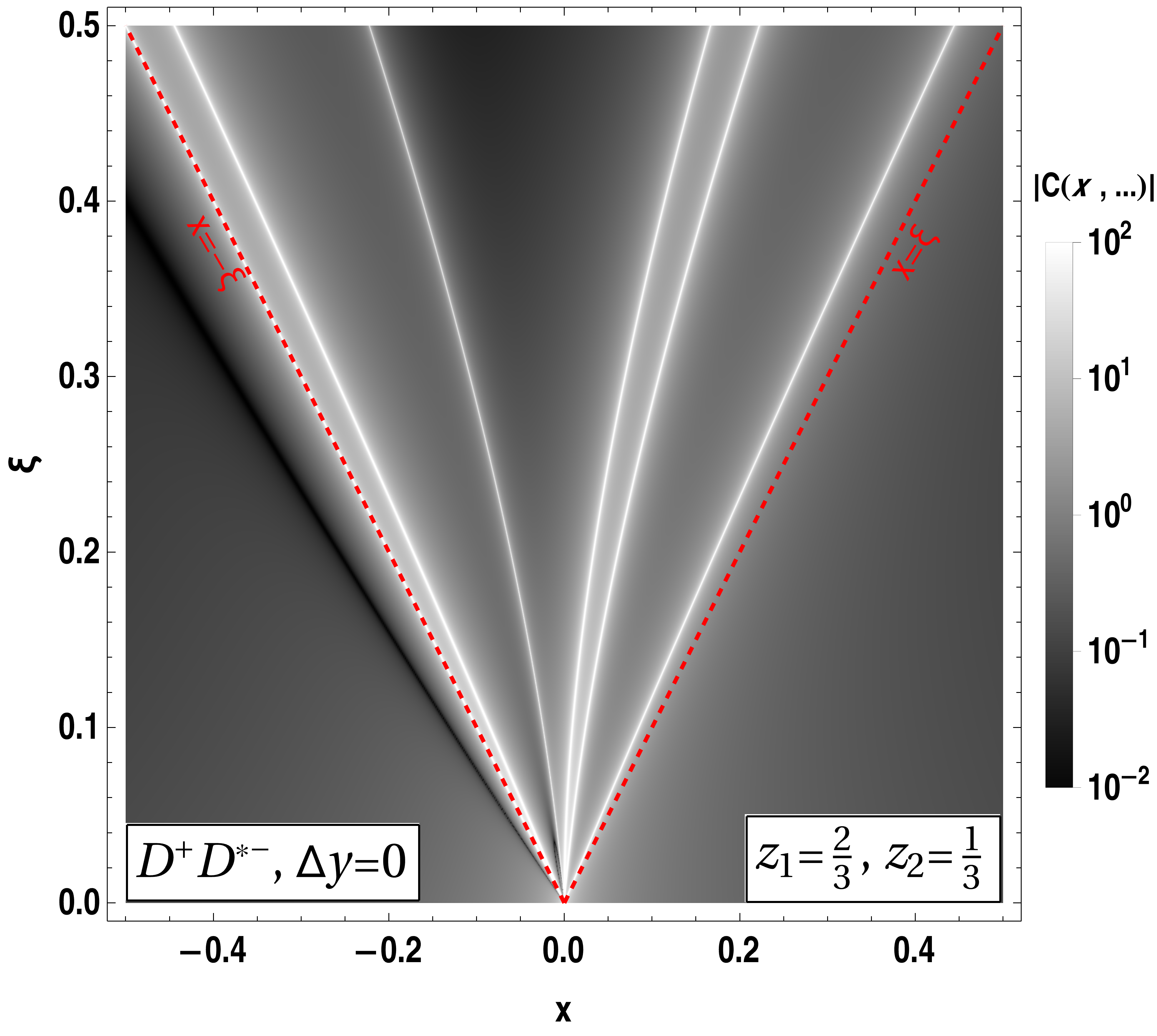}\includegraphics[width=8.5cm]{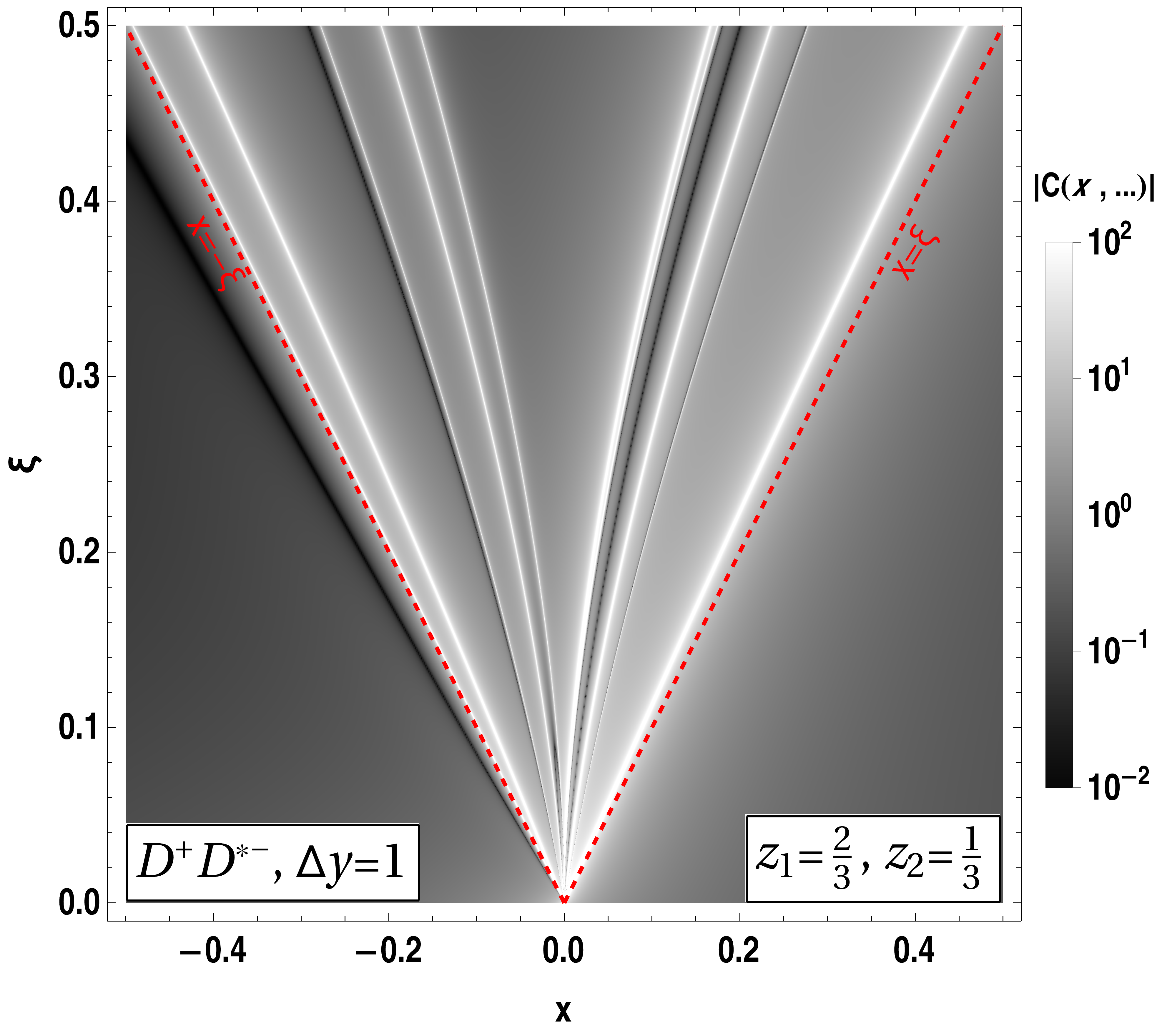}

\caption{\label{fig:CoefFunction} Density plot which illustrates the light
quark coefficient function $C_{T}^{q}$ (in relative units) as a function
of the variables $x$ and skewedness $\xi$, at fixed rapidity difference
$\Delta y$, between the heavy $D$-mesons and the fixed quark light-cone
fractions $z_{1},z_{2}$ in $D$-mesons. The first and the second
row differ by the choice of the values of $z_{1},\,z_{2}$; the left
and the right columns differ by the values of $\Delta y$. For the
sake of definiteness we consider $D^{+}D^{*-}$ in all plots). Thick
white lines effectively demonstrate the position of the poles $x_{k}^{\ell}$
of the coefficient function~(\ref{eq:Monome}). For reference, we
also added red dashed lines $x=\pm\xi$, which separate DGLAP and
ERBL regions.}
\end{figure}

The quark and gluon GPDs of the target contribute in the amplitudes~(\ref{eq:Ha},~\ref{eq:ETildeA})
integrated over the light-cone fraction $x$, and it is important
to understand which region gives the dominant contribution in this
convolution. The $x$-dependence of each individual Feynman diagram
has a form of the rational function of the variable $x$, and for
this reason the coefficient functions $C_{\mathfrak{a}}^{(\kappa)},\,\tilde{C}_{\mathfrak{a}}^{(\kappa)}$
can be represented as a sum of such contributions, 
\begin{equation}
C_{\mathfrak{a}}\left(x,\,y_{1},\,y_{2}\right)\sim\sum_{\ell}\frac{\mathcal{P}_{\ell}\left(x\right)}{\mathcal{Q}_{\ell}\left(x\right)},\label{eq:Monome}
\end{equation}
where the functions $\mathcal{P}_{\ell}\left(x\right),\,\mathcal{Q}_{\ell}\left(x\right)$
are polynomials of the variable $x$. The polynomials $\mathcal{Q}_{\ell}\left(x\right)$
in the denominators can include up to $n_{\ell}$ nodes $x_{k}^{(\ell)}$
in the region of integration, where $n_{\ell}$ is the number of free
propagators in the corresponding Feynman diagram~\footnote{For some diagrams the number of poles might be smaller due to accidental
cancellations of the $x$-dependence.}. The integral near the poles exists only in the principal value sense
and should be evaluated using 
\begin{equation}
\frac{1}{x-x_{k}^{(\ell)}\pm i0}={\rm P.V.}\left(\frac{1}{x-x_{k}^{(\ell)}}\right)\mp i\pi\,\delta\left(x-x_{k}^{(\ell)}\right).
\end{equation}
In Figure~\ref{fig:CoefFunction} we provide the density plots which
show the dependence of the light quark coefficient function $C_{T}^{q}\left(x,\,\xi,\,\Delta y,\,z_{1},\,z_{2}\right)$
on some of its arguments. At fixed $z_{1},\,z_{2}$, the poles show
up as bright lines on a dark background. As could be seen from the
analytic expressions in Appendix~~\ref{sec:CoefFunction}, in the
small-$\xi$ limit all the poles scale as $x_{k}^{(\ell)}\sim\xi$,
with proportionality coefficient which depends on $\Delta y,\,z_{1},\,z_{2}$;
for this reason, all the pole trajectories (bright lines in the density
plot~\ref{fig:CoefFunction}) are nearly straight. However, in the
final result the coefficient functions contribute in the convolution
with relatively broad $D$-meson distribution amplitudes, via the
effective (integrated) coefficient function 
\begin{align}
C_{{\rm int}}^{q}\left(x,\,\xi,\,\Delta y\right) & \equiv\int_{0}^{1}dz_{1}\int_{0}^{1}dz_{2}\,\,\varphi_{D_{1}}\left(z_{1}\right)\varphi_{D_{2}}\left(z_{2}\right)C_{T}^{q}\left(x,\,\xi,\,\Delta y,\,z_{1},\,z_{2}\right).\label{eq:Cintq}
\end{align}
As we can see from the next Figure,~\ref{fig:CoefFunction-smeared},
the function $C_{{\rm int}}^{q}$ does not have any singularities.
The imaginary part of $C_{{\rm int}}^{q}$ appears due to a deformation
of the integration contours in the integrals over $z_{1},\,z_{2}$
near the poles, which is carried out using the conventional $\xi\to\xi-i0$
prescription~\footnote{Due to the sophisticated structure of the expressions, we perform
this evaluation numerically, replacing $\xi\to\xi\left(1-i\varepsilon\right)$
with $\varepsilon\sim10^{-2}-10^{-3}$ (we checked that the result
has a very mild dependence on the choice of $\varepsilon$).}. The integrated function $C_{{\rm int}}^{q}$ is mostly concentrated
in the region $|x|\le\xi$, which suggests that the process is mainly
sensitive to the behavior of GPDs in that domain (the so-called ERBL
region). For the gluonic coefficient function, we observe a similar
behavior: there are poles for fixed $z_{1},\,z_{2}$, alghough they
are smeared after convolution with the distribution amplitudes.

\begin{figure}
\includegraphics[height=7cm]{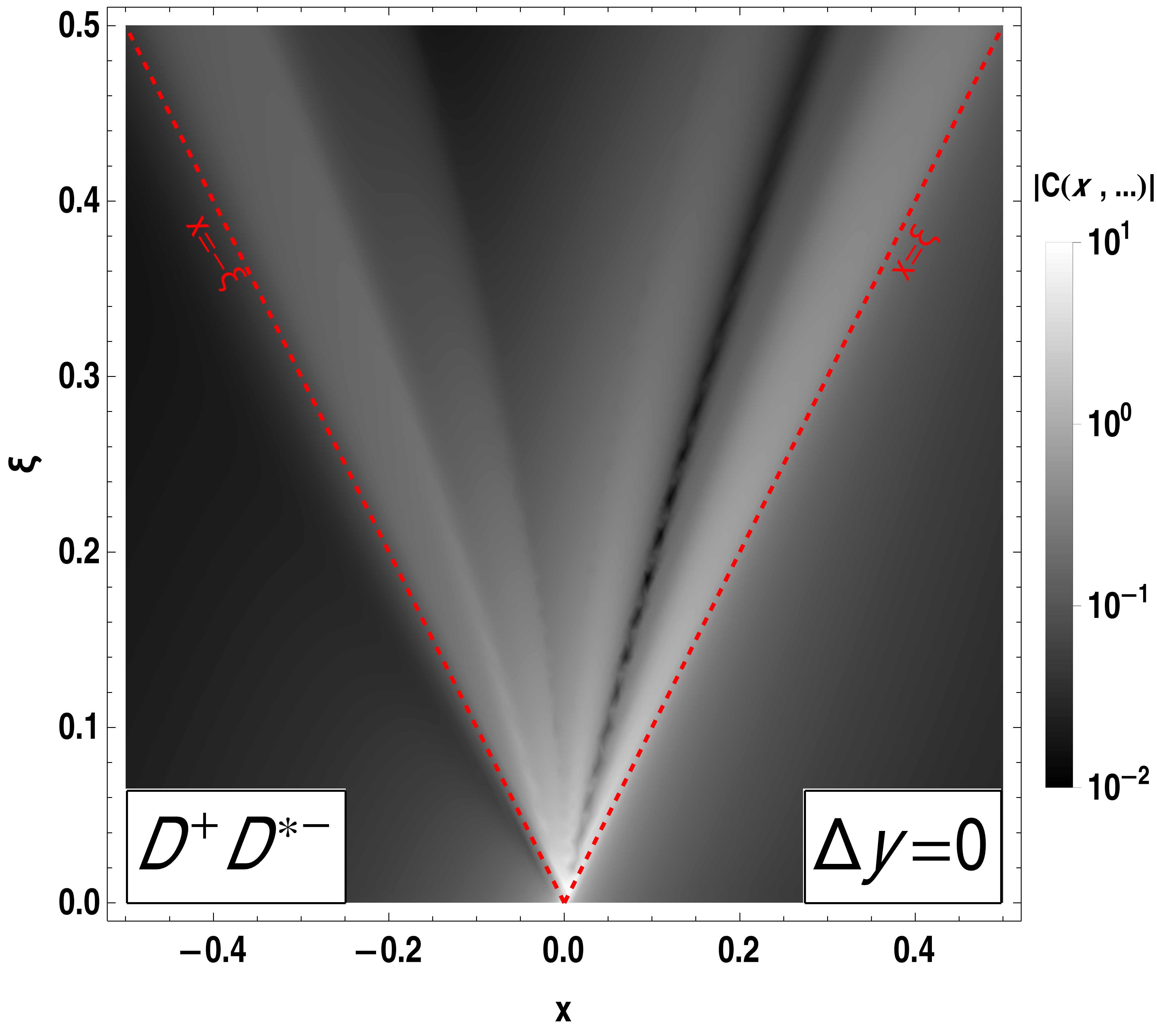}\includegraphics[height=7cm]{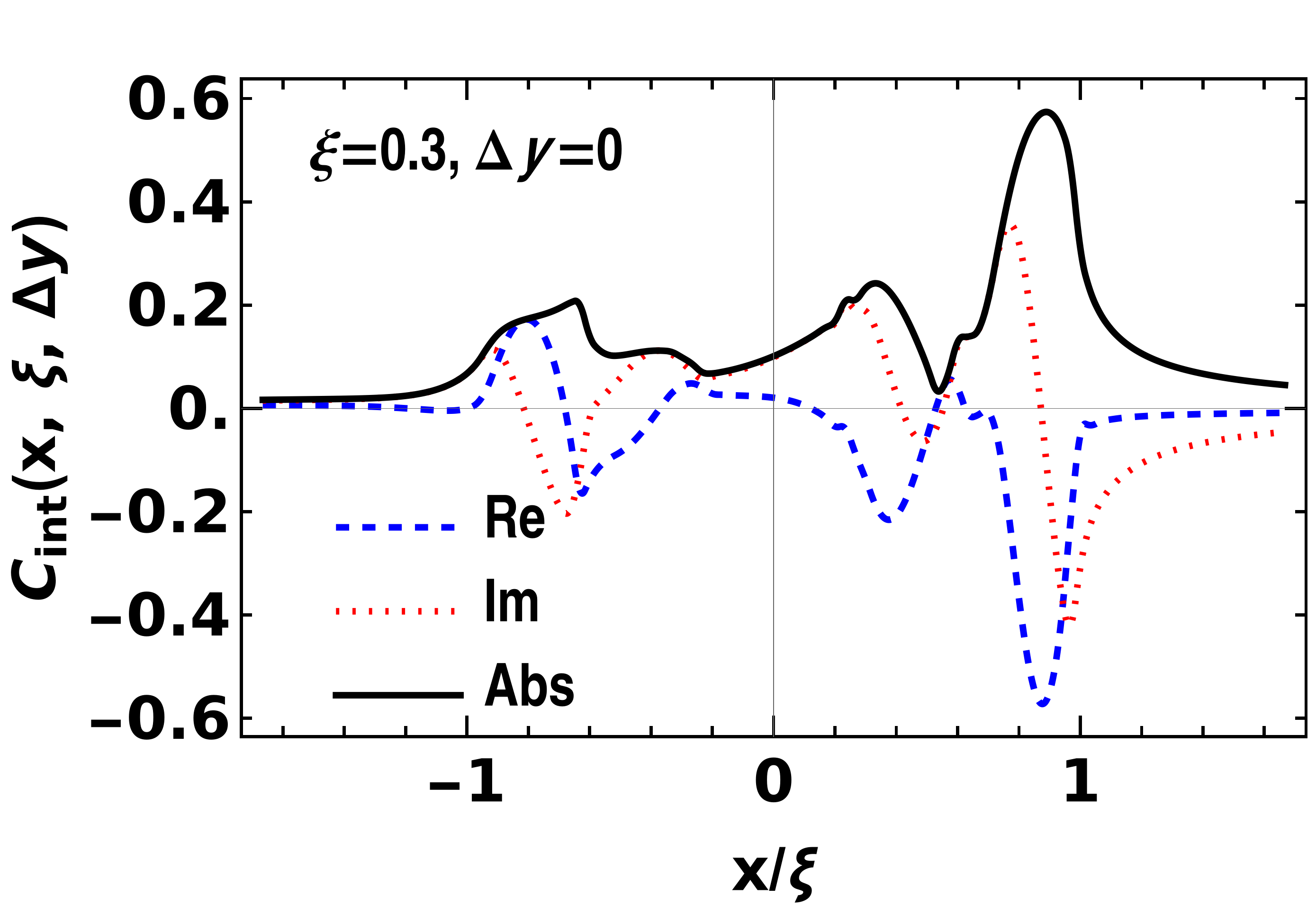}

\caption{\label{fig:CoefFunction-smeared} Left: Density plot which shows the
\emph{integrated} coefficient function $C_{{\rm int}}^{q}$ (\ref{eq:Cintq}),
as a function of the variables $x$ and skewedness $\xi$, at fixed
rapidity difference $\Delta y$ between the heavy $D$-mesons. Right:
The real, imaginary and absolute parts of the coefficient function
$C_{{\rm int}}^{q}$ as a function of $x/\xi$, at fixed $\xi,\,\Delta y$.
For the sake of definiteness we consider $D^{+}D^{*-}$ in all plots.
Both plots illustrate that the singularities were smeared after convolution
with the final-state $D$-meson distribution amplitudes. Similar behavior
is observed for other choices of kinematics and other mesons.}
\end{figure}

\section{Numerical results}

\label{sec:Numer} 
\begin{figure}
\includegraphics[width=5.7cm]{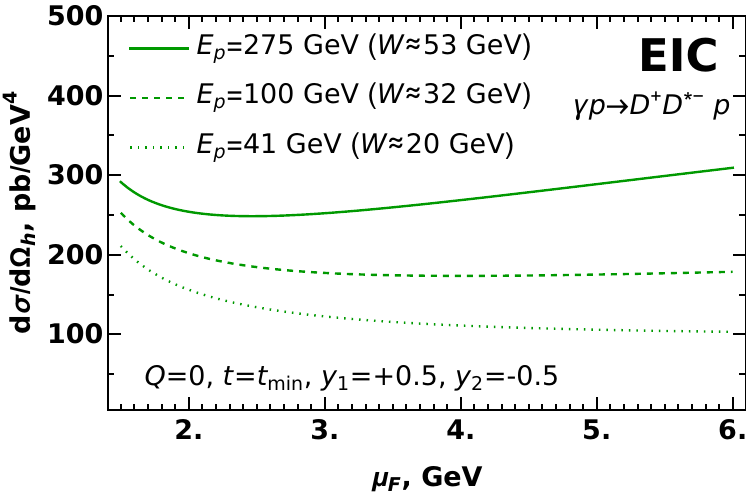} \includegraphics[width=5.7cm]{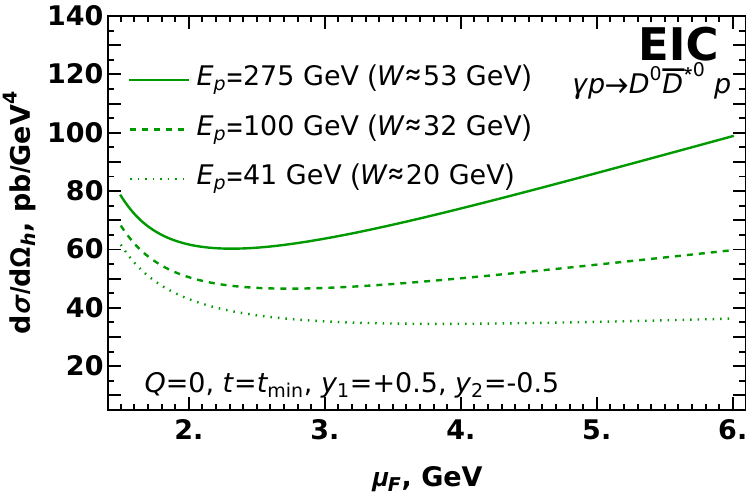}
\includegraphics[width=5.7cm]{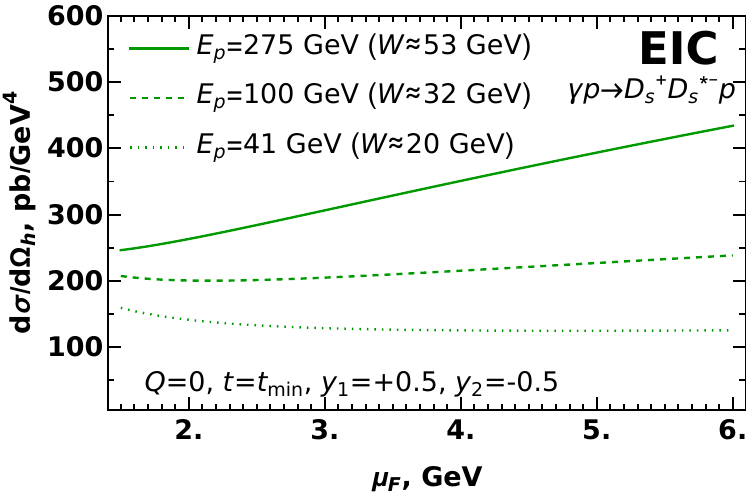} \caption{\label{fig:muFDependenceAll} The factorization scale dependence of
the cross-section, for different $D$-meson pairs. For the sake of
definiteness we consider the production at central rapidities ($Y=(y_{1}+y_{2})/2\approx0$),
for different energies $E_{p}$ of the proton beam. All these values
in the photoproduction regime correspond to small $x_{B}\lesssim5\times10^{-2}\ll1$.
All frame-dependent variables are given in the reference frame described
in Section~\ref{subsec:Kinematics}.}
\end{figure}

For the sake of definiteness we will make predictions
using the Kroll-Goloskokov parametrization of the GPDs~\cite{Goloskokov:2006hr,Goloskokov:2007nt,Goloskokov:2008ib,Goloskokov:2009ia,Goloskokov:2011rd,Goloskokov:2013mba}.
This parametrization effectively incorporates the evolution of the
generalized parton distributions, introducing a mild dependence of
the model parameters on the factorization scale $\mu_{F}$. In Figure~\ref{fig:muFDependenceAll}
we show how the cross-sections depend on the choice of this factorization
scale. This dependence is mild at moderate energies, but becomes very
pronounced at very high energies (small $x_{B}$). Such behavior is
not surprising: it is known from studies of other channels~\cite{DVMPcc1,DVMPcc2,DVMPcc3,DVMPcc4}
that this dependence exists, due to the omitted higher order corrections,
which become especially important in the kinematics of small-$x_{B}$.
in what follows, for the sake of definiteness, we will choose the
factorization scale $\mu_{F}=\mu_{R}=4\,{\rm GeV}\approx2M_{D}$.

\begin{figure}
\includegraphics[width=9cm]{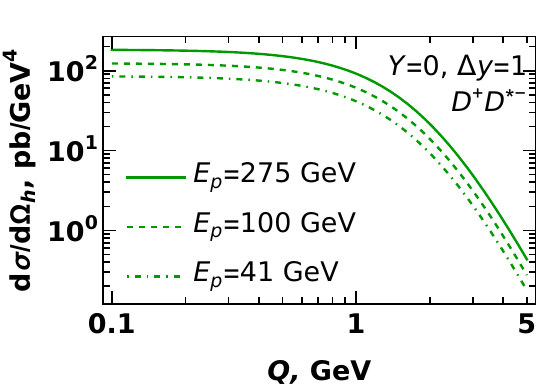}\caption{\label{fig:QDep}Dependence of the photoproduction cross-section~(\ref{eq:Photo-1})
on the virtuality $Q$ of the photon. In the left and right plots
we compare predictions for different rapidities $y_{J/\psi},y_{\eta_{c}}$
and different proton energies $E_{p}$. Both plots clearly illustrates
the transition from photoproduction to Bjorken regime in the region
$Q\sim1-2\,M_{J/\psi}$. In both plots the photon energy is evaluated
from~(\ref{eq:qPhoton-2},\ref{qPlus-1}). All frame-dependent variables
are given in the reference frame described in Section~\ref{subsec:Kinematics}.}
\end{figure}

In Figure~\ref{fig:QDep} we can see the $Q$-dependence of the cross-section
of the photoproduction subprocess $\gamma^{(*)}p\to M_{1}M_{2}p$.
The dependence is very mild up to $Q^{2}\lesssim\left(M_{1}+M_{2}\right)^{2}$,
since in the hard amplitudes the contribution of the $\mathcal{O}\left(Q^{2}\right)$
terms is negligibly compared to the $\mathcal{O}\left(\mathcal{M}_{12}^{2}\right)$
contributions. However, for large $Q^{2}\gg\left(M_{1}+M_{2}\right)^{2}$
the virtuality $Q^{2}$ turns into the hard scale and leads to strong
suppression of the cross-section. In electroproduction experiments
the flux of equivalent photons decreases rapidly as a function of
$Q^{2}$, and thus the kinematics where $Q$-dependence becomes pronounced,
is hardly achievable in the foreseeable future. For this reason in
what follows we will focus only on the photoproduction regime $Q\approx0$.

\begin{figure}
\includegraphics[width=5.5cm]{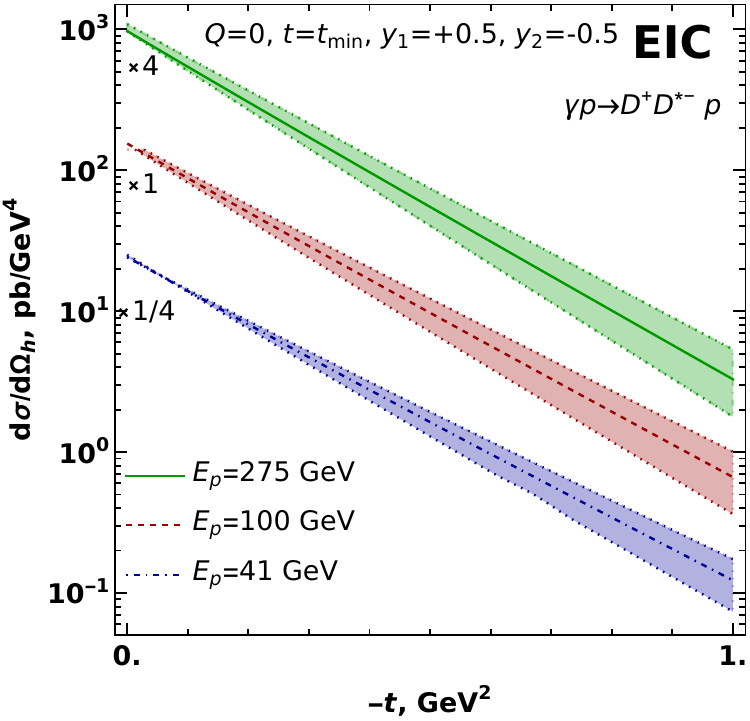}\includegraphics[width=5.5cm]{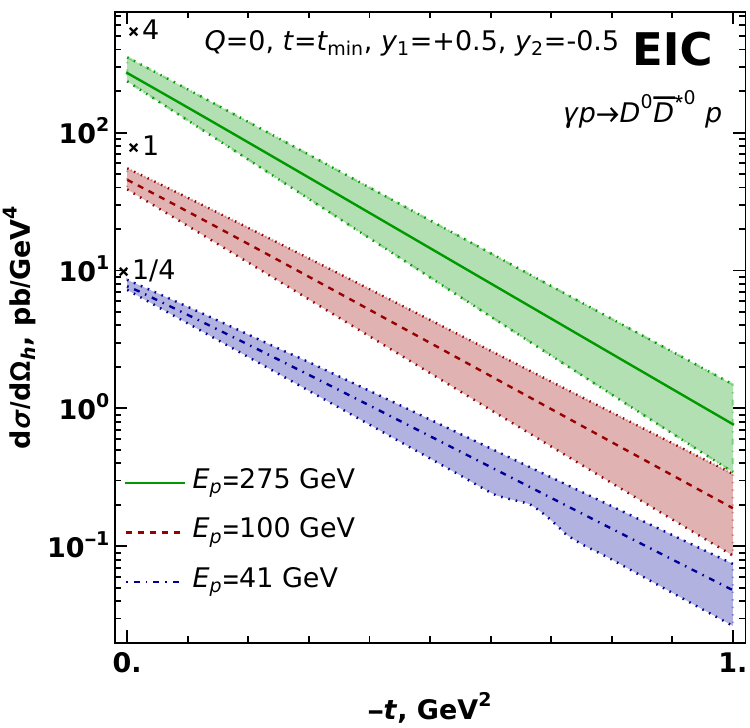}\includegraphics[width=5.5cm]{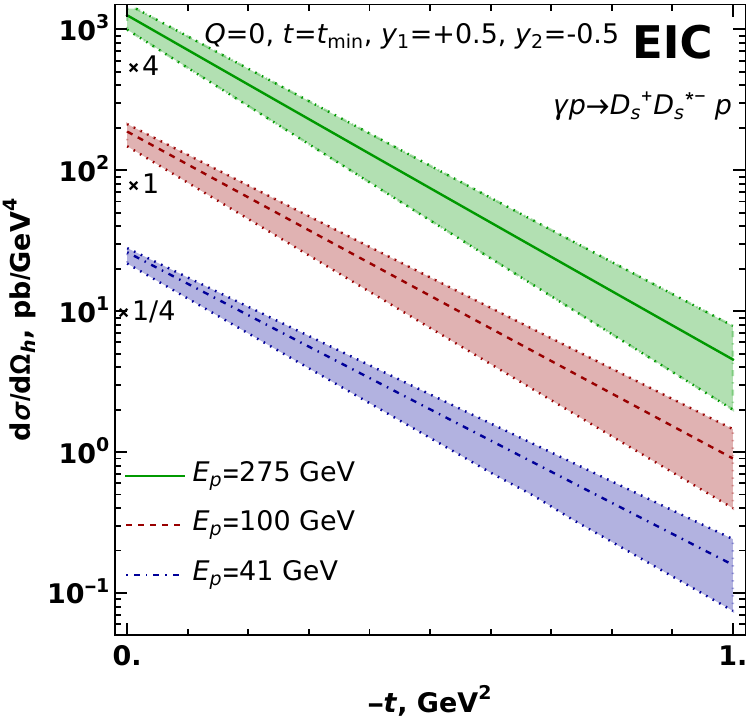}

\caption{\label{fig:tDep-1} The dependence of the production cross-sections
on the invariant momentum transfer $t$ for different meson pairs.
In all plots the colored bands reflect the uncertainty due to choice
of the factorization scale $\mu_{F}$; the central line corresponds
to factorization scale $\mu_{F}=4\,{\rm GeV}\approx2m_{D}$, whereas
the upper and lower limits of the colored bands correspond to $\mu_{F}=2\,{\rm GeV}$
and $\mu_{F}=8\,{\rm GeV}$, respectively. For better legibility the
cross-sections for $E_{p}={\rm 275}\,{\rm GeV}$ and $E_{p}=41\,{\rm GeV}$
are multiplied by constant factors $\times4$ and $\times1/4$ respectively
(shown near the left edge of each plot).}
\end{figure}

\begin{figure}
\includegraphics[width=8cm]{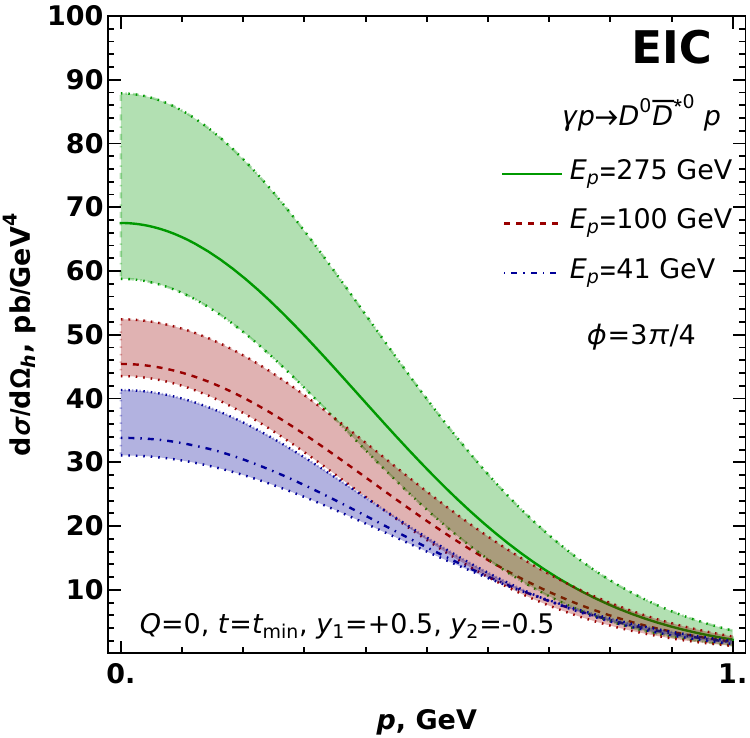}\includegraphics[width=8cm]{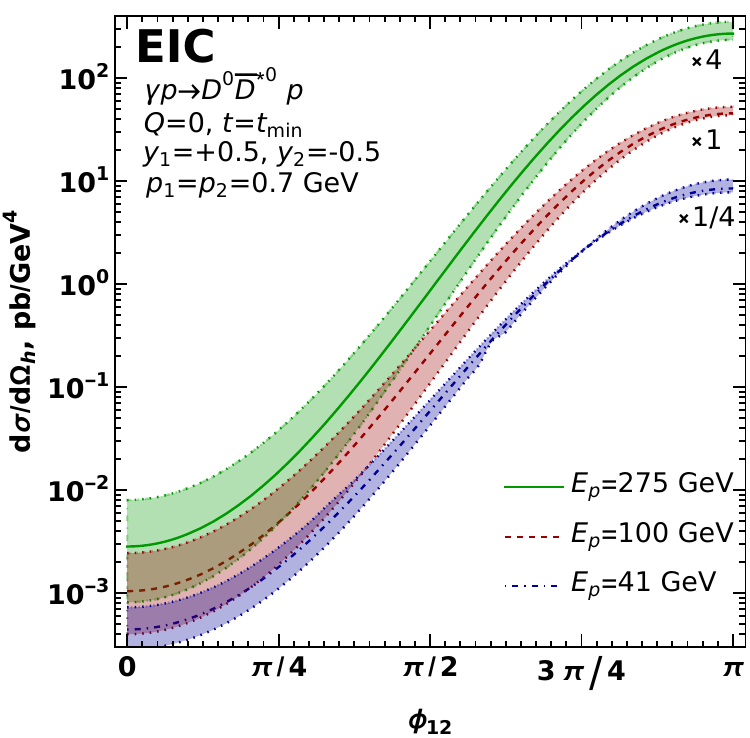}

\caption{\label{fig:tDep} Left plot: Dependence of the cross-section on the
transverse momentum $p_{T}$ of produced $D$-mesons, at fixed azimuthal
angle $\phi_{12}$ between the transverse momenta. Right plot: Dependence
on the azimuthal angle $\phi_{12}$, at fixed transverse momentum
$p_{T}$ of the produced $D$-mesons. The colored bands reflect the
uncertainty due to choice of the factorization scale $\mu_{F}$: the
central line corresponds to factorization scale $\mu_{F}=4\,{\rm GeV}\approx2m_{D}$,
whereas the upper and lower limits were evaluated with $\mu_{F}=8\,{\rm GeV}$
and $\mu_{F}=2\,{\rm GeV}$ respectively. In the right plot, for better
legibility, the cross-sections for $E_{p}={\rm 275}\,{\rm GeV}$ and
$E_{p}=41\,{\rm GeV}$ are multiplied by constant factors $\times4$
and $\times1/4$ respectively (shown near the right edge of each plot).
A sharp peak at large angles $\phi_{12}\approx\pi$ and relatively
small transverse momenta corresponds to the so-called back-to-back
kinematics, which minimizes the invariant momentum transfer $|t|$,
as can be seen from Eq~(\ref{eq:tDef}).}
\end{figure}

In the Figures~(\ref{fig:tDep-1},\ref{fig:tDep}) we show the dependence
of the cross-section~(\ref{eq:Photo-1}) on the momentum transfer
$t$ to the target, and the related distributions of the produced
$D$-mesons on transverse momenta and angle between them in the photon-proton
frame. In the collinear factorization picture, the transverse momenta
are disregarded in evaluation of the coefficient function, for this
reason this dependence stems entirely from the $t$-dependence implemented
in GPDs. The phenomenological analyses suggest that this dependence
should exhibit a very pronounced (nearly exponential) suppression
as a function of $|t|$. For this reason, the $D$-meson pairs are
produced predominantly with small oppositely directed ($\phi\approx\pi$)
transverse momenta, the so-called back-to-back kinematics which minimizes
the momentum transfer $t$ to the target.

\begin{figure}
\includegraphics[width=6cm]{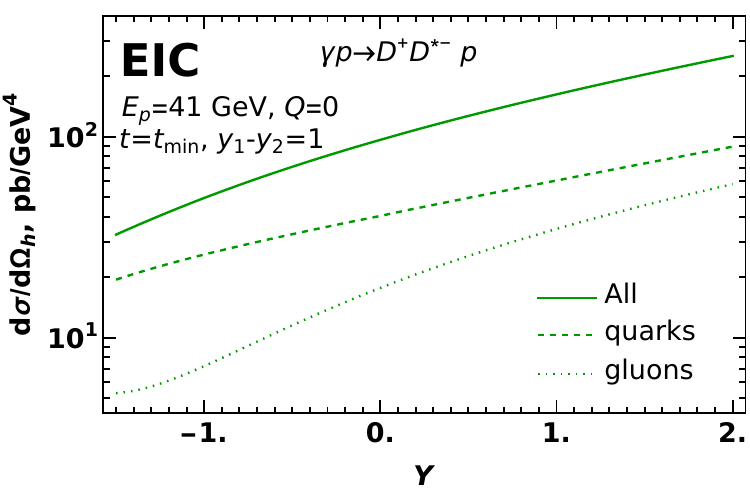}\includegraphics[width=6cm]{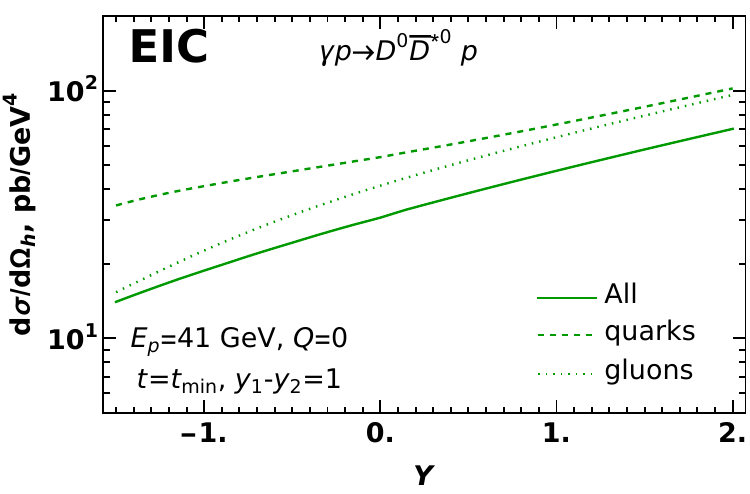}\includegraphics[width=6cm]{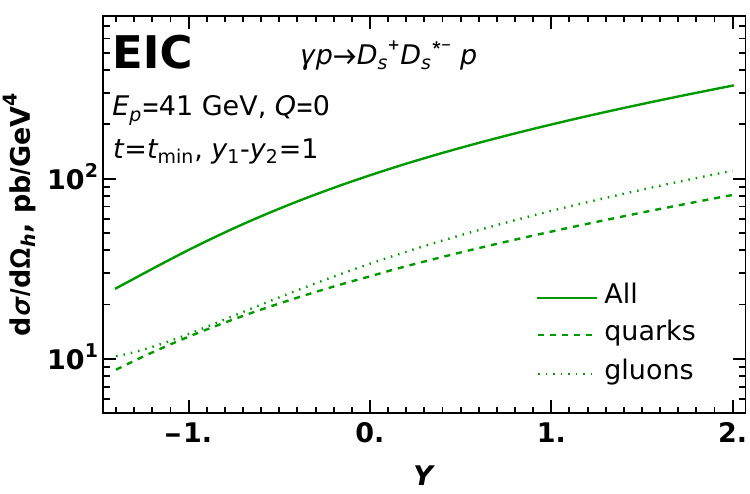}

\includegraphics[width=6cm]{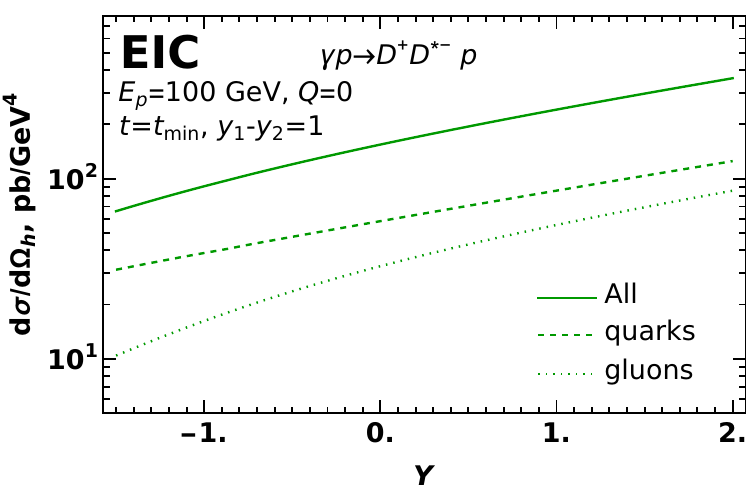}\includegraphics[width=6cm]{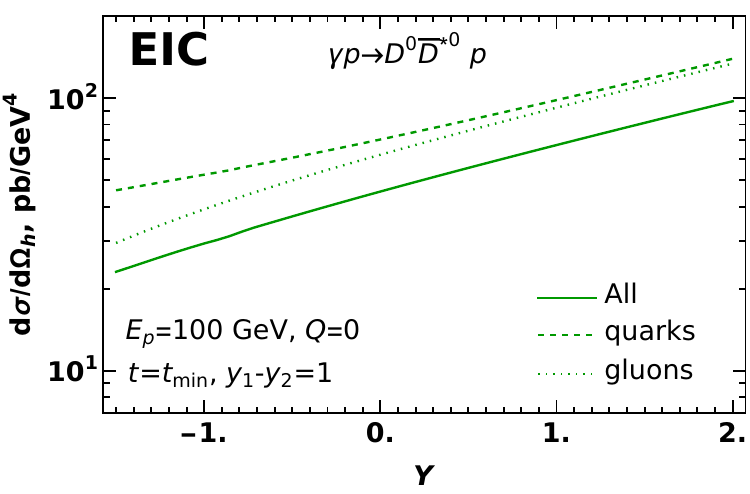}\includegraphics[width=6cm]{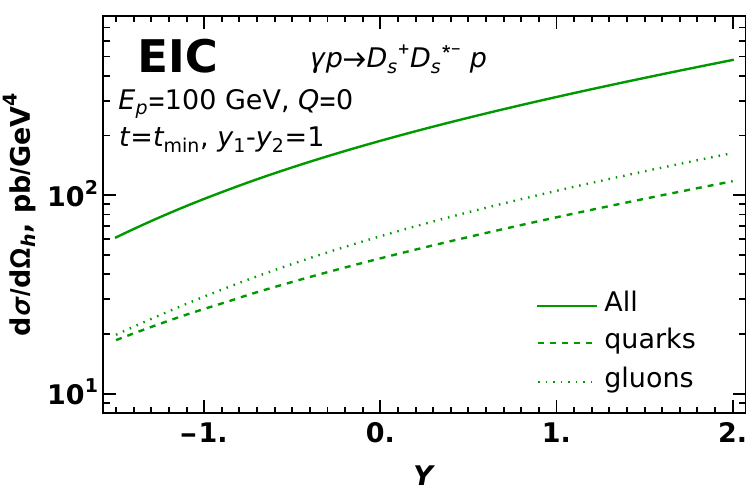}

\includegraphics[width=6cm]{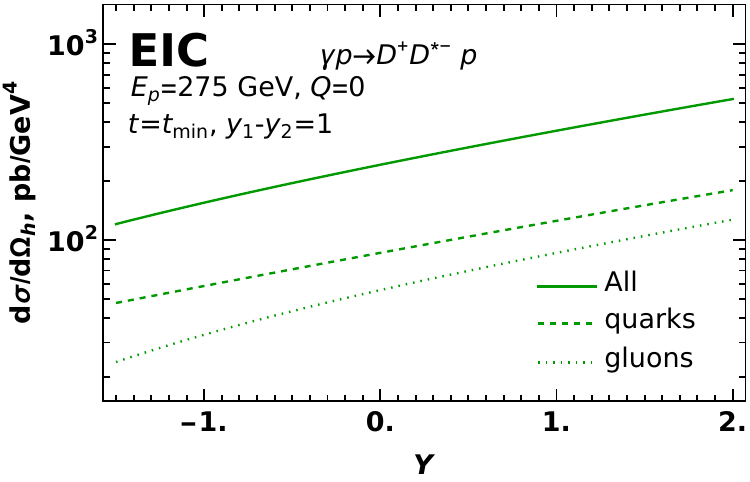}\includegraphics[width=6cm]{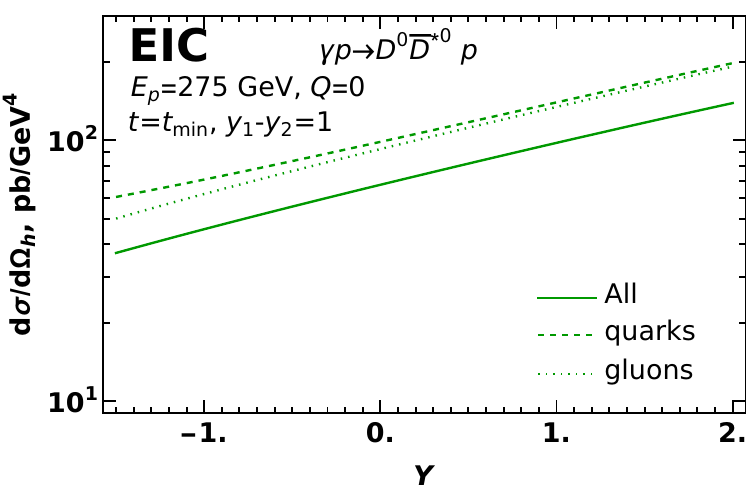}\includegraphics[width=6cm]{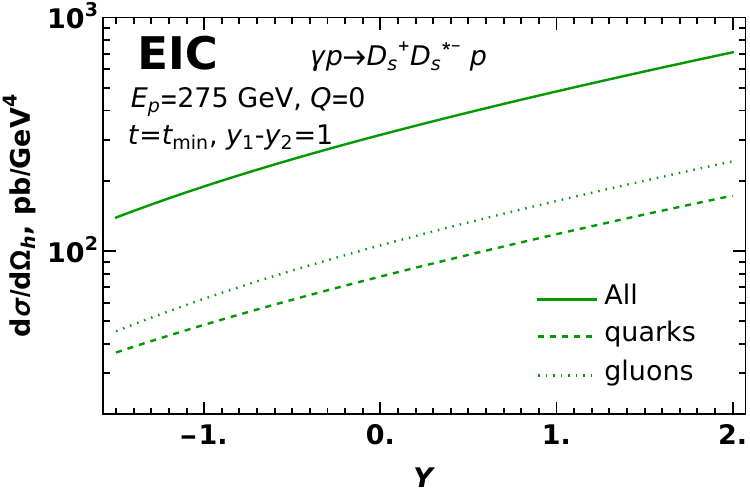}

\caption{\label{fig:yDep}Dependence of the cross-section on the average rapidity
$Y=\left(y_{1}+y_{2}\right)/2$, at fixed rapidity difference $y_{1}-y_{2}$
of the two mesons (positive rapidity is in direction of the photon/electron).
The upper row corresponds to proton energy $E_{p}\approx41$~GeV,
the middle row is for energy $E_{p}\approx100\,{\rm GeV}$, and the
lower row is for $E_{p}\approx275$~GeV. The curves marked as \textquotedblleft quarks\textquotedblright{}
and \textquotedblleft gluons\textquotedblright{} correspond to contributions
of only quark or only gluon GPDs. The curve marked as \textquotedblleft All\textquotedblright{}
takes them all into account (as well as additional  term due to interference
of quark and gluon contributions in the cross-section). See the text
for more detailed discussion.}
\end{figure}

\begin{figure}
\includegraphics[width=6cm]{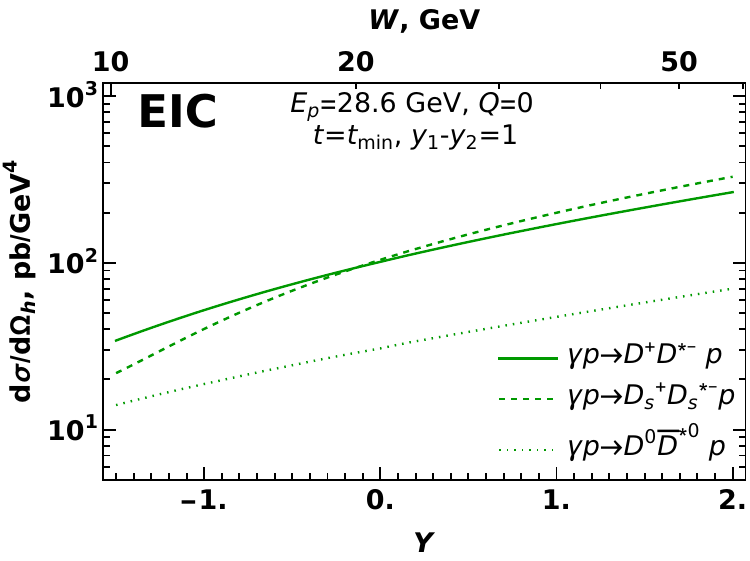}\includegraphics[width=6cm]{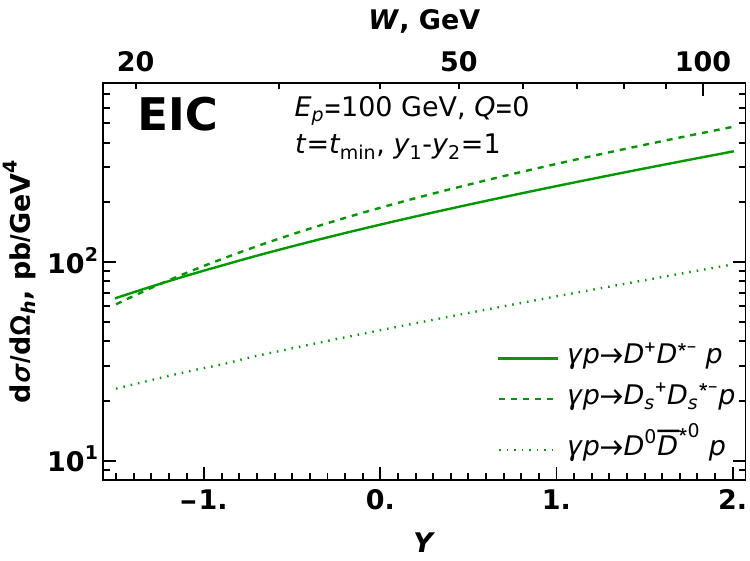}\includegraphics[width=6cm]{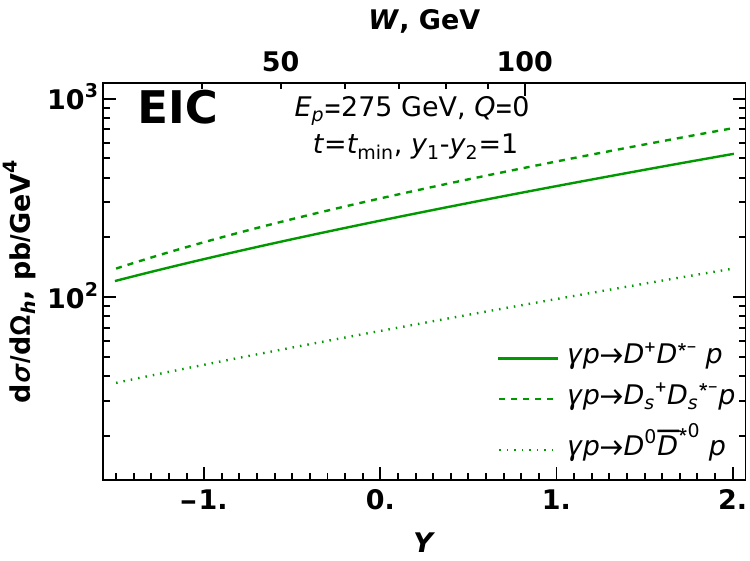}

\caption{\label{fig:yDep-1}Side-by-side comparison of the cross-sections,
for different mesons and different proton energies $E_{p}$. The lower
horizontal axis shows the average rapidity $Y=\left(y_{1}+y_{2}\right)/2$,
whereas in the upper axis we show the corresponding values of invariant
energy $W\equiv\sqrt{s_{\gamma p}}$. A comparison of different plots
shows that the cross-sections for the same $W$, but different proton
energy $E_{p}$ and $Y$, coincide with each other.}
\end{figure}

In Figure~\ref{fig:yDep} we show the dependence of the cross-section
on the rapidities of the produced quarkonia. For the sake of definiteness,
we assumed that the $D$-mesons are kinematically separated by a constant
rapidity gap $\text{\ensuremath{\Delta y=1}}$ units. The growth of
the cross-section with $Y$ can be understood if we take into account
that in the chosen kinematics, the increase of $Y$ leads to an increase
of the invariant energy $W^{2}$, a corresponding decrease of $x_{B},\,\xi$
and a growth of the gluon GPDs. For the same reason, the cross-section
has a mild dependence on proton energy $E_{p}$ at constant average
rapidity $Y=(y_{1}+y_{2})/2$. The magnitude of the cross-section
depends significantly on the quantum numbers (flavor content) of the
produced $D$-mesons. It is instructive to understand the main sources
of this dependence. As we discussed earlier in Section~\ref{subsec:Kinematics},
our choice of kinematics corresponds to relatively small values of
$x_{B},\xi\ll1$, and therefore we expect that all quark distributions
are dominated by sea quarks, which have a mild dependence on flavor.
The mass of the light quarks might be disregarded in the collinear
approximation. However, some terms in the coefficient function, namely
the diagrams which correspond to coupling of the photon to light quarks,
include flavor-dependent electric charges $e_{\ell}$ in prefactors.
Their interference with diagrams which correspond to the coupling
of the photon to heavy quarks can be constructive or destructive.
Furthermore, there is additional flavor dependence introduced by the
meson decay constant $f_{D}\approx209\,{\rm MeV},f_{D_{s}}\approx249\,{\rm MeV}$,
which can change the result for the cross-section by up to a factor
of two, since the decay constant contributes as $\sim f_{D}^{4}$.
For the $D^{+}D^{*-}$ mesons at small $(Y,W)$, the dominant contributions
stem from the quark sector, and the contribution of the gluons becomes
more pronounced at higher $Y$. For charged strange mesons $D_{s}^{+}D_{s}^{*-}$,
the contribution of the quark sector is slightly smaller and is on
par with the contribution of the gluons. Finally, for neutral $D^{0}\overline{D}^{*0}$
pairs, due to destructive interference of the contributions of quark
and gluon GPDs, the cross-section is smaller than for the other mesons.
In Figure~\ref{fig:yDep-1} we compare side-by-side the cross-sections
for different mesons, as well as show (in the upper horizontal axis)
the dependence on the invariant energy $W_{\gamma p}$. We may observe
that the cross-sections have the same value for the same invariant
energy $W_{\gamma p}$, independently of the proton energy $E_{p}$
or rapidity $Y$.

\begin{figure}
\includegraphics[width=6cm]{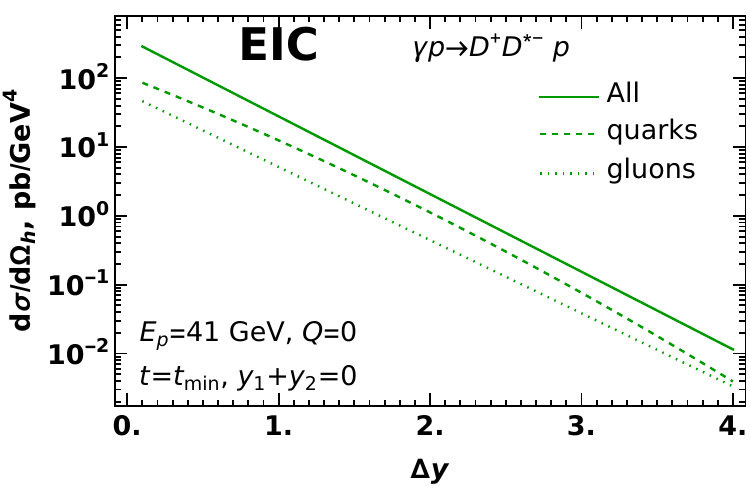}\includegraphics[width=6cm]{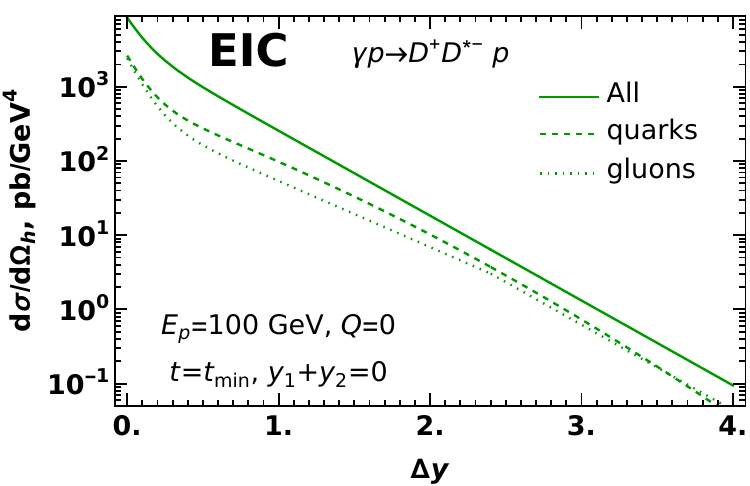}\includegraphics[width=6cm]{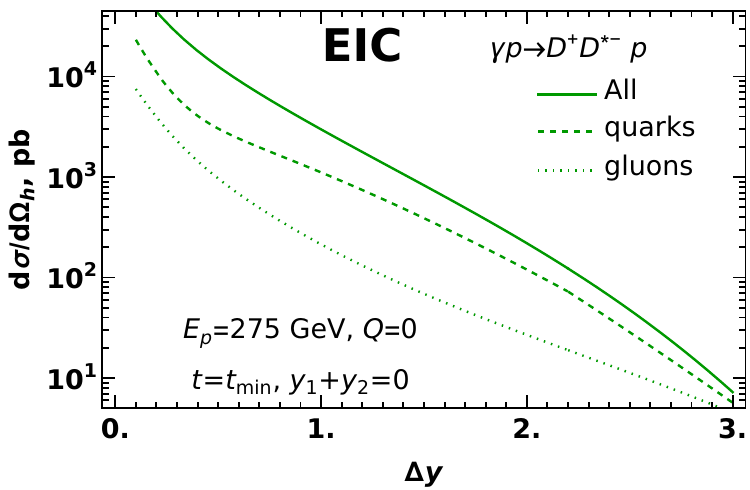}

\caption{\label{fig:dyDep}Dependence of the $D^{+}D^{*-}$ production cross-section
on the rapidity difference $\Delta y$, at fixed average rapidity
$Y=\left(y_{1}+y_{2}\right)/2$ and different proton energies $E_{p}$.
For other mesons we observe a similar behavior.}
\end{figure}

In Figure~\ref{fig:dyDep} we show the dependence of the cross-section
on the rapidity difference $\Delta y$ between the two $D$-mesons
at central rapidities. The cross-section decreases rapidly as function
of $\Delta y$, and similarly for the quark and gluon contributions.
This behavior might be explained by an increase of the variables $x_{B},\xi$,
and of the (minimal, longitudinal) momentum transfer $\left|t_{{\rm min}}\right|$
at large $\Delta y$ and fixed $Y$; this leads to a suppression of
both quark and gluon contributions, due to $t$-dependence encoded
in the partonic GPDs. In the limit $\Delta y\to0$ the cross-sections
remain finite, although numerically grow up to large values. Since
our approach is justified only for the kinematically separated $D$-meson
pairs, we do not consider that region.

\begin{figure}
\includegraphics[width=9cm]{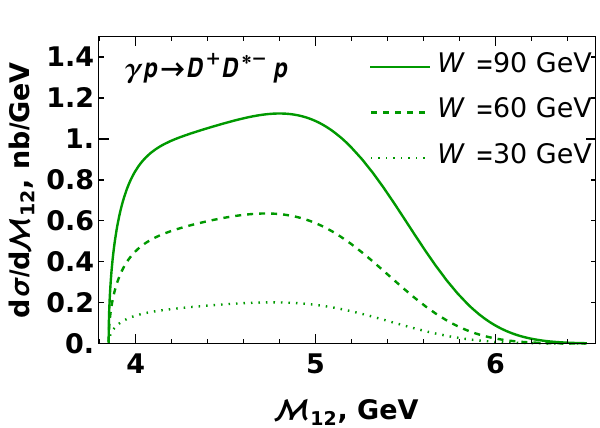}\includegraphics[width=9cm]{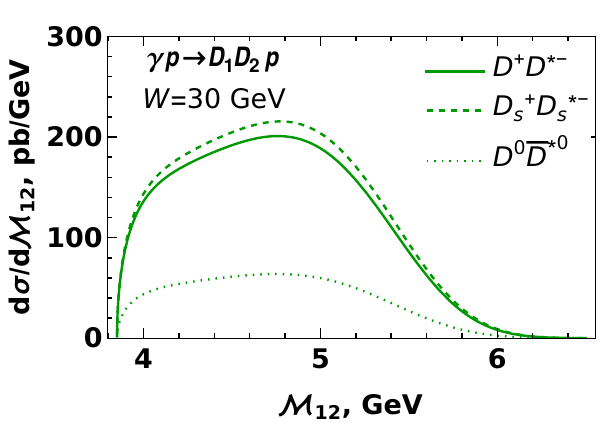}\caption{\label{fig:M12Dep} The distribution of the produced $D$-meson pairs
over their invariant mass ${\mathcal{M}}_{12}$, for several fixed
invariant energies $W$ of the $\gamma p$ collision (left plot) and
for different flavors at fixed $W$ (right plot). In the near-threshold
region $\mathcal{M}_{12}\approx2M_{D}\,(\text{\ensuremath{\mathcal{M}}}_{12}\lesssim4\,{\rm GeV})$
the relative velocity $v_{{\rm rel}}$ of the produced $D$-mesons
is small, and thus our approach might be not reliable (see the text
for more explanation).}
\end{figure}

Finally, in Figure~\ref{fig:M12Dep} we provide predictions for the
distribution of the produced $D$-meson pairs over their invariant
mass ${\mathcal{M}}_{12}$. For all flavors of $D$-mesons and all
energies $W$ of $\gamma p$ pairs, the distributions have a very
similar shape. Near the threshold, the cross-section grows due to
increase of the phase volume. However, in this region the relative
velocity $v_{{\rm rel}}$ of the $D$-mesons is small, and thus potentially
sizeable corrections might appear due to formation of the bound states
(so-called tetraquarks). For higher values of $\text{\ensuremath{\mathcal{M}_{12}\gtrsim4\,{\rm GeV}}}$
the values of $v_{{\rm rel}}$ become large enough to exclude such
soft near-threshold effects, and the collinear theory becomes well-justified.
The region of very large values of $\mathcal{M}_{12}\gtrsim6$~GeV
requires large rapidity difference $\Delta y$ or large transverse
momenta of the heavy mesons. In this kinematics the momentum transfer
$t$ to the proton is large, so the cross-section becomes suppressed,
in agreement with results shown in Figures~(\ref{fig:dyDep},\ref{fig:tDep}).

We also analyzed a possibility to study the suggested channels in
the kinematics of the proposed 22 GeV upgrade at JLab~\cite{Accardi:2023chb}.
However, we found that the cross-sections are extremely small in that
kinematics and beyond the reach of experimental studies. This happens
because in the kinematics of the proposed upgrade, the production
would occur with relatively large values $x_{B},\xi\sim\mathcal{M}_{12}^{2}/W^{2}\sim0.5-1$,
where the GPDs are strongly suppressed, both due to endpoint behavior
of the underlying parton distributions $\sim\left(1-x\right)^{n},\quad n=3-5$,
and additional suppression due to increase of the (longitudinal) momenta
transfer to the target, $|t|\gtrsim|t|_{{\rm min}}=m_{N}^{2}x_{B}^{2}/(1-x_{B})$.
We also made similar estimates for photoproduction of $B$-meson pairs,
yet found that even in the EIC kinematics the cross-sections are extremely
small (sub-picobarn level). This could be understood from the structure
of the coefficient functions, which scale as $\sim1/m_{Q}^{5}$ in
the heavy mass limit and thus are strongly suppressed. For this reason
we do not include predictions neither for JLab 22 GeV kinematics,
nor for $B$-meson pairs.

\section{Summary and conclusions}

\label{sec:Conclusions}In this paper, we analyzed the\emph{ }potential
of exclusive photoproduction of $D$-meson pairs for studies of the
GPDs of the target. We analyzed the photoproduction of pseudoscalar-vector
pairs with opposite $C$-parities and different flavor content ($D^{+}D^{*-},$~$D^{0}\bar{D}^{*0}$and
$D_{s}^{+}D_{s}^{*-}$), which get their dominant contribution from
the chiral-even GPDs. We focused on the kinematics of large invariant
masses of the produced meson pairs, moderate values of $x_{B}\in\left(10^{-3},\,10^{-1}\right)$
and small photon virtuality $Q^{2}$ ($Q^{2}\ll m_{Q}^{2}$), achievable
with low- and middle-energy $ep$ beams at the Electron Ion Collider.
We performed evaluations in the collinear factorization approach in
leading order over the strong coupling~$\alpha_{s}\left(m_{Q}\right)$.
In all channels the amplitude of the process obtains comparable contributions
from one of the light quark flavors and gluons. This feature might
present a special interest for phenomenological attempts to disentangle
the flavor structure of the light quark GPDs. The sensitivity of the
process to different GPD kinematics is controlled by the so-called
coefficient functions (partonic level amplitudes), which have nontrivial
behavior in the so-called ERBL region $|x|<\xi$, yet vanish rapidly
outside of it. Due to convolution with sufficiently broad distribution
amplitudes of $D$-mesons, the $x$-dependence of the coefficient
functions in the region $|x|<\xi$ is relatively moderate, with a
mild peak around $x\approx\xi$. The inverse deconvolution apparently
is not possible in view of complexity of the coefficient function,
however we believe that experimental study of this process might present
new constraints for phenomenological models of GPDs, especially in
the ERBL region. The pronounced $t$-dependence of the cross-section,
which stems from the phenomenological GPD models, implies that the
$D$-meson pairs are produced predominantly in back-to-back kinematics,
with relatively small and oppositely directed transverse momenta of
the mesons.

The results of this study complement our previous analyses~\cite{Andrade:2022rbn,Siddikov:2022bku}
of exclusive quarkonia pair production, which is sensitive only to
gluonic GPDs. If compared in the same kinematics, the cross-sections
for $D$-meson pairs are larger due to contribution of quark GPDs,
and different structure of the gluonic coefficient functions due to
replacement of one of the heavy quarks with nearly massless light
quark. This fact should facilitate experimental studies of the suggested
channel. The cross-section of the suggested process is comparable,
by order of magnitude, to the cross-sections of similar $2\to3$ processes
($\gamma^{*}p\to\gamma Mp$, $M=\pi,\,\rho$) suggested recently in
the literature~\cite{GPD2x3:9,GPD2x3:8,GPD2x3:7,GPD2x3:6,GPD2x3:5,GPD2x3:4,GPD2x3:3,GPD2x3:2,GPD2x3:1,Duplancic:2022wqn,ElBeiyad:2010pji,Boussarie:2016qop}
for exhaustive studies of the light quark GPDs. This happens because
the suppression due to heavy quark mass in $D$-meson pair production
numerically is on par with a suppression by the additional fine-structure
constant $\alpha_{{\rm em}}$ in processes which include additional
emitted photon. For this reason both $\gamma^{*}p\to\gamma Mp$ and
$D$-meson production could be used as complementary tools for the
study of both quark and gluon GPDs.

\section*{Acknowldgements}

We thank our colleagues at UTFSM university for encouraging discussions.
This research was partially supported by Proyecto ANID PIA/APOYO AFB220004
(Chile) and Fondecyt (Chile) grants 1220242 and 1230391.

\appendix

\section{Distribution amplitudes of the $D$-mesons}

\label{sec:DMesonDAs}In this appendix, for the sake of completeness,
we provide definitions and parametrizations of the $D$-meson distribution
amplitudes. For light mesons the distribution amplitudes are conventionally
characterized (ordered) by the twist of the corresponding quark-antiquark
operator~\cite{LightQuarkDA:1,LightQuarkDA:2,LightQuarkDA:3,LightQuarkDA:4,LightQuarkDA:5}.
For spinless pseudoscalar meson $M$ at leading twist, there is only
one distribution amplitude, defined as 
\begin{align}
\Phi^{(P)}\left(z\right) & =\int\frac{d\eta}{2\pi}e^{izp^{+}\eta}\left\langle \left|\bar{\psi}\left(-\frac{\eta}{2}\right)\gamma^{+}\gamma_{5}\mathcal{L}\left(-\frac{\eta}{2},\,\frac{\eta}{2}\right)\psi\left(\frac{\eta}{2}\right)\right|M(p)\right\rangle ,\label{eq:DAdefEta}\\
 & \mathcal{L}\left(-\frac{\eta}{2},\,\frac{\eta}{2}\right)\equiv\mathcal{P}{\rm exp}\left(i\int_{-\eta/2}^{\eta/2}d\zeta\,A^{+}\left(\zeta\right)\right).
\end{align}
where $p$ is the momentum of the meson (we assume that the meson
moves in the plus-direction), $z$ is the fraction of the momentum
carried by the quark, and $\mathcal{L}$ is the standard path-ordered
gauge link. Similarly, for the vector (spin-1) mesons there are 2
independent leading-twist distributions $\Phi_{||}^{(V)}$ and $\Phi_{\perp}^{(V)}$
defined as

\begin{align}
\Phi_{||}^{(V)}\left(z\right) & =\int\frac{d\eta}{2\pi}e^{izp^{+}\eta}\left\langle 0\left|\bar{\psi}\left(-\frac{\eta}{2}\right)\gamma^{+}\mathcal{L}\left(-\frac{\eta}{2},\,\frac{\eta}{2}\right)\psi\left(\frac{\eta}{2}\right)\right|M(p)\right\rangle ,\label{eq:DAdefPsiLL}\\
\Phi_{\perp}^{(V)}\left(z\right) & =\int\frac{d\eta}{2\pi}e^{izp^{+}\eta}\left\langle 0\left|\bar{\psi}\left(-\frac{\eta}{2}\right)\left(-i\sigma^{+\alpha}\varepsilon_{M,\,\alpha}^{*}(p)\right)\mathcal{L}\left(-\frac{\eta}{2},\,\frac{\eta}{2}\right)\psi\left(\frac{\eta}{2}\right)\right|M(p)\right\rangle ,\label{eq:DAdefPsiTT}
\end{align}
where $\varepsilon_{M}(p)$ is the (transverse) polarization vector
of the meson, which satisfies $n\cdot\varepsilon_{J/\psi}=p\cdot\varepsilon_{J/\psi,\,\mu}^{*}(p)=0$.
The definitions~(\ref{eq:DAdefEta}-\ref{eq:DAdefPsiTT}) may sometimes
include distribution amplitudes normalized to unity, taking out the
explicit meson decay constants $f_{M}$ in the right hand side. To
avoid ambiguity, we will use the notation $\varphi(z)$ (with corresponding
other indices) for such normalized DAs. These findings imply that
during perturbative evaluation of the coefficient function, the meson
formation from perturbative quarks might be described using the effective
vertices 
\begin{equation}
f_{P}\varphi^{(P)}\left(z\right)\gamma^{+}\gamma_{5},\quad f_{V}\varphi_{||}^{(V)}\left(z\right)\gamma^{+},\quad{\rm and}\quad-if_{V}\varphi_{\perp}^{(V)}\left(z\right)\sigma^{+\alpha}\varepsilon_{M,\,\alpha}^{*}\label{eq:TwVertex}
\end{equation}

\begin{figure}
\includegraphics[width=8cm]{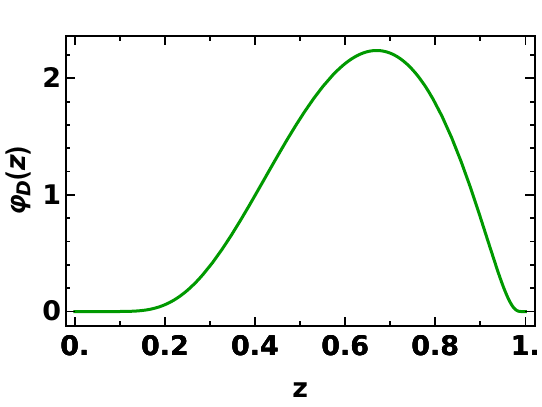}\caption{\label{fig:DMesonDA}Typical shape of the $D$-meson distribution
amplitude $\varphi_{D}(z)$. For definiteness this plot was done using
\textquotedblleft model 2\textquotedblright{} from~\cite{Zuo:2006re};
other models described in~\cite{Zhong:2022ugk,Beneke:2023nmj,Dhiman:2019ddr}
have very similar shapes.}
\end{figure}

Sometimes this scheme is also extended to heavy $D$-mesons. However,
for processes involving heavy quarks, the twist-based classification
of distribution amplitudes requires adjustments, since the heavy quark
mass limit breaks twist-based suppression of higher twist contributions
in physical amplitudes. Instead of this, the distribution amplitudes
are ordered by the inverse powers of the heavy quark mass $m_{Q}$,
using the Heavy Quark Effective Theory (HQET) framework~\cite{Neubert:1993mb}.
For the pseudoscalar and vector $D$-mesons containing $c$-quark,
this implies that we should replace~(\ref{eq:TwVertex}) with~\cite{Neubert:1993mb,Baek:1994kj,Luszczak:2011js,Beneke:2023nmj}
\begin{align}
 & f_{D}\varphi_{D}^{(P)}\left(z,\mu^{2}\right)\left(\frac{1+\hat{v}}{2}\gamma_{5}\right),\quad f_{D}\varphi_{D}^{(V)}\left(z,\mu^{2}\right)\left(\frac{1+\hat{v}}{2}\right)\hat{\varepsilon}\left(p\right)\label{eq:DADef-1}
\end{align}
where $v^{\mu}=p^{\mu}/M_{D}$ is the 4-vector $D$-meson velocity,
$\varepsilon^{\mu}\left(p\right)$ is the 4-vector polarization for
the vector mesons, $f_{D}$ is the corresponding decay constant. For
the charge-conjugate $D$-mesons (containing $\bar{c}$ quark), the
definitions~(\ref{eq:DADef-1}) should be modified as $v^{\mu}\to-v^{\mu}$.
Due to heavy quark spin-flavor symmetry~\cite{Georgi:1990um,Isgur:1989vq,Neubert:1993mb}
it is expected that the functional form $\varphi_{D}$ should be approximately
the same for pseudoscalar and vector mesons. While there is a plethora
of different phenomenological and model-based parametrizations~\cite{Zhong:2022ugk,Beneke:2023nmj,Dhiman:2019ddr,Zuo:2006re},
they all suggest that $\varphi_{D}\left(z\right)$ should have a broad
peak near $z\approx2/3$ and vanish at the extremes, as shown in Figure~\ref{fig:DMesonDA}.
This shape suggests that up to 1/3 of the momentum of the $D$-meson
is carried by the light-quark. 

\section{Evaluation of the coefficient functions}

\label{sec:CoefFunction} The coefficient functions (partonic amplitudes)
may be evaluated using standard light--cone rules, which may be found
in~\cite{Lepage:1980fj,Brodsky:1997de,Diehl:2000xz,Diehl:2003ny,Diehl:1999cg,Ji:1998pc}.
In the leading order, this requires the evaluation of all the diagrams
shown in Figures~\ref{fig:Photoproduction-B},\ref{fig:Photoproduction-A}.
This evaluations drastically simplify in the collinear factorization
approach, assuming that in our hierarchy of scales the mass of the
heavy quark $m_{Q}$ and photon virtuality $Q$ are large parameters,
$Q\sim m_{Q}\sim W\equiv\sqrt{s_{\gamma p}}$, and omitting the mass
of the proton $m_{N}$ and all the transverse momenta in the coefficient
functions evaluation. Below, in subsections~\ref{subsec:Light-quark-contribution},~\ref{subsec:Gluonic-contribution}
we discuss some technical details and provide final results for the
coefficient functions of the light quarks and gluons.

\subsection{Light quark contribution}

\label{subsec:Light-quark-contribution}The leading twist chiral even
quark GPDs $H^{q},E^{q},\tilde{H}^{q},\tilde{E}^{q}$, which are expected
to give the dominant contributions, are conventionally defined from
quark-antiquark correlators~\cite{Diehl:2003ny} 
\begin{align}
F^{q} & =\frac{1}{2}\int\frac{dz}{2\pi}\,e^{ix\bar{P}^{+}}\left\langle P'\left|\bar{\psi}\left(-\frac{z}{2}n\right)\gamma_{+}\mathcal{L}\left(-\frac{z}{2},\,\frac{z}{2}\right)\psi\left(\frac{z}{2}n\right)\right|P\right\rangle =\label{eq:defH}\\
 & =\frac{1}{2\bar{P}^{+}}\left(\bar{U}\left(P'\right)\gamma_{+}U\left(P\right)H^{q}\left(x,\xi,t\right)+\bar{U}\left(P'\right)\frac{i\sigma^{+\alpha}\Delta_{\alpha}}{2m_{N}}U\left(P\right)E^{q}\left(x,\xi,t\right)\right),\nonumber \\
\tilde{F}^{q} & =\frac{1}{2}\int\frac{dz}{2\pi}\,e^{ix\bar{P}^{+}}\left\langle P'\left|\bar{\psi}\left(-\frac{z}{2}n\right)\gamma_{+}\gamma_{5}\mathcal{L}\left(-\frac{z}{2},\,\frac{z}{2}\right)\psi\left(\frac{z}{2}n\right)\right|P\right\rangle =\label{eq:defHTilde}\\
 & =\frac{1}{2\bar{P}^{+}}\left(\bar{U}\left(P'\right)\gamma_{+}\gamma_{5}U\left(P\right)\tilde{H}^{q}\left(x,\xi,t\right)+\bar{U}\left(P'\right)\frac{\Delta^{+}\gamma_{5}}{2m_{N}}U\left(P\right)\tilde{E}^{q}\left(x,\xi,t\right)\right).\nonumber \\
 & \quad\mathcal{L}\left(-\frac{z}{2},\,\frac{z}{2}\right)\equiv\mathcal{P}{\rm exp}\left(i\int_{-z/2}^{z/2}d\zeta\,A^{+}\left(\zeta\right)\right).
\end{align}
In what follows we work in the light-cone gauge, so the gauge link
becomes trivial, $\mathcal{L}\left(\zeta_{1},\,\zeta_{2}\right)=1$.
The skewedness variable $\xi=\left(P'^{+}-P^{+}\right)/\left(P'^{+}-P^{+}\right)$
ca be related to the rapidities of the heavy mesons using~(\ref{eq:XiDef}).
For the sake of brevity we do not show explicitly the dependence on
the factorization scale~$\mu$, which is described by the conventional
DGLAP evolution equations. The factorization theorems allow to rewrite
the amplitudes of physical processes as convolution of these GPDs
with cross sections of partonic processes $\gamma q\to M_{1}M_{2}q$,
where the quarks before and after interactions have light-cone momenta
$k_{1,2}^{+}\sim\left(x\pm\xi\right)\bar{P}^{+}$. The transverse
part of the quark momentum $\Delta_{\perp}$ is small compared to
typical hard scales ($Q,m_{Q},\mathcal{M}_{12}$) and thus may be
disregarded in the evaluation of the coefficient function. The evaluation
of the amplitude requires calculation of all the diagrams shown in
Figure~\ref{fig:Photoproduction-B}, together with all possible permutations
of photon vertices, assuming that gluon vertices are fixed. The explicit
expressions for the quark spinors and projectors onto $F^{q}$ and
$\tilde{F}^{q}$ may be found in~\cite{Diehl:2003ny}; the spinor
algebra was done using FeynCalc package for \emph{Mathematica}~\cite{FeynCalc1,FeynCalc2}.
While the total number of diagrams is huge, in the collinear factorization
picture many diagrams vanish. This happens because the transverse
momenta of partons are disregarded in the collinear approach, so all
the light quark propagators are given by 
\begin{equation}
S(k)=\frac{\hat{k}-m}{k^{2}-m^{2}+i0}\approx\frac{k^{+}\gamma_{+}+k^{-}\gamma_{-}}{2k^{+}k^{-}+i0},
\end{equation}
the gluon propagators in light-cone gauge satisfy the light-cone gauge
condition $n_{\mu}G^{\mu\nu}=0,$ and for the $\gamma_{+},\,\gamma_{-}$
matrices in the numerator we have standard relations of Dirac algebra
$\gamma_{+}\gamma_{+}=\gamma_{-}\gamma_{-}=0,$ $\gamma_{+}\gamma_{-}=2-\gamma_{-}\gamma_{+}$,
$\gamma_{\pm}\boldsymbol{\gamma_{\perp}}=-\boldsymbol{\gamma_{\perp}}\gamma_{\pm}$.
For example, this leads to a cancellation of all diagrams with 3-gluon
vertices, whose representative is shown in the second diagram in the
first row of Figure~\ref{fig:Photoproduction-B}: the gluon vertex
itself is proportional to the difference of gluon momenta $\ell_{i}-\ell_{j}$,
which in the collinear limit has only light cone $\pm$ components.
The gluon propagator for collinear gluons is proportional to the transverse
part of the metric tensor, $g_{\mu\nu}^{\perp}=g_{\mu\nu}-\left(n_{\mu}^{(+)}n_{\nu}^{(-)}+n_{\mu}^{(-)}n_{\nu}^{(+)}\right)/n^{(+)}\cdot n^{(-)}$,
where $n_{\mu}^{(\pm)}$ are the light-cone vectors in the $\pm$
direction. After contraction of such propagators with the $\pm$ components
of $\ell_{i}^{\mu}-\ell_{j}^{\nu}$, we can see that the 3-gluon contribution
vanishes (beyond collinear approximation, giving corrections $\sim\mathcal{O}\left(t/M^{2}\right)$
to the coefficient function). The evaluation of the remaining diagrams
is straightforward and was done using the FeynCalc package for \emph{Mathematica}~\cite{FeynCalc1,FeynCalc2}.
In the evaluation of the coefficient function, we replaced the final-state
$D$-mesons with a system of collinear heavy-light quark-antiquarks
($\bar{q}Q$ and $\bar{Q}q$ respectively), adding appropriate spinor
projectors, as required by the definitions~\ref{eq:DADef-1} (see
Appendix~\ref{sec:DMesonDAs} for details). To avoid confusion, we
will use the notations $z_{1},z_{2}$ for the light-cone momenta fractions
of quarks and $\bar{z}_{1}\equiv1-z_{1},\bar{z}_{2}\equiv1-z_{2}$
for antiquarks in the first and the second mesons. In what follows
we focus only on the $Q\approx0$ limit and provide the results for
the transversely polarized photons (extensions for the case $Q\not=0$
is straightforward, although tthe final expressions become too lengthy
for publication). The final result for the light quark coefficient
functions are 
\begin{align}
C_{T}^{(q)}\left(x,\,\xi,\,\Delta y,z_{1},\,z_{2}\right) & =\kappa\left[\frac{N_{c}^{2}-1}{2N_{c}}\left(e_{\ell}\sum_{k=1}^{4}a_{k}^{(q)}+e_{H}\sum_{k=1}^{3}b_{k}^{(q)}\right)-\frac{1}{2N_{c}}\left(e_{\ell}\sum_{k=1}^{4}c_{k}^{(q)}+e_{H}d_{1}^{(q)}\right)\right]\\
\tilde{C}_{T}^{(q)}\left(x,\,\xi,\,\Delta y,z_{1},\,z_{2}\right) & =\kappa\left[\frac{N_{c}^{2}-1}{2N_{c}}\left(e_{\ell}\sum_{k=1}^{4}\tilde{a}_{k}^{(q)}+e_{H}\sum_{k=1}^{3}\tilde{b}_{k}^{(q)}\right)-\frac{1}{2N_{c}}\left(e_{\ell}\sum_{k=1}^{4}\tilde{c}_{k}^{(q)}+e_{H}\tilde{d}_{1}^{(q)}\right)\right]
\end{align}
where 
\begin{equation}
\kappa=\left(4\pi\alpha_{s}\right)^{2}\sqrt{4\pi\alpha_{{\rm em}}}\left(\varepsilon_{D^{*}}^{*}\cdot\varepsilon_{T}^{(\gamma)}\right),\label{eq:kappa_def}
\end{equation}
$\varepsilon_{T}^{(\gamma)}$ and $\varepsilon_{D^{*}}$ are the polarization
vectors of the incident photon and produced vector meson, $e_{\ell},\,e_{H}$,
are the charges of the light and heavy quarks (in units of proton
charge), and the constants $a_{k},b_{k},c_{k},d_{k},\tilde{a}_{k},\tilde{b}_{k},\tilde{c}_{k},\tilde{d}_{k}$
are given by

\begin{align}
a_{1}^{(q)} & =\tilde{a}_{1}^{(q)}=\frac{e^{3\Delta y}\xi}{2m_{Q}^{5}z_{1}(1+\xi)\left(1+e^{\Delta y}z_{2}\right)\left(1+2\xi+e^{\Delta y}(1+z_{1}+2\xi)\right)}\times\\
 & \times\left[\left(x-\xi+\frac{\xi}{1+\xi}\frac{\bar{z}_{2}}{1+e^{\Delta y}}\right)\left(\left(1+e^{\Delta y}\right)x-\frac{\xi}{1+\xi}\left(\xi+e^{\Delta y}(z_{1}+\xi)\right)\right)\right]^{-1},\nonumber \\
 & \bar{z}_{\alpha}\equiv1-z_{\alpha},\quad\alpha=1,2,
\end{align}
\begin{align}
a_{2}^{(q)} & =-\tilde{a}_{2}^{(q)}=\frac{e^{2\Delta y}\left(1-e^{\Delta y}\right)}{2m_{Q}^{5}z_{1}\left(e^{\Delta y}\bar{z}_{1}+z_{2}\right)\left(1+e^{\Delta y}z_{2}\right)\left(\bar{z}_{1}+e^{\Delta y}z_{2}\right)}\times\\
 & \times\left[\left(2(1+\xi)\left(1+e^{\Delta y}\right)-\bar{z}_{2}\right)\left(x-\xi+\frac{\xi}{1+\xi}\frac{\bar{z}_{2}}{1+e^{\Delta y}}\right)\right]^{-1}\nonumber 
\end{align}

\begin{align}
a_{3}^{(q)} & =\tilde{a}_{3}^{(q)}=\frac{e^{2\Delta y}\left(e^{\Delta y}-1\right)}{2m_{Q}^{5}z_{1}\left(1+e^{\Delta y}\bar{z}_{1}\right)\left(e^{\Delta y}\bar{z}_{1}+z_{2}\right)\left(\bar{z}_{1}+e^{\Delta y}z_{2}\right)}\times\\
 & \times\left[\left(x-\frac{\xi^{2}}{1+\xi}-\frac{e^{\Delta y}}{e^{\Delta y}+1}\frac{\xi}{1+\xi}z_{1}\right)\left(2(1+\xi)\left(1+e^{\Delta y}\right)-\bar{z}_{2}\right)\right]^{-1}\nonumber 
\end{align}

\begin{align}
a_{4}^{(q)} & =-\tilde{a}_{4}^{(q)}=-\frac{e^{\Delta y}\left(e^{\Delta y}-1\right)}{2m_{Q}^{5}\left(\bar{z}_{1}+e^{\Delta y}\right)z_{1}\left(e^{\Delta y}\bar{z}_{1}+z_{2}\right)\left(\bar{z}_{1}+e^{\Delta y}z_{2}\right)}\times\\
 & \times\left[\left(2\left(1+e^{-\Delta y}\right)(1+\xi)-z_{1}\right)\left(x+\xi-\frac{e^{\Delta y}}{1+e^{\Delta y}}\frac{\xi}{1+\xi}z_{1}\right)\right]^{-1}\nonumber 
\end{align}

\begin{align}
b_{1}^{(q)} & =-\tilde{b}_{1}^{(q)}=\frac{e^{2\Delta y}\xi\left(\xi(1+2\xi)\left(e^{2\Delta y}-1\right)-e^{\Delta y}\left(x(1+\xi)\left(1+e^{\Delta y}\right)-\xi\left(-2+\xi+e^{\Delta y}(2+\xi)\right)\right)z_{1}\right)}{2m_{Q}^{5}\left(x+x\xi-\xi^{2}\right)z_{1}\left(1+e^{\Delta y}z_{2}\right)\left(\left(1+e^{\Delta y}\right)x(1+\xi)-\xi\left(\xi\bar{z}_{2}+e^{\Delta y}(1+\xi)\right)\right)}\times\\
 & \times\left[\left(1+e^{\Delta y}\right)\left(x+\xi-\frac{e^{\Delta y}}{1+e^{\Delta y}}\frac{\xi}{1+\xi}z_{1}\right)\left(\left(1+2\xi+e^{\Delta y}(3+2\xi)\right)z_{1}-e^{\Delta y}z_{1}^{2}-\left(1+e^{\Delta y}\right)(1+2\xi)\right)\right]^{-1}\nonumber 
\end{align}

\begin{align}
b_{2}^{(q)} & =\tilde{b}_{2}^{(q)}=\frac{e^{3\Delta y}\left(1+e^{\Delta y}\right)(1+\xi)\left(1+2\xi-z_{1}\right)}{2m_{Q}^{5}\left(2\left(1+e^{\Delta y}\right)(1+\xi)-\bar{z}_{2}\right)\left(2\left(1+e^{\Delta y}\right)(1+\xi)-e^{\Delta y}z_{1}-\bar{z}_{2}\right)\left(1+e^{\Delta y}z_{2}\right)\left(\bar{z}_{1}+e^{\Delta y}z_{2}\right)}\times\\
 & \times\left[\left((1+2\xi)\bar{z}_{1}+\frac{e^{\Delta y}}{1+e^{\Delta y}}z_{1}\left(z_{1}-2\right)\right)\left(x(1+\xi)\left(1+e^{\Delta y}\right)-\xi\left(\xi+e^{\Delta y}(1+\xi)\right)-\xi z_{2}\right)\right]^{-1}\nonumber 
\end{align}

\begin{align}
b_{3}^{(q)}= & -\tilde{b}_{3}^{(q)}=\frac{e^{2\Delta y}\left(1+2\xi-z_{1}\right)}{2m_{Q}^{5}z_{1}\left(2(1+\xi)-\frac{e^{\Delta y}}{1+e^{\Delta y}}z_{1}\right)\left(x+\xi-\frac{e^{\Delta y}}{1+e^{\Delta y}}\frac{\xi}{1+\xi}z_{1}\right)\left(\bar{z}_{1}+e^{\Delta y}z_{2}\right)}\\
 & \times\left[\left((1+2\xi)\left(1+e^{\Delta y}\right)\bar{z}_{1}+e^{\Delta y}z_{1}\left(z_{1}-2\right)\right)\left((1+2\xi)\left(1+e^{\Delta y}\right)+e^{\Delta y}\bar{z}_{1}+z_{2}\right)\right]^{-1}\nonumber 
\end{align}

\begin{align}
c_{1}^{(q)} & =-\tilde{c}_{1}^{(q)}=-\frac{e^{2\Delta y}\xi\left(x+x\xi+\xi^{2}-\xi z_{1}-\xi z_{2}\right)}{2m_{Q}^{5}z_{1}\left(x+\xi-\frac{e^{\Delta y}}{1+e^{\Delta y}}\frac{\xi}{1+\xi}z_{1}\right)\left(\left(1+e^{\Delta y}\right)(1+2\xi)+\bar{z}_{2}\right)\left(1+e^{\Delta y}z_{2}\right)\left(x(1+\xi)+\xi^{2}+\xi\frac{\bar{z}_{2}}{1+e^{\Delta y}}\right)}\times\\
 & \times\left[\left(1+e^{\Delta y}\right)(1+\xi)(x+\xi)\left(z_{1}+z_{2}e^{\Delta y}\right)-\xi z_{1}z_{2}\left(1+e^{2\Delta y}\right)+e^{\Delta y}\xi\left(1-z_{1}^{2}-z_{2}^{2}\right)\right]^{-1}\nonumber 
\end{align}

\begin{align}
c_{2}^{(q)} & =-\tilde{c}_{2}^{(q)}=-\frac{e^{2\Delta y}\left(1-e^{\Delta y}\right)\xi}{2m_{Q}^{5}\left(x(1+\xi)-\xi^{2}\right)\left(\bar{z}_{1}e^{\Delta y}+z_{2}\right)\left(\bar{z}_{1}+e^{\Delta y}z_{2}\right){}^{2}\left(1+e^{\Delta y}z_{2}\right)}\times\\
 & \times\left(\left(1+e^{\Delta y}\right)x-\frac{\xi}{1+\xi}\left(\xi+e^{\Delta y}(1+\xi)\right)-\frac{\xi}{1+\xi}z_{2}\right)^{-1}\nonumber 
\end{align}

\begin{align}
c_{3}^{(q)}=\tilde{c}_{3}^{(q)} & =-\frac{e^{2\Delta y}\left(e^{\Delta y}-1\right)\xi^{2}}{2m_{Q}^{5}\left(x+x\xi+\xi^{2}\right)\left(\bar{z}_{1}e^{\Delta y}+z_{2}\right)\left(1+e^{\Delta y}z_{2}\right)\left(\bar{z}_{1}+e^{\Delta y}z_{2}\right)}\times\\
 & \times\left[\left(1+e^{\Delta y}\right)\left(x+\xi-\frac{\xi}{1+\xi}\frac{e^{\Delta y}\bar{z}_{1}+z_{2}}{1+e^{\Delta y}}\right)\left(\left(1+e^{\Delta y}\right)x(1+\xi)+\xi\left(1+\xi+e^{\Delta y}\xi\right)-\xi z_{2}\right)\right]^{-1}\nonumber 
\end{align}

\begin{align}
c_{4}^{(q)} & =-\tilde{c}_{4}^{(q)}=-\frac{e^{2\Delta y}\left(1-e^{\Delta y}\right)\xi}{2m_{Q}^{5}z_{1}\left(x+\xi-\frac{e^{\Delta y}}{1+e^{\Delta y}}\frac{\xi}{1+\xi}z_{1}\right)\left(z_{1}+e^{\Delta y}\bar{z}_{2}\right)\left(\bar{z}_{1}e^{\Delta y}+z_{2}\right)\left(\bar{z}_{1}+e^{\Delta y}z_{2}\right)}\times\\
 & \left((x-\xi)(1+\xi)\left(1+e^{\Delta y}\right)+e^{\Delta y}\xi\bar{z}_{1}+\xi z_{2}\right)^{-1}\nonumber 
\end{align}

\begin{align}
d_{1}^{(q)} & =-\tilde{d}_{1}^{(q)}=\frac{e^{2\Delta y}\xi}{2m_{Q}^{5}z_{1}(1+\xi)\left(1+e^{\Delta y}\right)\left(z_{2}^{2}-1+\left(1+e^{\Delta y}\right)(1+2\xi)z_{2}\right)\left(x+\xi-\frac{e^{\Delta y}}{1+e^{\Delta y}}\frac{\xi}{1+\xi}z_{1}\right)\left(1+e^{\Delta y}z_{2}\right)}\times\\
 & \times\left[\left(x+x\xi+\xi^{2}\bar{z}_{2}\right)+\frac{2\xi}{1+e^{\Delta y}}+\left(x(1+\xi)+\xi e^{\Delta y}\left(1+2e^{\Delta y}\right)\right)z_{2}\right.-\nonumber \\
 & \left.\qquad-\tanh\left(\frac{\Delta y}{2}\right)\xi z_{2}^{2}+\frac{\xi(1+\xi)z_{1}}{1+e^{\Delta y}}\left(2\left(1+e^{\Delta y}\frac{\xi}{1+\xi}\right)-\left(e^{\Delta y}-1\right)\frac{z_{2}}{1+\xi}\right)\right]\times\nonumber \\
 & \times\left[\left(\left(1+e^{\Delta y}\right)x-\frac{\xi}{1+\xi}\left(\xi+e^{\Delta y}(1+\xi)\right)-\frac{\xi}{1+\xi}z_{2}\right)\right.\times\nonumber \\
 & \times\left.\qquad\left(\xi z_{1}z_{2}\left(1+e^{2\Delta y}\right)-e^{\Delta y}\xi\left(1+z_{2}\bar{z}_{2}-z_{1}^{2}\right)-\left(z_{1}+z_{2}e^{\Delta y}\right)\left(1+e^{\Delta y}\right)\left(x(1+\xi)-\xi^{2}\right)\frac{}{}\right)\right]^{-1}\nonumber 
\end{align}
We may observe that, as a function of variable $x$, each of the contributions
has at most one or two poles, whose position depends on the skewedness
$\xi$, rapidity difference $\Delta y$, and the light-cone fractions
$z_{1},z_{2}$ carried by quarks in the corresponding $D$-mesons.
The endpoint singularities at $z_{1}=0$, $z_{1}=1$, $z_{2}=0$ and
$z_{2}=1$ do not present any difficulties, neither for the factorization
property at conceptual level, nor for numerical integration, since
in the amplitude these coefficient functions contribute multiplied
by the distribution amplitudes of the $D$-mesons, which vanish rapidly
at these points.

\subsection{Gluonic contribution}

\label{subsec:Gluonic-contribution}The definition of the leading
twist gluon GPDs is very similar to the quark GPDs definitions and
differ only by the structure of the operators inserted between the
initial and final states, namely~\cite{Diehl:2003ny,DVMPcc1} 
\begin{align}
F^{g}\left(x,\xi,t\right) & =\frac{1}{\bar{P}^{+}}\int\frac{dz}{2\pi}\,e^{ix\bar{P}^{+}}\left\langle P'\left|G^{+\mu\,a}\left(-\frac{z}{2}n\right)\mathcal{L}\left(-\frac{z}{2},\,\frac{z}{2}\right)G_{\,\,\mu}^{+\,a}\left(\frac{z}{2}n\right)\right|P\right\rangle =\label{eq:defF}\\
 & =\left(\bar{U}\left(P'\right)\gamma_{+}U\left(P\right)H^{g}\left(x,\xi,t\right)+\bar{U}\left(P'\right)\frac{i\sigma^{+\alpha}\Delta_{\alpha}}{2m_{N}}U\left(P\right)E^{g}\left(x,\xi,t\right)\right),\nonumber \\
\tilde{F}^{g}\left(x,\xi,t\right) & =\frac{-i}{\bar{P}^{+}}\int\frac{dz}{2\pi}\,e^{ix\bar{P}^{+}}\left\langle P'\left|G^{+\mu\,a}\left(-\frac{z}{2}n\right)\mathcal{L}\left(-\frac{z}{2},\,\frac{z}{2}\right)\tilde{G}_{\,\,\mu}^{+\,a}\left(\frac{z}{2}n\right)\right|P\right\rangle =\label{eq:defFTilde}\\
 & =\left(\bar{U}\left(P'\right)\gamma_{+}\gamma_{5}U\left(P\right)\tilde{H}^{g}\left(x,\xi,t\right)+\bar{U}\left(P'\right)\frac{\Delta^{+}\gamma_{5}}{2m_{N}}U\left(P\right)\tilde{E}^{g}\left(x,\xi,t\right)\right).\nonumber \\
 & \tilde{G}^{\mu\nu,\,a}\equiv\frac{1}{2}\varepsilon^{\mu\nu\alpha\beta}G_{\alpha\beta}^{a},\quad\mathcal{L}\left(-\frac{z}{2},\,\frac{z}{2}\right)\equiv\mathcal{P}{\rm exp}\left(i\int_{-z/2}^{z/2}d\zeta\,A^{+}\left(\zeta\right)\right).
\end{align}
where $\mathcal{L}$ is the standard path-ordered gauge link. In what
follows we'll use the standard light-cone gauge $A^{+}=0$, in which
$\mathcal{L}=1$, and the remaining two-gluon operators in~(\ref{eq:defF},~\ref{eq:defFTilde})
may be rewritten in the form 
\begin{align}
 & G^{+\mu_{\perp}\,a}\left(z_{1}\right)G_{\,\,\mu_{\perp}}^{+\,a}\left(z_{2}\right)=g_{\mu\nu}^{\perp}\left(\partial^{+}A^{\mu_{\perp},a}(z_{1})\right)\left(\partial^{+}A^{\nu_{\perp}a}(z_{2})\right),\\
 & G^{+\mu_{\perp}\,a}\left(z_{1}\right)\tilde{G}_{\,\,\mu_{\perp}}^{+\,a}\left(z_{2}\right)=G^{+\mu_{\perp}\,a}\left(z_{1}\right)\tilde{G}_{-\mu_{\perp}}^{\,a}\left(z_{2}\right)=\frac{1}{2}\varepsilon_{-\mu_{\perp}\alpha\nu}G^{+\mu_{\perp}\,a}\left(z_{1}\right)G^{\alpha\nu,\,a}\left(z_{2}\right)=\\
 & =\varepsilon_{-\mu_{\perp}+\nu_{\perp}}G^{+\mu_{\perp}\,a}\left(z_{1}\right)G^{+\nu_{\perp},\,a}\left(z_{2}\right)=\varepsilon_{\mu\nu}^{\perp}G^{+\mu_{\perp}\,a}\left(z_{1}\right)G^{+\nu_{\perp},\,a}\left(z_{2}\right)=\varepsilon_{\mu\nu}^{\perp}\left(\partial^{+}A^{\mu,a}(z_{1})\right)\left(\partial^{+}A^{\nu,\,a}(z_{2})\right).\nonumber 
\end{align}
The integration over $z$ in~(\ref{eq:defF},~\ref{eq:defFTilde})
effectively corresponds to a transition to the momentum space, where
the derivatives $\partial_{z_{1}}^{+},\,\partial_{z_{2}}^{+}$ convert
into the multiplicative factors $k_{1,2}^{+}\sim\left(x\pm\xi\right)\bar{P}^{+}$,
and thus~(\ref{eq:defF},~\ref{eq:defFTilde}) may be rewritten
in a joint form as~\cite{DVMPcc1} 
\begin{align}
\frac{1}{\bar{P}^{+}}\int\frac{dz}{2\pi}\,e^{ix\bar{P}^{+}}\left.\frac{\frac{}{}}{}\left\langle P'\left|A_{\mu}^{a}\left(-\frac{z}{2}n\right)A_{\nu}^{b}\left(\frac{z}{2}n\right)\right|P\right\rangle \right|_{A^{+}=0\,{\rm gauge}} & =\frac{\delta^{ab}}{N_{c}^{2}-1}\left(\frac{-g_{\mu\nu}^{\perp}F^{g}\left(x,\xi,t\right)-\varepsilon_{\mu\nu}^{\perp}\tilde{F}^{g}\left(x,\xi,t\right)}{2\,\left(x-\xi+i0\right)\left(x+\xi-i0\right)}\right).\label{eq:defF-1}
\end{align}
The factors $\pm i0$ in the denominator of~(\ref{eq:defF-1}) reflect
the conventional $\xi\to\xi-i0$ prescription used to define the deformation
of the integration contour near the poles of the amplitude. The structure
of the numerator of Eq.~(\ref{eq:defF-1}) implies that the coefficient
functions $C_{a}$ and $\tilde{C}_{a}$ may be extracted from the
corresponding partonic amplitude of $\gamma g\to D_{1}D_{2}g$, contracting
the Lorentz indices of the gluons with $g_{\mu\nu}^{\perp}$ and $\varepsilon_{\mu\nu}^{\perp}$
respectively.

\begin{figure}
\includegraphics[scale=0.6]{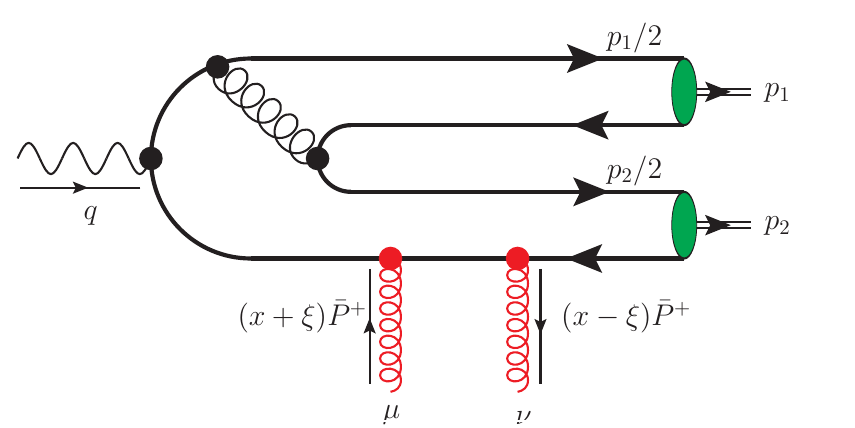}\includegraphics[scale=0.6]{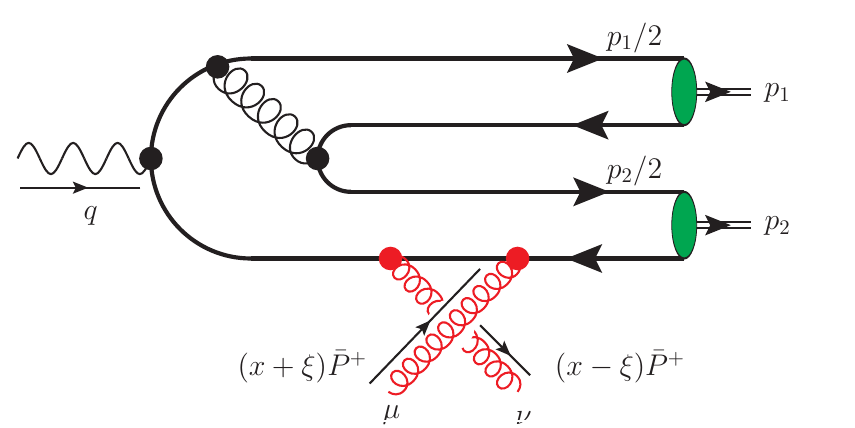}

\caption{\label{fig:Photoproduction-Permute}Schematic illustration of the
diagrams with direct and permuted $t$-channel gluons, which are related
to each other by an inversion of sign in front of light-cone fraction
$x\leftrightarrow-x$, and permutation of the Lorentz indices $\mu\leftrightarrow\nu$.}
\end{figure}

In the leading order over $\alpha_{s}$, the partonic amplitude $\gamma g\to D_{1}D_{2}g$
gets contribution from the Feynman diagrams shown in the Figure~\ref{fig:Photoproduction-A}.
Each diagram in that Figure should be understood as a sum of two contributions,
with photon coupled to heavy or light quark line, and complemented
with diagrams which differ by permutation of the $t$-channel gluons,
as demonstrated in Figure~\ref{fig:Photoproduction-Permute}. The
latter permutation leads to contributions which differ only by the
sign in front of the light-cone variable $x$ and interchange of the
Lorentz indices $\mu\leftrightarrow\nu$. As explained earlier, the
coefficient functions $C_{\mathfrak{a}}$, $\tilde{C}_{\mathfrak{a}}$,
may be found contracting the free Lorentz indices $\mu,\nu$ with
symmetric $g_{\mu\nu}^{\perp}$ or antisymmetric $\varepsilon_{\mu\nu}^{\perp}$;
for this reason $C_{\mathfrak{a}}$ and $\tilde{C}_{\mathfrak{a}}$
will be even or odd functions of the variable $x$ respectively. In
the collinear picture we assume that the quarks and antiquarks in
the produced $D$-mesons carry the light-cone momenta $z_{a}p_{a}$
and $\left(1-z_{a}\right)p_{a}$ respectively, so all the parton momenta
in leading-order diagrams of Figure~\ref{fig:Photoproduction-A}
may be fixed by energy-momentum conservation and represented as linear
combinations of the momenta of the $D$-mesons and $t$-channel gluons.
Using~(\ref{eq:defF-1}), we may rewrite the coefficient functions
of the transversely polarized photons as

\begin{align}
C_{T}^{(g)}\left(x,\,\xi,\,\Delta y,z_{1},\,z_{2}\right) & =\kappa\frac{\mathcal{C}_{T}\left(x,\,\xi,\,\Delta y,z_{1},\,z_{2}\right)+\mathcal{C}_{T}\left(x,\,\xi,\,\Delta y,z_{1},\,z_{2}\right)}{\left(x-\xi+i0\right)\left(x+\xi-i0\right)}\label{eq:CDef1}\\
\tilde{C}_{T}^{(g)}\left(x,\,\xi,\,\Delta y,z_{1},\,z_{2}\right) & =\kappa\frac{\tilde{\mathcal{C}}_{T}\left(x,\,\xi,\,\Delta y,z_{1},\,z_{2}\right)-\tilde{\mathcal{C}}_{T}\left(x,\,\xi,\,\Delta y,z_{1},\,z_{2}\right)}{\left(x-\xi+i0\right)\left(x+\xi-i0\right)},\label{eq:CDef2}
\end{align}

where the constant $\kappa$ was defined earlier in~(\ref{eq:kappa_def}),
and the factors $\left(x\pm\xi\mp i0\right)^{-1}$ in~(\ref{eq:CDef1},~\ref{eq:CDef2}),
come from~(\ref{eq:defF-1}). Since the gluon GPDs $H^{g},\,E^{g}$,
are even functions of the variable $x$, and $\tilde{H}^{g},\,\tilde{E}^{g}$
are odd functions~\cite{Diehl:2003ny}, in tthe convolution over
$x$ both terms in the numerators of~(\ref{eq:CDef1},~\ref{eq:CDef2})
give equal nonzero contributions. Numerically the dominant contribution
to unpolarized cross-section comes from the GPD $H^{g}$, whereas
the contribution of $\tilde{H}^{g}$ is negligibly small. The evaluation
of the diagrams from Figure~\ref{fig:Photoproduction-A} was done
using the FeynCalc package for \emph{Mathematica}~\cite{FeynCalc1,FeynCalc2}
(see the beginning of the previous Appendix~\ref{subsec:Light-quark-contribution}
for some technical details); the contributions to $\mathcal{C}_{\mathfrak{a}}$
and $\tilde{\mathcal{C}}_{\mathfrak{a}}$ were extracted, contracting
the free Lorentz indices $\mu,\nu$ of the partonic amplitude with
$g_{\mu\nu}^{\perp}$ or $\varepsilon_{\mu\nu}^{\perp}$, respectively.

Explicit expressions for the functions $\mathcal{C}_{T},\,\tilde{\mathcal{C}}_{T}$
are given by 
\begin{align}
\mathcal{C}_{T} & =e_{\ell}\left(\frac{N_{c}^{2}-1}{4N_{c}}\sum_{k=1}^{7}a_{k}-\frac{1}{4N_{c}}\sum_{k=1}^{3}b_{k}\right)+e_{H}\left(\frac{N_{c}^{2}-1}{4N_{c}}\sum_{k=1}^{7}A_{k}-\frac{1}{4N_{c}}\sum_{k=1}^{3}B_{k}\right)
\end{align}
where $e_{\ell},e_{H}$, are the charges of the light and heavy quark
(in units of the proton charge), and the expressions in front of them
correspond to a sum of the diagrams in which photon is connected to
the light or heavy quark lines respectively. Explicitly, these contributions
are given by

\begin{align}
a_{1} & =\frac{2e^{2\Delta y}\left(e^{2\Delta y}-1\right)\left(-e^{\Delta y}(\xi+1)(x-\xi)+(\xi+1)z_{2}\left(2e^{\Delta y}\xi+\xi-x\right)+\xi(2\xi+1)+\xi z_{2}^{2}\right)}{m_{Q}^{5}\left(e^{\Delta y}+z_{2}\right)\left(e^{\Delta y}\bar{z}_{1}+z_{2}\right)\left(e^{\Delta y}z_{2}+1\right){}^{2}\left(e^{\Delta y}z_{2}+\bar{z}_{1}\right)\left(2e^{\Delta y}(\xi+1)+2\xi+z_{2}+1\right)}\times\label{eq:a1}\\
 & \times\left(\xi\left(e^{\Delta y}(\xi+1)+\xi\right)-\left(\left(e^{\Delta y}+1\right)(\xi+1)x\right)+\xi z_{2}\right)^{-1}\nonumber 
\end{align}
\begin{align}
A_{1}= & \frac{2e^{2\Delta y}\left(e^{2\Delta y}-1\right)}{m_{Q}^{5}z_{2}\left(z_{2}+e^{\Delta y}\bar{z}_{1}\right)\left(e^{\Delta y}z_{2}+e^{2\Delta y}+1\right)\left(e^{\Delta y}z_{2}+\bar{z}_{1}\right)}\times\label{eq:A1}\\
 & \times\left[(x+\xi)z_{2}(1+\xi)\left(e^{\Delta y}z_{2}+e^{2\Delta y}+1\right)\right.-\nonumber \\
 & \left.-\xi\left(z_{2}+2\right)\left(\left(e^{\Delta y}+1\right)(2\xi+1)+z_{2}\left(2\left(e^{\Delta y}+e^{2\Delta y}\right)(1+\xi)+1\right)+e^{\Delta y}z_{2}\bar{z}_{2}\right)\right]\times\nonumber \\
 & \times\left[\left(\left(e^{\Delta y}+1\right)(2\xi+1)+z_{2}\left(2e^{2\Delta y}(\xi+1)+e^{\Delta y}\left(2\xi+z_{2}+1\right)+1\right)\right)\times\right.\nonumber \\
 & \left.\times\left((x-\xi)(1+\xi)\left(e^{\Delta y}+1\right)\left(e^{\Delta y}z_{2}+1\right)+\xi\left(e^{\Delta y}(1+z_{2}\bar{z}_{2})+\bar{z}_{2}\right)\right)\right]^{-1}\nonumber 
\end{align}
\begin{align}
a_{2} & =\frac{2\left(e^{\Delta y}+1\right)\left(-2\xi+z_{2}-2\right)}{m_{Q}^{5}\bar{z}_{2}\left(e^{\Delta y}\bar{z}_{2}+z_{1}\right)\left(e^{\Delta y}(1+2\xi)+2(\xi+1)-z_{2}\right)\left(e^{\Delta y}\left(2\xi+z_{1}+1\right)+2\xi-z_{2}+2\right)}\times\\
 & \times\left(\xi\left(e^{\Delta y}(\xi+2)+\xi\right)+\left(e^{\Delta y}+1\right)(\xi+1)\left(x\left(1-z_{2}e^{\Delta y}\right)+z_{2}\xi\left(e^{\Delta y}-2\right)\right)-\left(\left(e^{\Delta y}-1\right)\xi z_{2}^{2}\right)\right)\times\nonumber \\
 & \times\left[\frac{}{}\left(2\left(e^{\Delta y}+1\right)(\xi+1)z_{2}-z_{2}^{2}+1\right)\left(\left(e^{\Delta y}+1\right)(\xi+1)z_{2}(x-\xi)-\xi+\xi z_{2}^{2}\right)\right]^{-1}\nonumber 
\end{align}
\begin{align}
A_{2}= & \frac{2\left(e^{\Delta y}+1\right)\left(2\xi+\bar{z}_{2}\right)\left(\left(e^{\Delta y}+1\right)(\xi+1)\left(e^{\Delta y}(x-\xi)+2\xi\right)+\left(e^{\Delta y}-1\right)\xi z_{2}\right)}{m_{Q}^{5}z_{2}\left(e^{\Delta y}\bar{z}_{2}+z_{1}\right)\left(z_{2}-2\left(e^{\Delta y}+1\right)(\xi+1)\right)\left(e^{\Delta y}\left(2\xi+z_{1}+1\right)+\left(2\xi+\bar{z}_{2}+1\right)\right)}\times\\
 & \times\left[\frac{}{}\left(\left(e^{\Delta y}+1\right)(2\xi+1)+z_{2}\left(-e^{\Delta y}(2\xi+1)-2\xi+z_{2}-3\right)\right)\left(\left(e^{\Delta y}+1\right)(\xi+1)(x-\xi)+\xi z_{2}\right)\right]^{-1}\nonumber 
\end{align}
\begin{align}
a_{3} & =-\frac{2e^{\Delta y}\left(e^{2\Delta y}-1\right)}{m_{Q}^{5}z_{1}\left(z_{2}-1\right)\left(e^{\Delta y}+\bar{z}_{1}\right)\left(e^{\Delta y}\bar{z}_{1}+z_{2}\right)\left(e^{\Delta y}z_{2}+\bar{z}_{1}\right)\left(e^{\Delta y}\left(2\xi+1+\bar{z}_{1}\right)+2(\xi+1)\right)}\times\\
 & \times\left[\frac{}{}(x-\xi)(\xi+1)\left(\left(4\xi-z_{2}+3\right)z_{1}e^{\Delta y}+\bar{z}_{1}\left(2\xi-z_{2}+2\right)+\left(e^{2\Delta y}-z_{1}e^{\Delta y}(2\xi+1)\right)\right)\right.\nonumber \\
 & +\left.e^{2\Delta y}\xi z_{1}(2\xi+1)-e^{\Delta y}z_{1}\xi\left((3+2\xi)\left(z_{2}+2(\xi+1)\right)+z_{1}\left(-2\xi+z_{2}-2\right)\right)\frac{}{}\right]\nonumber \\
 & \times\left[\frac{}{}\left((2\xi+1)e^{\Delta y}\xi+2\xi+1+\bar{z}_{2}\right)\left(\left(e^{\Delta y}+1\right)(\xi+1)(x-\xi)+e^{\Delta y}\xi z_{1}\right)\right]^{-1},\nonumber 
\end{align}
\begin{align}
A_{3}= & -\frac{2e^{\Delta y}\left(e^{2\Delta y}-1\right)}{m_{Q}^{5}\left(e^{\Delta y}\bar{z}_{1}+z_{2}\right)\left(e^{\Delta y}z_{2}+\bar{z}_{1}\right)\left(e^{2\Delta y}\left(2\xi-z_{1}+2\right)+e^{\Delta y}\left(4\xi-z_{1}\bar{z}_{1}+3\right)+2(1+\xi)(1-2z_{1})\right)}\times\\
 & \times\left[\frac{}{}(x-\xi)(\xi+1)\left(\left(e^{\Delta y}+1\right)^{2}(2\xi+1)-\left(e^{\Delta y}+1\right)(2\xi+1)z_{1}-\left(2e^{\Delta y}+1\right)z_{2}+z_{1}z_{2}\right)\right.\nonumber \\
 & +\left.e^{\Delta y}\xi\left(z_{1}+1\right)\left(\left(e^{\Delta y}+2\right)(2\xi+1)-2(\xi+2)z_{2}+z_{1}\left(-2\xi+z_{2}-1\right)\right)\frac{}{}\right]\times\nonumber \\
 & \times\left[\frac{}{}\left(\left(e^{\Delta y}+1\right)(2\xi+1)-z_{2}\left(e^{\Delta y}(2\xi+1)+2\xi-z_{2}+3\right)\right)\left(e^{\Delta y}\xi+\left(e^{\Delta y}+1\right)(\xi+1)z_{1}(\xi-x)-e^{\Delta y}\xi z_{1}^{2}\right)\right]^{-1},\nonumber 
\end{align}

\begin{align}
a_{4} & =\frac{2e^{\Delta y}\left(e^{\Delta y}+1\right)\left(2\xi-z_{2}+2\right)}{m_{Q}^{5}\bar{z}_{1}\bar{z}_{2}\left(e^{\Delta y}\left(e^{\Delta y}+1\right)(2\xi+z_{1}+1)+z_{1}\left(e^{\Delta y}\left(2\xi+z_{1}\right)+2(\xi+1)\right)\right)\left((2\xi+1)\left(e^{\Delta y}\xi+1\right)+\bar{z}_{2}\right)}\\
 & \times\left[\frac{}{}(x-\xi)\left(e^{\Delta y}+1\right)(\xi+1)z_{1}+e^{\Delta y}\xi\left(z_{1}-1\right)\left(\left(e^{\Delta y}+1\right)(2\xi+1)+\left(e^{\Delta y}-1\right)z_{1}\right)\right]\times\nonumber \\
 & \times\left[\frac{}{}\left(z_{1}+e^{\Delta y}\bar{z}_{2}\right)\left(\xi\left(1+z_{1}e^{\Delta y}\right)-\left(e^{\Delta y}+1\right)\left(x(1+\xi)-\xi^{2}\right)\right)\left(e^{\Delta y}\left(2\xi+z_{1}+1\right)+\left(2\xi+1+\bar{z}_{2}\right)\right)\right]^{-1}\nonumber 
\end{align}
\begin{align}
A_{4} & =-\frac{2e^{\Delta y}\left(e^{\Delta y}+1\right)\left(2\xi+\bar{z}_{2}\right)}{m_{Q}^{5}\left(e^{\Delta y}+z_{1}\right)\left(e^{\Delta y}\bar{z}_{2}+z_{1}\right)\left(e^{\Delta y}\left(2\xi+z_{1}+1\right)+2(\xi+1)\right)\left(e^{\Delta y}\left(2(1+\xi)-\bar{z}_{1}\right)+2(1+\xi)-z_{2}\right)}\times\\
 & \left[(x-\xi)\left(e^{\Delta y}+1\right)(\xi+1)\left(e^{\Delta y}+z_{1}\right)+e^{\Delta y}\xi\left(z_{1}-2\right)\left(2\left(e^{\Delta y}\xi+\xi+1\right)+\left(e^{\Delta y}-1\right)z_{1}\right)\right]\times\nonumber \\
 & \left[\frac{}{}\left(\bar{z}_{2}\left(e^{\Delta y}+1\right)(2\xi+1)+z_{2}\left(z_{2}-2\right)\right)\left(\left(e^{\Delta y}+1\right)(\xi+1)\bar{z}_{1}(x-\xi)+e^{\Delta y}\xi\left(z_{1}-2\right)z_{1}\right)\right]^{-1}\nonumber 
\end{align}

\begin{align}
a_{5} & =\frac{2e^{2\Delta y}\left(e^{\Delta y}+1\right)\xi\left(x(1+\xi)+\xi\left(\xi+z_{1}+2\bar{z}_{2}\right)\right)}{m_{Q}^{5}z_{1}\bar{z}_{2}\left(z_{1}+e^{\Delta y}\bar{z}_{2}\right)\left((2\xi+1)(e^{\Delta y}+1)+\bar{z}_{2}\right)\left(\left(e^{\Delta y}+1\right)(1+\xi)(x-\xi)+e^{\Delta y}\xi z_{1}\right)}\times\\
 & \times\left(-\left(e^{\Delta y}+1\right)(\xi+1)\bar{z}_{2}(x-\xi)+\xi z_{1}\left(2\left(e^{\Delta y}\xi+\xi+1\right)+\left(e^{\Delta y}-1\right)z_{2}\right)\right)\times\nonumber \\
 & \times\left[\left(\left(e^{\Delta y}+1\right)\left(\xi^{2}+x(1+\xi)\right)+\xi\left(e^{\Delta y}z_{1}+\bar{z}_{2}\right)\right)\times\right.\nonumber \\
 & \left.\left(\left(e^{\Delta y}+1\right)(1+\xi)\left(e^{\Delta y}\left(2z_{2}-1\right)-z_{1}\right)(x+\xi)-\xi\left(e^{2\Delta y}\bar{z}_{1}\left(2z_{2}-1\right)-e^{\Delta y}\left(1+z_{1}\bar{z}_{1}+2z_{2}\bar{z}_{2}\right)-2z_{1}z_{2}\right)\right)\right]^{-1}\nonumber 
\end{align}

\begin{align}
A_{5} & =\frac{2e^{2\Delta y}\left(e^{\Delta y}+1\right)\xi\left((1+\xi)(x+\xi)+\xi\left(z_{1}-2z_{2}\right)\right)}{m_{Q}^{5}\left(e^{\Delta y}\left(z_{2}-1\right)-z_{1}\right)\left(e^{\Delta y}\left(2z_{2}-1\right)-z_{1}\right)\left(\left(e^{\Delta y}\bar{z}_{2}+1\right)(2\xi+1)-z_{2}\left(2(1+\xi)+\bar{z}_{2}\right)\right)}\times\\
 & \times\left(\left(e^{\Delta y}+1\right)(1+\xi)z_{2}(x-\xi)+\xi\left(z_{1}+1\right)\left(\left(e^{\Delta y}+1\right)(2\xi+1)+\left(e^{\Delta y}-1\right)z_{2}\right)\right)\times\nonumber \\
 & \times\left[\frac{}{}\left(e^{\Delta y}\xi-\left(e^{\Delta y}+1\right)(1+\xi)z_{1}(x-\xi)-e^{\Delta y}\xi z_{1}^{2}\right)\left(\left(e^{\Delta y}+1\right)\left(\xi^{2}+x(1+\xi)\right)+\xi\left(e^{\Delta y}z_{1}-1+2\bar{z}_{2}\right)\right)\times\right.\nonumber \\
 & \times\left.\left(\left(e^{\Delta y}+1\right)\left(\xi^{2}+x(1+\xi)\right)+\xi\left(e^{\Delta y}z_{1}+\bar{z}_{2}\right)\right)\frac{}{}\right]^{-1}\nonumber 
\end{align}
\begin{align}
a_{6} & =\frac{2e^{\Delta y}\left(e^{\Delta y}+1\right)\xi\left(x(1+\xi)-\xi^{2}-2\xi+\xi z_{2}\right)}{m_{Q}^{5}\bar{z}_{2}\left(e^{\Delta y}\bar{z}_{2}+2z_{1}\right)\left(e^{\Delta y}\bar{z}_{2}+z_{1}\right)\left((1+2\xi)(e^{\Delta y}+1)+\bar{z}_{2}\right)\left(\left(e^{\Delta y}+1\right)\left(x(1+\xi)-\xi^{2}\right)-\xi\left(e^{\Delta y}z_{1}+\bar{z}_{2}\right)\right)}\\
 & \times\left[\frac{}{}\left(e^{\Delta y}+1\right)(\xi+1)\bar{z}_{2}(x-\xi)+\right.\nonumber \\
 & +\left.\frac{}{}\xi\bar{z}_{2}\left(e^{\Delta y}\left(\left(e^{\Delta y}+1\right)(2\xi+1)+1\right)-\left(e^{\Delta y}-1\right)z_{2}\right)+2\xi z_{1}\left(2\left(\left(e^{\Delta y}+1\right)\xi+1\right)+\left(e^{\Delta y}-1\right)z_{2}\right)\right]\nonumber \\
 & \times\left[\left(\left(e^{\Delta y}+1\right)\left(x(1+\xi)-\xi^{2}\right)-2e^{\Delta y}\xi z_{1}-\xi\bar{z}_{2}\right)\left(\left(e^{\Delta y}+1\right)(\xi+1)z_{2}(\xi-x)+\xi(1-z_{2}^{2})\right)\frac{}{}\right]^{-1}\nonumber 
\end{align}

\begin{align}
A_{6} & =-\frac{2e^{\Delta y}\left(e^{\Delta y}+1\right)\xi\left((1+\xi)(x-\xi)+\xi z_{2}\right)}{m_{Q}^{5}z_{2}\left(e^{\Delta y}\bar{z}_{2}+z_{1}\right)\left(\left(e^{\Delta y}+1\right)(2\xi+1)+z_{2}\left((2\xi+1)\left(e^{\Delta y}+1\right)+z_{2}-2\right)\right)}\times\\
 & \times\left[\left(e^{\Delta y}+1\right)(1+\xi)z_{2}(x-\xi)-\frac{}{}\right.\nonumber \\
 & \left.\quad\frac{}{}-\xi\left((1+2\xi)\left(e^{\Delta y}+1\right)^{2}-\left(e^{2\Delta y}-1\right)(1+2\xi)z_{2}+2z_{1}(1+2\xi)\left(e^{\Delta y}+1\right)+\left(e^{\Delta y}-1\right)z_{2}(1+2z_{1})\right)\right]\nonumber \\
 & \times\left[\left(\left(e^{\Delta y}+1\right)(1+\xi)(x-\xi)+\xi z_{2}\right)\left(\frac{}{}\left(e^{\Delta y}+1\right)\left(x(1+\xi)-\xi^{2}\right)-\xi\left(e^{\Delta y}z_{1}+\bar{z}_{2}\right)\right)\times\frac{}{}\right.\nonumber \\
 & \times\left.\left(\frac{}{}\left(e^{\Delta y}+1\right)(1+\xi)\left(e^{\Delta y}\bar{z}_{2}+2z_{1}\right)(x-\xi)+e^{2\Delta y}\xi\bar{z}_{2}\left(1-2z_{1}\right)-e^{\Delta y}\xi\left(4z_{1}^{2}-2z_{1}-z_{2}\bar{z}_{2}-1\right)+2\xi z_{1}z_{2}\right)\right]^{-1}\nonumber 
\end{align}
\begin{align}
a_{7} & =\frac{2e^{\Delta y}\left(e^{2\Delta y}-1\right)}{m_{Q}^{5}\bar{z}_{2}\left(e^{\Delta y}\bar{z}_{1}+z_{2}\right)\left(e^{\Delta y}+z_{2}\right)\left(e^{\Delta y}z_{2}+1\right)\left(e^{\Delta y}z_{2}+\bar{z}_{1}\right)}\times\\
 & \times\frac{2e^{\Delta y}\left(e^{2\Delta y}-1\right)\left(\xi\left(e^{\Delta y}+z_{2}\right)\left(2\xi+\bar{z}_{2}\right)-\left((\xi+1)\left(e^{\Delta y}z_{2}+1\right)\right)(x-\xi)\right)}{\left(e^{\Delta y}z_{2}+\bar{z}_{1}\right)\left((2\xi+1)\left(e^{\Delta y}+1\right)+\bar{z}_{2}\right)\left(\xi\bar{z}_{2}-\left(e^{\Delta y}+1\right)\left(x(1+\xi)-\xi^{2}\right)\right)},\nonumber 
\end{align}
\begin{align}
A_{7} & =\frac{2e^{\Delta y}\left(e^{2\Delta y}-1\right)}{m_{Q}^{5}z_{2}\left(e^{\Delta y}z_{2}+e^{2\Delta y}+1\right)\left(\left(e^{\Delta y}+1\right)(2\xi+1)-z_{2}\left(\left(e^{\Delta y}+1\right)(1+2\xi)+\bar{z}_{2}+1\right)\right)}\times\\
 & \times\frac{\left(\xi\left(z_{2}+2\right)\left(-\bar{z}_{2}\left(e^{\Delta y}+1\right)(2\xi+1)-\bar{z}_{2}^{1}+1\right)-\left((\xi+1)z_{2}\left(e^{\Delta y}z_{2}+2\right)\right)(x+\xi)\right)}{\left(z_{2}+e^{\Delta y}\bar{z}_{1}\right)\left(e^{\Delta y}z_{2}+\bar{z}_{1}\right)\left(\xi\left(e^{\Delta y}\bar{z}_{2}+z_{2}\bar{z}_{2}+1\right)+\bar{z}_{2}(1+\xi)\left(e^{\Delta y}+1\right)\left(x-\xi\right)\right)},\nonumber 
\end{align}

\begin{align}
b_{1} & =\frac{2e^{2\Delta y}\left(e^{\Delta y}+1\right)\xi\left(\xi^{2}+x(1+\xi)-\xi z_{1}\right)}{m_{Q}^{5}z_{1}z_{2}\left(e^{\Delta y}\left(2\xi+z_{1}+1\right)+2\xi+1\right)\left(e^{\Delta y}\xi\left(1-z_{1}^{2}\right)+\left(e^{\Delta y}+1\right)(1+\xi)z_{1}(x+\xi)\right)}\times\\
 & \times\frac{\left(z_{1}\left(\left(e^{\Delta y}+1\right)(\xi+1)(\xi-x)+\left(e^{\Delta y}-1\right)\xi z_{2}\right)+\left(e^{\Delta y}+1\right)\xi(2\xi+1)z_{2}\right)}{\left(e^{\Delta y}\bar{z}_{2}+z_{1}\right)\left(\left(e^{\Delta y}+1\right)\left(\xi^{2}+x(1+\xi)\right)+e^{\Delta y}\xi z_{1}+\xi\bar{z}_{2}\right)\left(\left(e^{\Delta y}+1\right)(\xi+1)(x-\xi)+\xi z_{2}\right)},\nonumber 
\end{align}
\begin{align}
B_{1} & =\frac{2e^{2\Delta y}\left(e^{\Delta y}+1\right)\xi\left((1+\xi)(x+\xi)-\xi z_{1}\right)}{m_{Q}^{5}z_{1}\left(e^{\Delta y}\left(z_{2}-1\right)-z_{1}\right)\left(z_{1}\left(e^{\Delta y}\left(2\xi+z_{1}+1\right)+2\xi+1\right)-e^{\Delta y}\right)\left(\left(e^{\Delta y}+1\right)(\xi+1)(x+\xi)-e^{\Delta y}\xi z_{1}\right)}\times\\
 & \times\frac{\left(\xi\bar{z}_{2}\left(2\left(e^{\Delta y}(\xi+1)+\xi\right)+\left(e^{\Delta y}-1\right)z_{1}\right)+(1+\xi)\left(e^{\Delta y}+1\right)\left(z_{1}+1\right)\left(x-\xi\right)\right)}{\left(\left(e^{\Delta y}+1\right)(\xi+1)x+\xi\left(e^{\Delta y}\left(\xi+z_{1}\right)+\xi+\bar{z}_{2}\right)\right)\left(\left(e^{\Delta y}+1\right)(\xi+1)z_{2}(x-\xi)-\xi\left(1-z_{2}^{2}\right)\right)},\nonumber 
\end{align}
\begin{align}
b_{2} & =\frac{2e^{\Delta y}\left(e^{2\Delta y}-1\right)\xi}{m_{Q}^{5}z_{1}z_{2}\left(e^{\Delta y}\bar{z}_{1}+z_{2}\right)\left(e^{\Delta y}z_{2}+\bar{z}_{1}\right)\left(e^{\Delta y}\xi z_{1}-\left(e^{\Delta y}+1\right)(\xi+1)(\xi+x)\right)}\times\\
 & \times\frac{\left(\xi^{2}z_{2}\left(\bar{z}_{1}e^{\Delta y}+2\xi+2\right)-\xi(\xi+1)\left(e^{\Delta y}z_{1}-\left(e^{\Delta y}+1\right)z_{2}\right)(x-\xi)\right)}{\left(e^{\Delta y}\xi z_{1}-\left(e^{\Delta y}+1\right)\left(-\xi^{2}+x(1+\xi)\right)\right)\left(\left(e^{\Delta y}+1\right)(1+\xi)(x-\xi)+\xi z_{2}\right)},\nonumber 
\end{align}

\begin{align}
 & B_{2}=\frac{2\xi e^{\Delta y}\left(e^{2\Delta y}-1\right)}{m_{Q}^{5}\left(e^{\Delta y}\bar{z}_{1}+z_{2}\right)\left(e^{\Delta y}z_{2}+\bar{z}_{1}\right)\left(e^{\Delta y}\xi\left(1-z_{1}^{2}\right)+\left(e^{\Delta y}+1\right)(1+\xi)z_{1}(x+\xi)\right)}\times\\
 & \times\left[\left(e^{\Delta y}+1\right)(1+\xi)^{2}z_{1}(x-\xi)^{2}+\xi(\xi+1)\left(\left(e^{\Delta y}+1\right)z_{1}\left(2\xi+z_{2}+1\right)+e^{\Delta y}\left(1-z_{1}^{2}\right)\right)(x-\xi)\frac{}{}\right.\nonumber \\
 & \left.+\xi^{2}\bar{z}_{2}\left(2e^{\Delta y}\xi-2z_{1}\left(1+\xi\right)-e^{\Delta y}z_{1}\bar{z}_{1}\right)\frac{}{}\right]\times\nonumber \\
 & \times\left[\left(e^{\Delta y}\xi\left(1-z_{1}^{2}\right)+\left(e^{\Delta y}+1\right)z_{1}\left(x(1+\xi)-\xi^{2}\right)\right)\left(\left(e^{\Delta y}+1\right)(1+\xi)z_{2}(x-\xi)-\xi\left(1-z_{2}^{2}\right)\right)\right]^{-1},\nonumber 
\end{align}
\begin{align}
 & b_{3}=\frac{2e^{2\Delta y}\left(e^{\Delta y}+1\right)\xi\left((1+\xi)(x-\xi)+\xi\left(z_{2}-2z_{1}\right)\right)}{m_{Q}^{5}z_{1}\left(e^{\Delta y}z_{2}+\bar{z}_{1}\right)\left(\left(e^{\Delta y}+1\right)\left(2\xi+1\right)+z_{1}e^{\Delta y}\right)\left(e^{\Delta y}\xi z_{1}-\left(e^{\Delta y}+1\right)\left(x(1+\xi)-\xi^{2}\right)\right)}\times\\
 & \times\frac{\left(e^{\Delta y}+1\right)\left(e^{\Delta y}\left(x(1+\xi)-\xi^{2}\right)+\xi(2\xi+1)\right)-e^{\Delta y}\left(e^{\Delta y}-1\right)\xi z_{1}}{\left(e^{\Delta y}+1\right)(1+\xi)\left(e^{\Delta y}z_{2}+1-2z_{1}\right)(x-\xi)+\xi\left(e^{\Delta y}\left(z_{2}^{2}-4z_{1}\bar{z}_{1}\right)+z_{2}\left(1-2z_{1}\right)\left(e^{2\Delta y}+1\right)\right)}\times\nonumber \\
 & \times\left(\left(e^{\Delta y}+1\right)\left(x(1+\xi)-\xi^{2}\right)-\xi\left(e^{\Delta y}z_{1}+\bar{z}_{2}\right)\right)^{-1},\nonumber 
\end{align}

\begin{align}
B_{3} & =-\frac{2e^{2\Delta y}\left(e^{\Delta y}+1\right)\xi\left(x(1+\xi)+\xi\left(z_{2}-2z_{1}-\xi\right)\right)}{m_{Q}^{5}\left(e^{\Delta y}z_{2}+1-2z_{1}\right)\left(e^{\Delta y}z_{2}+\bar{z}_{1}\right)\left(z_{1}\left(\left(e^{\Delta y}+1\right)\left(2\xi+1\right)+z_{1}e^{\Delta y}\right)-e^{\Delta y}\right)}\times\label{eq:B3}\\
 & \times\frac{\left(e^{\Delta y}\left(e^{\Delta y}+1\right)(\xi+1)\left(z_{1}+1\right)(x+\xi)-\xi\left(e^{\Delta y}-1\right)\left(\left(e^{\Delta y}+1\right)(2\xi+1)z_{1}-e^{\Delta y}\left(1-z_{1}^{2}\right)\right)\right)}{\left(\left(e^{\Delta y}+1\right)\left(x(1+\xi)-\xi^{2}\right)-\xi\left(e^{\Delta y}z_{1}+\bar{z}_{2}\right)\right)\left(\left(e^{\Delta y}+1\right)\left(x(1+\xi)-\xi^{2}\right)-\xi\left(2e^{\Delta y}z_{1}+\bar{z}_{2}\right)\right)}\times\nonumber \\
 & \times\left(\left(e^{\Delta y}+1\right)z_{1}\left(x(1+\xi)-\xi^{2}\right)+e^{\Delta y}\xi\left(1-z_{1}^{2}\right)\right)^{-1}.\nonumber 
\end{align}

Similar to the case of quark coefficient functions, all the contributions
can have up to three non-coinciding poles as a function of variable
$x$. The position of the poles depends on the kinematics of the produced
$D$-mesons (variables $\Delta y,\xi$), as well as the light-cone
fractions $z_{1},z_{2}$ carried by the quarks inside the $D$-mesons.
As we mentioned in the main text, all the convolutions (integrals)
which include these coefficient functions, should be interpreted in
the principal value sense, using the standard $\xi\to\xi-i0$ prescription~\cite{DVMPcc1}
for contour deformation near the poles.


\begin{thebibliography}{10}
\bibitem{Diehl:2000xz} M.~Diehl, T.~Feldmann, R.~Jakob and P.~Kroll,
Nucl.\ Phys.\ B \textbf{596}, 33 (2001) {[}Erratum-ibid.\ B \textbf{605},
647 (2001){]} {[}arXiv:hep-ph/0009255{]}.

\bibitem{Goeke:2001tz} K.~Goeke, M.~V.~Polyakov and M.~Vanderhaeghen,
Prog.\ Part.\ Nucl.\ Phys.\ \textbf{47}, 401 (2001) {[}arXiv:hep-ph/0106012{]}.

\bibitem{Diehl:2003ny} M.~Diehl, Phys.\ Rept.\ \textbf{388}, 41
(2003) {[}arXiv:hep-ph/0307382{]}.

\bibitem{Guidal:2013rya}M.~Guidal, H.~Moutarde and M.~Vanderhaeghen,
``\emph{Generalized Parton Distributions in the valence region from
Deeply Virtual Compton Scattering},'' Rept. Prog. Phys. \textbf{76}
(2013), 066202 {[}arXiv:1303.6600 {[}hep-ph{]}{]}.

\bibitem{Boer:2011fh} D.~Boer, M.~Diehl, R.~Milner, R.~Venugopalan,
W.~Vogelsang, D.~Kaplan, H.~Montgomery, S.~Vigdor, A.~Accardi
and E.~C.~Aschenauer, \emph{et al}. ``\emph{Gluons and the quark
sea at high energies: Distributions, polarization, tomography},''
{[}arXiv:1108.1713 {[}nucl-th{]}{]}.

\bibitem{Burkert:2022hjz} V.~Burkert, L.~Elouadrhiri, A.~Afanasev,
J.~Arrington, M.~Contalbrigo, W.~Cosyn, A.~Deshpande, D.~Glazier,
X.~Ji and S.~Liuti, \emph{et al}. ``\emph{Precision Studies of
QCD in the Low Energy Domain of the EIC},'' {[}arXiv:2211.15746 {[}nucl-ex{]}{]}.

\bibitem{Egerer:2021ymv}C.~Egerer~\emph{et al}. {[}HadStruc{]},
``\emph{Towards high-precision parton distributions from lattice
QCD via distillation},'' JHEP \textbf{11} (2021), 148 {[}arXiv:2107.05199
{[}hep-lat{]}{]}.

\bibitem{Karpie:2021pap}J.~Karpie \emph{et al}. {[}HadStruc{]},
``\emph{The continuum and leading twist limits of parton distribution
functions in lattice QCD},'' JHEP \textbf{11} (2021), 024 {[}arXiv:2105.13313
{[}hep-lat{]}{]}.

\bibitem{Bhattacharya:2022aob}S.~Bhattacharya, K.~Cichy, M.~Constantinou,
J.~Dodson, X.~Gao, A.~Metz, S.~Mukherjee, A.~Scapellato, F.~Steffens
and Y.~Zhao, ``\emph{Generalized parton distributions from lattice
QCD with asymmetric momentum transfer: Unpolarized quarks,}'' Phys.
Rev. D \textbf{106} (2022) no.11, 114512 {[}arXiv:2209.05373 {[}hep-lat{]}{]}.

\bibitem{Bhattacharya:2023ays}S.~Bhattacharya, K.~Cichy, M.~Constantinou,
X.~Gao, A.~Metz, J.~Miller, S.~Mukherjee, P.~Petreczky, F.~Steffens
and Y.~Zhao, ``\emph{Moments of proton GPDs from the OPE of nonlocal
quark bilinears up to NNLO},'' Phys. Rev. D \textbf{108} (2023) no.1,
014507 {[}arXiv:2305.11117 {[}hep-lat{]}{]}.

\bibitem{Bhattacharya:2023nmv}S.~Bhattacharya, K.~Cichy, M.~Constantinou,
J.~Dodson, A.~Metz, A.~Scapellato and F.~Steffens, ``\emph{Chiral-even
axial twist-3 GPDs of the proton from lattice QCD},'' Phys. Rev.
D \textbf{108} (2023) no.5, 054501 {[}arXiv:2306.05533 {[}hep-lat{]}{]}.

\bibitem{Kumericki:2016ehc} K.~Kumericki, S.~Liuti and H.~Moutarde,
``\emph{GPD phenomenology and DVCS fitting: Entering the high-precision
era},'' Eur. Phys. J. A \textbf{52} (2016) no.6, 157 {[}arXiv:1602.02763
{[}hep-ph{]}{]}.

\bibitem{Pire:2015iza}B.~Pire and L.~Szymanowski, \emph{``Neutrino-production
of a charmed meson and the transverse spin structure of the nucleon,''}
Phys. Rev. Lett. \textbf{115} (2015) no.9, 092001 {[}arXiv:1505.00917
{[}hep-ph{]}{]}.

\bibitem{Pire:2017lfj}B.~Pire, L.~Szymanowski and J.~Wagner, \emph{``Exclusive
neutrino-production of a charmed meson,''} Phys. Rev. D \textbf{95}
(2017) no.9, 094001 {[}arXiv:1702.00316 {[}hep-ph{]}{]}.

\bibitem{Pire:2017yge}B.~Pire and L.~Szymanowski, \emph{``Exclusive
neutrino production of a charmed vector meson and transversity gluon
generalized parton distributions,''} Phys. Rev. D \textbf{96} (2017)
no.11, 114008 {[}arXiv:1711.04608 {[}hep-ph{]}{]}.

\bibitem{Pire:2021dad}B.~Pire, L.~Szymanowski and J.~Wagner, \emph{``Charged
current electroproduction of a charmed meson at an electron-ion collider,''}
Phys. Rev. D \textbf{104} (2021) no.9, 094002 {[}arXiv:2104.04944
{[}hep-ph{]}{]}.

\bibitem{GPD2x3:9} G. Duplan\v{c}i\'{c}, S. Nabeebaccus, K. Passek-Kumeri\v{c}ki,
B. Pire, L. Szymanowski and S. Wallon, \emph{``Accessing chiral-even
quark generalised parton distributions in the exclusive photoproduction
of a $\gamma\pi^{\pm}$ pair with large invariant mass in both fixed-target
and collider experiments,''} {[}arXiv:2212.00655 {[}hep-ph{]}{]}.

\bibitem{GPD2x3:8} G. Duplan\v{c}i\'{c}, K. Passek-Passek-Kumeri\v{c}ki,
B. Pire, L. Szymanowski and S. Wallon, JHEP 11 (2018) 179 {[}arXiv:1809.08104
{[}hep-ph{]}{]}.

\bibitem{GPD2x3:7} R. Boussarie, B. Pire, L. Szymanowski and S. Wallon,
JHEP 02 (2017) 054 {[}arXiv:1609.03830 {[}hep-ph{]}{]}.

\bibitem{GPD2x3:6} W. Cosyn and B. Pire, Phys. Rev. D 103 (2021)
114002 {[}arXiv:2103.01411 {[}hep-ph{]}{]}.

\bibitem{GPD2x3:5} A. Pedrak, B. Pire, L. Szymanowski and J. Wagner,
Phys. Rev. D 101 (2020) 114027 {[}arXiv:2003.03263 {[}hep-ph{]}{]}.

\bibitem{GPD2x3:4} B. Pire, L. Szymanowski and S. Wallon, Phys. Rev.
D \textbf{101} (2020) 074005 {[}arXiv:1912.10353 {[}hep-ph{]}{]}.

\bibitem{GPD2x3:3} A. Pedrak, B. Pire, L. Szymanowski and J. Wagner,
Phys. Rev. D \textbf{96} (2017) 074008 {[}arXiv:1708.01043 {[}hep-ph{]}{]}.

\bibitem{GPD2x3:2} M. El Beiyad, B. Pire, M. Segond, L. Szymanowski
and S. Wallon, Phys. Lett. B \textbf{688} (2010) 154 {[}arXiv:1001.4491
{[}hep-ph{]}{]}.

\bibitem{GPD2x3:1} D.Y. Ivanov, B. Pire, L. Szymanowski and O.V.
Teryaev, Phys. Lett. B \textbf{550} (2002) 65 {[}arXiv:hep-ph/0209300{]}.

\bibitem{Duplancic:2022wqn} G.~Duplan\v{c}i\'{c}, S.~Nabeebaccus,
K.~Passek-Kumeri\v{c}ki, B.~Pire, L.~Szymanowski and S.~Wallon,
``\emph{Accessing GPDs through the exclusive photoproduction of a
photon-meson pair with a large invariant mass},'' {[}arXiv:2212.01034
{[}hep-ph{]}{]}.

\bibitem{ElBeiyad:2010pji}M.~El Beiyad, B.~Pire, M.~Segond, L.~Szymanowski
and S.~Wallon, ``\emph{Photoproduction of a pi rhoT pair with a
large invariant mass and transversity generalized parton distribution},''
Phys. Lett. B \textbf{688} (2010), 154-167 {[}arXiv:1001.4491 {[}hep-ph{]}{]}.

\bibitem{Boussarie:2016qop}R.~Boussarie, B.~Pire, L.~Szymanowski
and S.~Wallon, \emph{``Exclusive photoproduction of a $\gamma\,\rho$
pair with a large invariant mass,''} JHEP \textbf{02} (2017), 054
{[}erratum: JHEP \textbf{10} (2018), 029{]} {[}arXiv:1609.03830 {[}hep-ph{]}{]}.

\bibitem{Pedrak:2017cpp}A.~Pedrak, B.~Pire, L.~Szymanowski and
J.~Wagner, ``\emph{Hard photoproduction of a diphoton with a large
invariant mass},'' Phys. Rev. D \textbf{96} (2017) no.7, 074008 {[}erratum:
Phys. Rev. D \textbf{100} (2019) no.3, 039901{]} {[}arXiv:1708.01043
{[}hep-ph{]}{]}.

\bibitem{Pedrak:2020mfm}A.~Pedrak, B.~Pire, L.~Szymanowski and
J.~Wagner, ``\emph{Electroproduction of a large invariant mass photon
pair},'' Phys. Rev. D \textbf{101} (2020) no.11, 114027 {[}arXiv:2003.03263
{[}hep-ph{]}{]}.

\bibitem{GPD2x3:10} J.-W. Qiu and Z. Yu, ``\emph{Exclusive production
of a pair of high transverse momentum photons in pion-nucleon collisions
for extracting generalized parton distributions}'', {[}arXiv:2205.07846
{[}hep-ph{]}{]}.

\bibitem{GPD2x3:11} J.-W. Qiu and Z. Yu, ``\emph{Single diffractive
hard exclusive processes for the study of generalized parton distributions}'',
{[}arXiv:2210.07995 {[}hep-ph{]}{]}.

\bibitem{LehmannDronke:1999vvq}B.~LehmannDronke, P.~V.~Pobylitsa,
M.~V.~Polyakov, A.~Schafer and K.~Goeke, ``\emph{Hard diffractive
electroproduction of two pions},'' Phys. Lett. B \textbf{475} (2000),
147-156 {[}arXiv:hep-ph/9910310 {[}hep-ph{]}{]}.

\bibitem{LehmannDronke:2000hlo}B.~Lehmann-Dronke, A.~Schafer, M.~V.~Polyakov
and K.~Goeke, ``\emph{Angular distributions in hard exclusive production
of pion pairs},'' Phys. Rev. D \textbf{63} (2001), 114001 {[}arXiv:hep-ph/0012108
{[}hep-ph{]}{]}.

\bibitem{Clerbaux:2000hb}B.~Clerbaux and M.~V.~Polyakov, ``\emph{Partonic
structure of pi and rho mesons from data on hard exclusive production
of two pions off nucleon},'' Nucl. Phys. A \textbf{679} (2000), 185-195
{[}arXiv:hep-ph/0001332 {[}hep-ph{]}{]}.

\bibitem{Goncalves:2015sfy}V.~P.~Goncalves, B.~D.~Moreira and
F.~S.~Navarra, ``\emph{Double vector meson production in $\gamma\gamma$
interactions at hadronic colliders},'' Eur. Phys. J. C \textbf{76}
(2016) no.3, 103 {[}arXiv:1512.07482 {[}hep-ph{]}{]}.

\bibitem{Goncalves:2019txs} V.P.~Goncalves and R.~Palota da Silva,
``\emph{Exclusive and diffractive quarkonium - pair production at
the LHC and FCC},'' Phys. Rev. D \textbf{101} (2020) no.3, 034025
{[}arXiv:1912.02720 {[}hep-ph{]}{]}.

\bibitem{Goncalves:2006hu}V.~P.~Goncalves and M.~V.~T.~Machado,
``Dipole model for double meson production in two-photon interactions
at high energies,'' Eur. Phys. J. C \textbf{49} (2007), 675-684 {[}arXiv:hep-ph/0605304
{[}hep-ph{]}{]}.

\bibitem{Baranov:2012vu}S.~Baranov, A.~Cisek, M.~Klusek-Gawenda,
W.~Schafer and A.~Szczurek, ``\emph{The $\gamma\gamma\to J/\psi J/\psi$
reaction and the $J/\psi J/\psi$ pair production in exclusive ultraperipheral
ultrarelativistic heavy ion collisions},'' Eur. Phys. J. C \textbf{73}
(2013) no.2, 2335 {[}arXiv:1208.5917 {[}hep-ph{]}{]}.

\bibitem{Yang:2020xkl}H.~Yang, Z.~Q.~Chen and C.~F.~Qiao, ``\emph{NLO
QCD corrections to exclusive quarkonium-pair production in photon-photon
collision},'' Eur. Phys. J. C \textbf{80} (2020) no.9, 806.

\bibitem{Goncalves:2016ybl}V.~P.~Goncalves, B.~D.~Moreira and
F.~S.~Navarra, ``\emph{Double vector meson production in photon-hadron
interactions at hadronic colliders,}'' Eur. Phys. J. C \textbf{76}
(2016) no.7, 388 {[}arXiv:1605.05840 {[}hep-ph{]}{]}.

\bibitem{Andrade:2022rbn}S.~Andradé, M.~Siddikov and I.~Schmidt,
``\emph{Exclusive photoproduction of heavy quarkonia pairs},'' {[}arXiv:2202.03288
{[}hep-ph{]}{]}.

\bibitem{Siddikov:2022bku}M.~Siddikov and I.~Schmidt, ``\emph{Exclusive
production of quarkonia pairs in collinear factorization framework},''
Phys. Rev. D \textbf{107} (2023) no.3, 034037 {[}arXiv:2212.14019
{[}hep-ph{]}{]}.

\bibitem{Luszczak:2011js}M.~Luszczak and A.~Szczurek, ``\emph{Exclusive
$D\bar{D}$ meson pair production in peripheral ultrarelativistic
heavy ion collisions},'' Phys. Lett. B \textbf{700} (2011), 116-121
{[}arXiv:1103.4268 {[}nucl-th{]}{]}.

\bibitem{Accardi:2012qut}A.~Accardi \emph{et al.}, Eur.~Phys.doi:10.1016/j.physletb.2009.11.040~J.~A
\textbf{52}, no. 9, 268 (2016) {[}arXiv:1212.1701 {[}nucl-ex{]}{]}.

\bibitem{DOEPR}Press release at the website of the United States
Department of Energy: \href{https://www.energy.gov/articles/us-department-energy-selects-brookhaven-national-laboratory-host-major-new-nuclear-physics}{https://www.energy.gov/articles/us-department-energy-selects-brookhaven-national-laboratory-host-major-new-nuclear-physics}.

\bibitem{BNLPR}Press-release at the website of the Brookhaven National
Laboratory (BNL): \href{https://www.bnl.gov/newsroom/news.php?a=116998}{https://www.bnl.gov/newsroom/news.php?a=116998}.

\bibitem{AbdulKhalek:2021gbh} R.~Abdul Khalek \emph{et al}. ``\emph{Science
Requirements and Detector Concepts for the Electron-Ion Collider:
EIC Yellow Report},'' {[}arXiv:2103.05419 {[}physics.ins-det{]}{]}.

\bibitem{Diehl:1999cg}M.~Diehl, T.~Gousset and B.~Pire, ``\emph{Polarization
in deeply virtual meson production},'' {[}arXiv:hep-ph/9909445 {[}hep-ph{]}{]}.

\bibitem{ZEUS:1998xpo}J.~Breitweg \emph{et al}. {[}ZEUS{]}, ``\emph{Exclusive
electroproduction of $\rho^{0}$ and $J/\psi$ mesons at HERA},''
Eur. Phys. J. C \textbf{6} (1999), 603-627 {[}arXiv:hep-ex/9808020
{[}hep-ex{]}{]}.

\bibitem{Neubert:1993mb} M.~Neubert, ``\emph{Heavy quark symmetry},''
Phys. Rept. \textbf{245} (1994), 259-396 {[}arXiv:hep-ph/9306320 {[}hep-ph{]}{]}.

\bibitem{Goloskokov:2013mba}S.~V.~Goloskokov and P.~Kroll, \emph{``Transversity
in exclusive vector-meson leptoproduction,''} Eur. Phys. J. C \textbf{74}
(2014), 2725 {[}arXiv:1310.1472 {[}hep-ph{]}{]}.

\bibitem{Belitsky:2001ns} A.~V.~Belitsky, D.~Mueller and A.~Kirchner,
Nucl.\ Phys.\ B \textbf{629}, 323 (2002) {[}arXiv:hep-ph/0112108{]}.

\bibitem{Belitsky:2005qn} A.~V.~Belitsky and A.~V.~Radyushkin,
Phys.\ Rept.\ \textbf{418}, 1 (2005) {[}arXiv:hep-ph/0504030{]}.

\bibitem{Goloskokov:2006hr}S.~V.~Goloskokov and P.~Kroll, 
 Eur.~Phys.~J.~C \textbf{50}, 829 (2007) {[}hep-ph/0611290{]}.

\bibitem{Goloskokov:2007nt}S.~V.~Goloskokov and P.~Kroll, 
 Eur.~Phys.~J.~C \textbf{53}, 367 (2008) {[}arXiv:0708.3569 {[}hep-ph{]}{]}.

\bibitem{Goloskokov:2008ib}S.~V. Goloskokov and P.~Kroll, Eur.~Phys.~J.~C
\textbf{59} (2009) 809 {[}arXiv:0809.4126 {[}hep-ph{]}{]}.

\bibitem{Goloskokov:2009ia}S.~V.~Goloskokov and P.~Kroll, Eur.~Phys.~J.~C
\textbf{65}, 137 (2010) {[}arXiv:0906.0460 {[}hep-ph{]}{]}.

\bibitem{Goloskokov:2011rd}S.~V.~Goloskokov and P.~Kroll, Eur.~Phys.~J.~A
\textbf{47}, 112 (2011) {[}arXiv:1106.4897 {[}hep-ph{]}{]}.

\bibitem{DVMPcc1} D.Yu. Ivanov, A. Schafer, L. Szymanowski and G.
Krasnikov, Eur. Phys. J. C\textbf{ 34} (2004) 297, {[}arXiv:hep-ph/0401131{]}.

\bibitem{DVMPcc2} M. Vanttinen and L. Mankiewicz, Phys. Lett. B \textbf{440}
(1998) 157, {[}arXiv:hep-ph/9807287{]}.

\bibitem{DVMPcc3} J. Koempel, P. Kroll, A. Metz and J. Zhou, Phys.
Rev. D \textbf{85} (2012) 051502(R) {[}arXiv:1112.1334 {[}hep-ph{]}{]}.

\bibitem{DVMPcc4} Z.L. Cui, M.C. Hu and J.P. Ma, Eur. Phys. J C \textbf{79}
(2019), 812 {[}aXiv:1804.05293 {[}hep-ph{]}{]}.

\bibitem{Accardi:2023chb}A.~Accardi, P.~Achenbach, D.~Adhikari,
A.~Afanasev, C.~S.~Akondi, N.~Akopov, M.~Albaladejo, H.~Albataineh,
M.~Albrecht and B.~Almeida-Zamora, \emph{et al}. ``\emph{Strong
Interaction Physics at the Luminosity Frontier with 22 GeV Electrons
at Jefferson Lab},'' {[}arXiv:2306.09360 {[}nucl-ex{]}{]}.

\bibitem{LightQuarkDA:1} V.M. Braun and I.B. Filyanov, Z. Phys. C48
(1990) 239,

\bibitem{LightQuarkDA:2} P. Ball, JHEP 01 (1999) 010.

\bibitem{LightQuarkDA:3} P. Ball and V.M. Braun, Phys.Rev. D54 (1996)
2182.

\bibitem{LightQuarkDA:4} P. Ball, V. M. Braun, Y. Koike and K. Tanaka,
Nucl.Phys. B529 (1998) 323.

\bibitem{LightQuarkDA:5} P. Ball and V.M. Braun, Nucl.Phys. B543
(1999) 201.

\bibitem{Baek:1994kj}M.~S.~Baek, S.~Y.~Choi and H.~S.~Song,
``\emph{Exclusive heavy meson pair production at large recoil},''
Phys. Rev. D \textbf{50} (1994), 4363-4371.

\bibitem{Beneke:2023nmj}M.~Beneke, G.~Finauri, K.~K.~Vos and
Y.~Wei, ``\emph{QCD Light-Cone Distribution Amplitudes of Heavy
Mesons from boosted HQET},'' {[}arXiv:2305.06401 {[}hep-ph{]}{]}.

\bibitem{Georgi:1990um}~H.~Georgi, ``\emph{An Effective Field
Theory for Heavy Quarks at Low-energies},'' Phys. Lett. B \textbf{240}
(1990), 447-450.

\bibitem{Isgur:1989vq}~N. lsgur and M.B. Wise, ``\emph{Weak Decays
of Heavy Mesons in the Static Quark Approximation}'' Phys. Left.
B \textbf{232} (1989) 113.

\bibitem{Zhong:2022ugk} T.~Zhong, D.~Huang and H.~B.~Fu, ``\emph{Revisiting
D-meson twist-2, 3 distribution amplitudes},'' Chin. Phys. C \textbf{47}
(2023) no.5, 053104 {[}arXiv:2212.04641 {[}hep-ph{]}{]}.

\bibitem{Dhiman:2019ddr}N.~Dhiman, H.~Dahiya, C.~R.~Ji and H.~M.~Choi,
``\emph{Twist-2 Pseudoscalar and Vector Meson Distribution Amplitudes
in Light-Front Quark Model with Exponential-type Confining Potential},''
Phys. Rev. D~\textbf{100} (2019) no.1, 014026 {[}arXiv:1902.09160
{[}hep-ph{]}{]}.

\bibitem{Zuo:2006re}F.~Zuo and T.~Huang, ``\emph{$B_{c}$ ($B$)$\to D\ell\bar{nu}$
form-factors in light-cone sum rules and the $D$ meson distribution
amplitude},'' Chin. Phys. Lett. \textbf{24} (2007), 61-64 {[}arXiv:hep-ph/0611113
{[}hep-ph{]}{]}.

\bibitem{Lepage:1980fj}G. P. Lepage and S. J. Brodsky, ``\emph{Exclusive
processes in perturbative quantum chromodynamics}'', Phys. Rev. D
\textbf{22} (1980) 2157.

\bibitem{Brodsky:1997de}S.~J.~Brodsky, H.~C.~Pauli and S.~S.~Pinsky,
``\emph{Quantum chromodynamics and other field theories on the light
cone},'' Phys. Rept. \textbf{301} (1998), 299-486 {[}arXiv:hep-ph/9705477
{[}hep-ph{]}{]}.

\bibitem{Ji:1998pc} X.~D.~Ji, J.\ Phys.\ G \textbf{24}, 1181
(1998) {[}arXiv:hep-ph/9807358{]}.

\bibitem{FeynCalc1}V. Shtabovenko, R. Mertig and F. Orellana, Comput.
Phys. Commun., 207, 432-444, 2016, arXiv:1601.01167.

\bibitem{FeynCalc2} R. Mertig, M. Böhm, and A. Denner, Comput. Phys.
Commun., 64, 345-359, 1991.

\end{thebibliography}
\end{document}